\pdfoutput=1
\documentclass[fleqn,usenatbib]{mnras}

\usepackage{newtxtext,newtxmath}

\usepackage[T1]{fontenc}
\usepackage{ae,aecompl}

\usepackage{graphicx}	
\usepackage{amsmath}	
\usepackage{amssymb}	

\usepackage{txfonts}

\usepackage{multirow}

\defcitealias{Dwarkadas2010a}{D10}

\title[X-raying SN1996cr in High Resolution]{The Exceptional X-ray Evolution of SN\,1996cr in High Resolution}

\author[J. Quirola-V\'asquez et al.]{
J. Quirola-V\'asquez,$^{1,2}$\thanks{E-mail: jquirola@astro.puc.cl}
F. E. Bauer,$^{2,1,3}$
V. V. Dwarkadas,$^{4}$ 
C. Badenes,$^{5}$
\newauthor{W. N. Brandt},$^{6,7,8}$
T. Nymark,$^{9}$
D. Walton$^{10}$
\\
$^{1}$Millennium Institute of Astrophysics (MAS), Nuncio Monse$\tilde{n}$or S\'otero Sanz 100, Providencia, Santiago, Chile\\
$^{2}$Instituto de Astrof\'isica, Pontificia Universidad Cat\'olica de Chile, Casilla 306, Santiago 22, Chile\\
$^{3}$Space Science Institute, 4750 Walnut Street, Suite 205, Boulder, Colorado 80301, USA\\
$^{4}$Department of Astronomy and Astrophysics, University of Chicago, 5640 S Ellis Ave, Chicago, IL 60637, USA\\
$^{5}$Department of Physics and Astronomy and Pittsburgh Particle Physics, Astrophysics and Cosmology Center, University of Pittsburgh,\\ Pittsburgh, PA 15260, USA\\
$^{6}$Department of Astronomy \& Astrophysics, 525 Davey Laboratory, The Pennsylvania State University, University Park, PA 16802, USA\\ 
$^{7}$Institute for Gravitation and the Cosmos, The Pennsylvania State University, University Park, PA 16802, USA\\
$^{8}$Department of Physics, 104 Davey Laboratory, The Pennsylvania State University, University Park, PA 16802, USA\\
$^{9}$Vetenskapens Hus, Kungliga Tekniska H\"{o}gskolan, SE-100 44 Stockholm\\
$^{10}$Institute of Astronomy, University of Cambridge, Madingley Road, Cambridge, CB3 0HA, United Kingdom
}

\date{Accepted XXX. Received YYY; in original form ZZZ}

\pubyear{2019}

\begin{document}
\label{firstpage}
\pagerange{\pageref{firstpage}--\pageref{lastpage}}
\maketitle

\begin{abstract}
We present X-ray spectra spanning 18 years of evolution for SN\,1996cr, one of the five nearest SNe detected in the modern era. {\it Chandra} HETG exposures in 2000, 2004, and 2009 allow us to resolve spectrally the velocity profiles of Ne, Mg, Si, S, and Fe emission lines and monitor their evolution as tracers of the ejecta-circumstellar medium (CSM) interaction. To explain the diversity of X-ray line profiles, we explore several possible geometrical models. Based on the highest signal-to-noise 2009 epoch, we find that a polar geometry with two distinct opening angle configurations and internal obscuration can successfully reproduce all of the observed line profiles. The best fit model consists of two plasma components: (1) a mildly absorbed (2$\times$10$^{21}$\,cm$^{-2}$), cooler ($\approx$2\,keV) with high Ne, Mg, Si, and S abundances associated with a wide polar interaction region (half-opening angle $\approx$58$^{\circ}$); (2) a moderately absorbed (2$\times$10$^{22}$\,cm$^{-2}$), hotter ($\ga$20\,keV) plasma with high Fe abundances and strong internal obscuration associated with a narrow polar interaction region (half-opening angle $\approx$20$^{\circ}$). We extend this model to seven further epochs with lower signal-to-noise ratio and/or lower spectral-resolution between 2000-2018, yielding several interesting trends in absorption, flux, geometry and expansion velocity. We argue that the hotter and colder components are associated with reflected and forward shocks, respectively, at least at later epochs. We discuss the physical implications of our results and plausible explosion scenarios to understand the X-ray data of SN\,1996cr.
\end{abstract}




\section{Introduction}\label{sec:Intro}

Core-Collapse supernovae (CCSNe) are powerful astrophysical events, generated by the explosion and death of massive stars \citep[$M>8M_\odot$;][]{Baade1934,Hoyle1960,Woosley2002a,Branch2017}. As a fundamental component in the evolution of the Universe, they enrich the interstellar medium (ISM) with heavy elements that are critical for forming new generations of stars and planets. At the same time, these events provide a unique window to study the still poorly understood physical processes that occur during the final stages of massive stars' lives, via photoionization and shock interaction between the ejecta and the circumstellar material (CSM)

Type IIn SNe are a relatively rare subclass of CCSNe \citep[<10\%;][]{Eldridge2013} which exhibit strong narrow Hydrogen and Helium emission lines in their optical spectra \citep[e.g.,][]{Schlegel1990a, Filippenko1997a}. They are often associated with explosions that occur in dense CSM \citep[up to $n$$\sim$$10^{6}$ cm$^{-3}$; e.g.,][]{Fransson2014a, Chandra2015a, Dwarkadas2016a}, which were produced by stellar winds and outflows during previous evolutionary phases; progenitors of SNe IIn are typically thought to have mass-loss rates in the range of $\dot{M}{\sim}10^{-4}$ to $0.3\,M_\odot$\,yr$^{-1}$ in the decades prior to explosion \citep[e.g.,][]{Woosley2002a, Smith2014a}. Type IIn SNe are generally X-ray (and less frequently radio) bright, owing to the shock interaction between the ejecta and CSM. The X-rays arise from thermal processes, while the radio emission comes from non-thermal synchrotron emission. Moreover, because they are masked by strong ongoing CSM interaction \citep{Smith2014b}, type IIn SNe rarely exhibit a classical nebular phase with a clear radioactive decay tail.

We focus here on the nearby SN\,1996cr, which was initially discovered in the disk of the Circinus Galaxy by \emph{Chandra} X-ray Observatory \citep{Sambruna2001a, Bauer2001a} and later observed as a type IIn \citep{Bauer2008a}, although its explosion epoch is only loosely constrained between 1995-02-28 and 1996-03-16, and its type at early epochs is yet to be established. SN\,1996cr has remained bright at X-ray, optical, and radio wavelengths for nearly two decades, placing it amongst the remarkable handful of long-lived CCSNe attributed to strong ejecta-CSM interactions: e.g., SNs 1978K, 1979C, 1986J, 1988Z, 1993J, 2005kd, 2007bg, 2010jl, 2009ip, 1998S, and  1987A \citep[e.g., respectively,][]{Chandra2012b, Smith2014b, Leonard2000a, Margutti2017a, Dwarkadas2016a, Michael2002a, Salas2013,Zhekov2006a, Dewey2008a}. Due to its relative proximity at $d\approx$3.7\,Mpc, SN\,1996cr affords us an exceptional opportunity to study its features \citep{Bauer2008a, Dwarkadas2010a, Dewey2011a, Meunier2013a} and evolution in great detail. 

\begin{figure}
    \includegraphics[scale=0.5]{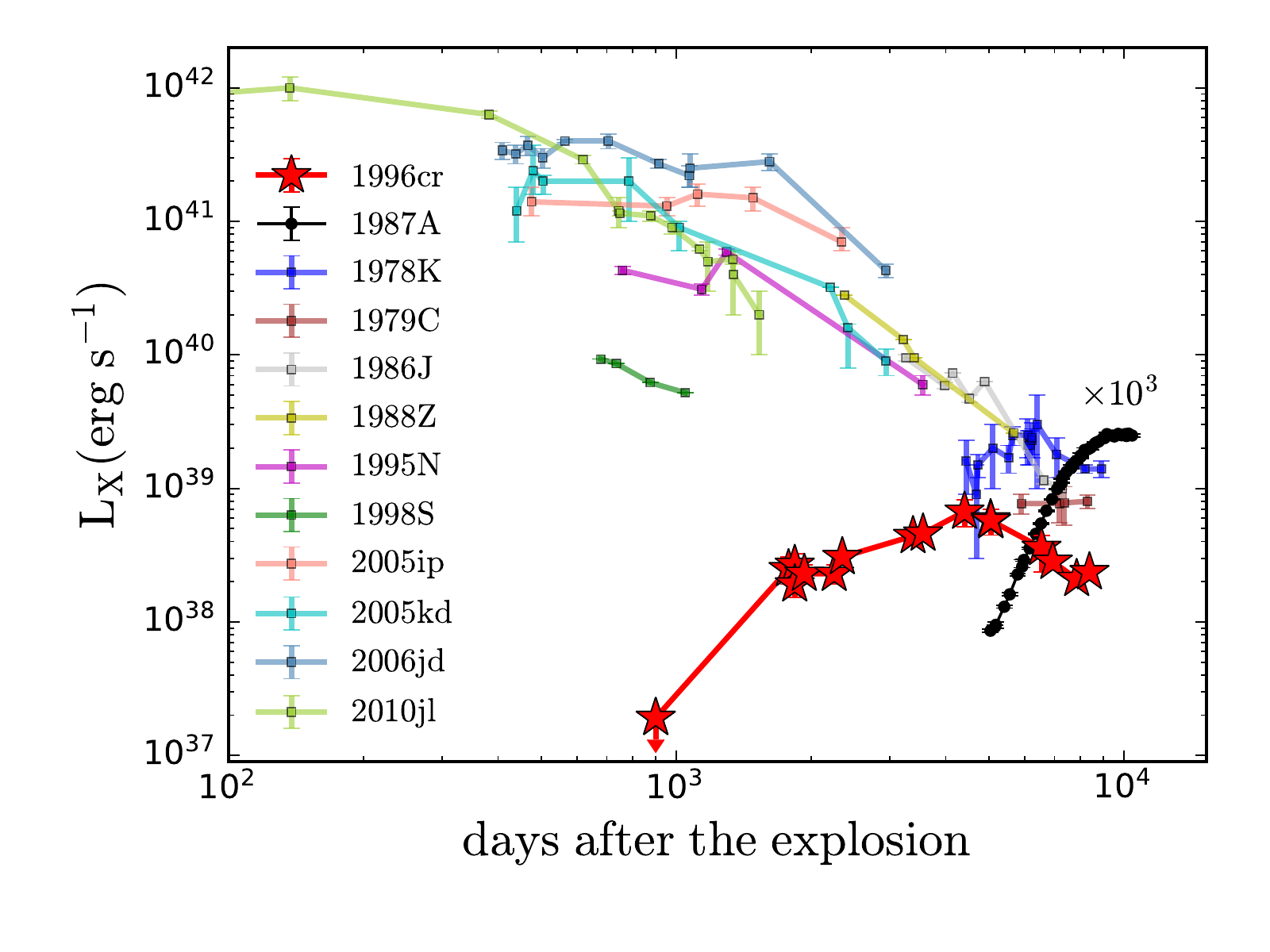}
    \caption{Representative X-ray light curves for a handful of strong CSM-interacting SNe (colour points), SN\,1987A (black points, multiplied by 10$^{3}$) and SN\,1996cr (\emph{red} stars). The data were taken from \citet{Ross2017a} and \citet[][SN\,1979C]{Immler2005}, with additional points added for SN\,1996cr from \emph{XMM-Newton}; no attempt has been made to regularize the X-ray band in which the luminosities from each SN are reported (e.g, some are reported as 0.5--2.0 keV, while others as 2.0--10 keV). The SNe appear to separate around $\sim$1000 days into early and late emitters. While several famous type IIn SNe start out strong and fade with time, SN\,1996cr increases with time, much like SN\,1987A. SNe\,1978K and 1979C may have had a similar evolution, as both exhibit flat X-ray evolution at late times, but lack early constraints to distinguish them as such.}
    \label{fig:light_curves}
\end{figure}

SN\,1996cr's radio emission shows an initial rise which is attributed to a combination of increasing CSM density and decreasing free-free absorption, which provides estimates of the CSM free electron density and hence insight into the ionization of SN\,1996cr's CSM \citep{Meunier2013a}. The X-ray luminosity likewise exhibits an initial increase with time, seen previously in the famous SN\,1987A \citep[e.g.,][]{Michael2002a, Frank2016a} and see later in SN\,2014C \citep{Margutti2017a}. This particular tendency, both in the radio and X-ray bands, is best explained by the interaction of ejecta material with a density enhancement (i.e., a dense shell) in the CSM; Fig.~\ref{fig:light_curves} compares SN\,1996cr's X-ray light curve to several other strong CSM-interacting SNe, including SN\,1987A (the latter multiplied by $10^3$ for easier comparison). The luminosity data used in Fig.~\ref{fig:light_curves} is a literature compilation with distinct energy ranges; for example, SN\,1996cr and SN\,1987A are shown for 0.5--2.0\,keV, while SN\,2010jl and SN\,2006jd, are for 0.2--10.0\,keV. The X-ray data for SN\,1996cr and other SNe are available in the Supernova X-ray Database\footnote{http://kronos.uchicago.edu/snex/} \citep[SNaX;][]{Ross2017a} and \citet[][SN\,1979C]{Immler2005}.

The optical spectrum of SN\,1996cr suggests that its progenitor was likely a massive star which shed many solar masses prior to explosion. Notably, the broad, high-velocity, multi-component Oxygen line complexes, in the optical range, hint at a possible concentric shell or ring-like morphology arising from the interaction of the forward shock and a dense shell produced by a wind-blown bubble \citep{Bauer2008a}. These unparalleled features supported a deep \emph{Chandra} campaign (PI Bauer) to obtain high resolution X-ray spectra taken between December 2008 and March 2009.

\citet[][hereafter D10]{Dwarkadas2010a} used hydrodynamical simulations to model the X-ray light curve and spectra at different epochs and thereby constrain the surrounding CSM structure of SN\,1996cr. Unlike most other Type IIns, SN 1996cr exploded in a low-density medium (see above), before interacting with a dense shell of material located at a distance of $d\lesssim0.03$ pc \citep[three times smaller than SN\,1987A's ring;][]{Dewey2012a}. \citetalias{Dwarkadas2010a} argued that the dense CSM shell likely formed due to the interaction of a fast Wolf-Rayet (WR, $M>$30$M_{\odot}$) or SN\,1987A-like blue supergiant (BSG, $M>$15--20$M_{\odot}$) wind \citep[$\dot{M}\sim$ 10$^{-5}$--10$^{-4}$$M_\odot$yr$^{-1}$;][]{Crowther2007a}, which turned on $\gtrsim$10$^{3}$--10$^{4}$\,yrs prior to explosion, and plowed up a previously existing red supergiant (RSG) wind ($\dot{M}\sim10^{-4}$ $M_\odot$yr$^{-1}$).\footnote{A luminous blue variable (LBV) stage was disfavored but could not be completely ruled out.} Under this scenario, SN\,1996cr should have presumably exploded as a SN type Ib/c or II peculiar \citep[e.g.][]{Stockdale2009a,Margutti2017a}.

In this paper, we revisit the X-ray spectral analysis of SN\,1996cr, focusing in particular on the unique high spectral resolution and high signal-to-noise data acquired by {\it Chandra} over the past two decades. The detailed velocity structure of strong X-ray emission lines detected in this object provides a window into the physical processes of young SNe and allow us to probe the ejecta dynamics and abundances with great detail \citep[e.g.,][]{Dewey2011a, Dewey2012a, Katsuda2014a}. We initially consider different geometrical and physical scenarios to explain the 2009 {\it Chandra} data, which offers the highest signal-to-noise and hence the firmest constraints. We then explore the physical nature and evolution of the SN by applying our best fit scenario to high-quality X-ray observations at other epochs (2000, 2001, 2004, 2013, 2014, 2016, 2018) obtained by \emph{Chandra} and the \emph{X-ray Multi-mirror Mission} (\emph{XMM-Newton}). Until now, only SN\,1987A\footnote{SN\,1993J has $>$200 ks of {\it Chandra} HETG exposure which has yet to be published.} has had high resolution X-ray spectroscopic campaigns using \emph{Chandra} or \emph{XMM-Newton} \citep{Burrows2000,Michael2002a,Zhekov2006a, Dewey2008a,Zhekov2009a,Sturm2010a}. The outline of the paper is as follows: $\S$2 presents the data reduction; $\S$3 explores how we build our source model, the physical implications, and the results from applying it to the 2009 and other epochs; $\S$4 explains the main outcomes and their interpretations to constrain its nature; and finally, $\S$5 presents our conclusions, final comments and future work. 

Following \citet{Bauer2008a}, we assume that the Circinus Galaxy is observed through a Galactic `window' with a neutral hydrogen column density of $N_{\rm H}$$=$(3.0$\pm$0.3)$\times$ 10$^{21}$~cm$^{-2}$, with possible additional internal obscuration \citep{Schlegel1998,Dickey1990,Bauer2001a}. Similar to \citetalias{Dwarkadas2010a}, we assume an explosion date of 1995.4 for SN\,1996cr throughout this paper. We adopt a position of $\alpha$=14$^h$13$^m$10$^s$.01, $\delta$=-65$^\circ$20$^\prime$44.$^{\prime\prime}$4 (J2000), for SN\,1996cr, determined from radio observations, which is 25$^{\prime\prime}$ to the south of the Circinus Galaxy nucleus. Errors are quoted at 1-$\sigma$ confidence unless stated otherwise.

\begin{table*}
   \caption{X-ray observations used in this work, ordered by date.  Designated epochs, as noted in $\S$\ref{sec:Data}, are separated by horizontal lines.
   {\it Column 1:} Observation ID. 
   {\it Column 2:} Observation UT date.
   {\it Column 3:} Cleaned, useful exposure time. When three values are listed, these denote the MOS1, MOS2 and \emph{pn} instruments, respectively.
   {\it Column 4:} X-ray instrument used.}
       \centering
       \begin{tabular}{lcrl}
       \hline
       \hline
        ObsID & Date (UT) & Exposure (ks) & Instruments \\ 
        (1) & (2) & (3) & (4) \\
       \hline
       \hline
        374 & 2000-06-15 & $7.1$ & \emph{Chandra} HETG\\
        62877 & 2000-06-16 & $60.2$ & \emph{Chandra} HETG\\
       \hline
        0111240101 & 2001-08-06 & $85.5/91.8/59.5$ & \emph{XMM-Newton} MOS1/MOS2/\emph{pn}\\
       \hline
        4770 & 2004-06-02 & $55.0$ & \emph{Chandra} HETG\\
        4771 & 2004-11-28 & $59.5$ & \emph{Chandra} HETG\\
       \hline
        10223 & 2008-12-15 & $102.9$ & \emph{Chandra} HETG\\
        10224 & 2008-12-23 & $77.1$ &\emph{Chandra} HETG\\
        10225 & 2008-12-26 & $67.9$ & \emph{Chandra} HETG\\
        10226 & 2008-12-08 & $19.7$ & \emph{Chandra} HETG\\
        10832 & 2008-12-18 & $20.6$ & \emph{Chandra} HETG\\
        10833 & 2008-12-22 & $28.4$ & \emph{Chandra} HETG\\
        10842 & 2008-12-27 & $36.7$ & \emph{Chandra} HETG\\
        10843 & 2008-12-29 & $57.0$ & \emph{Chandra} HETG\\
        10844 & 2008-12-24 & $27.2$ & \emph{Chandra} HETG\\
        10850 & 2009-03-03 & $16.5$ & \emph{Chandra} HETG\\
        10872 & 2009-03-04 & $13.9$ & \emph{Chandra} HETG\\
        10873 & 2009-03-01 & $18.1$ & \emph{Chandra} HETG\\
       \hline
        0701981001 & 2013-02-03 & $47.8/49.0/36.4$ & \emph{XMM-Newton} MOS1/MOS2/\emph{pn}\\
       \hline
        0656580601 & 2014-03-01 & $31.4/31.2/17.1$ & \emph{XMM-Newton} MOS1/MOS2/\emph{pn}\\
       \hline
        0792382701 & 2016-08-23 & $19.8/19.6/17.0$ & \emph{XMM-Newton} MOS1/MOS2/\emph{pn}\\
        \hline
        0780950201 & 2018-02-07 & $41.9/41.3/35.7$ & \emph{XMM-Newton} MOS1/MOS2/\emph{pn}\\
        \hline
       \end{tabular}
       \label{tab:data}
\end{table*}

\section{Data analysis}\label{sec:Data}

We use data obtained between 2000 and 2018, ergo 5 to 21 years after the explosion, respectively, taken by the \emph{Chandra} X-ray Observatory \citep[CXO;][]{Weisskopf2002a} and \emph{XMM-Newton} \citep[][]{Jansen2001a}. We describe the processing and data reduction of each below.

\subsection{Chandra X-ray Observatory}

As we are principally interested in modeling the high signal-to-noise, high spectral resolution data, we focus on the available Chandra X-ray observatory data taken using the High-Energy Transmission Grating \citep[HETG;][]{Canizares2005}, dispersed onto the Advanced CCD Imaging Spectrometer S-array \citep[ACIS-S;][]{Garmire2003}; see Table~\ref{tab:data}. The HETG instrument consists of the High Energy Grating (HEG) and the Medium Energy Grating (MEG) assemblies, which operate simultaneously and have spectral resolutions of $0.7$--$80$ eV (for $0.8$--$10.0$ keV) and $0.5$--$70$ eV (for $0.4$--$8.0$ keV), respectively. The gratings have different energy-dependent effective areas, such that the MEG is generally more efficient for observing lower energy lines ($\lesssim$3 keV) while the HEG better for higher energy ones ($\gtrsim$3 keV). The gratings disperse a fraction of incident photons along dispersion axes offset by 10 degrees, such that the first and second orders of the HEG and MEG form a narrow X-shaped pattern on the ACIS-S detector \citep{Canizares2005}. Roughly half of the photons pass through the gratings undispersed (preferentially higher-energy photons) and comprise the HETG 0th order image on ACIS-S, with a spectral resolution of $100$--$170$ eV between $0.4$--$8.0$ keV. For completeness, we extracted the low-resolution, 0th order data and retained it to help reduce uncertainties on some of the parameters of our model. With respect to the HETG extraction, SN\,1996cr is a point source and, due to the spatial and spectral photon selection, has negligible background and no obvious contamination from the AGN or other point source spectra (dispersed or undispersed).

The \emph{Chandra} data were reduced using \texttt{CIAO} (v4.9) and corresponding calibration files (CALDB v4.7.4). After standard processing and cleaning, we extracted each HEG/MEG spectrum as follows. We resolve the spectral orders making use of the procedures \texttt{tg\_create\_mask} and \texttt{tg\_resolve\_events}, and create response files (ARF and RMF) for each spectral order using the \texttt{mktgresp} tool; we consider only the $m=\pm1$ orders in this work. Finally, we combine spectra from the positive and negative orders and ObsIds for each epoch using the script \texttt{combine\_grating\_spectra}. For the zeroth order data, we adopt source and background extraction regions with radii of 3\farcs44 and 9\farcs84, respectively, and use the \texttt{specextract} script to extract spectra and create response files, considering a point source aperture correction for the ancillary files. We combine the 0th order spectra with the \texttt{combine\_spectra} script.
To produce the 2009 epoch, we combined all 12 ObsIDs taken between 2008-12-8 and 2009-3-4, for a total combined exposure of $\sim$485\,ks. For the 2000 epoch, we combined HETG spectra for ObsIDs 374 and 62877 for a total exposure time of $\sim$67.3\,ks, while for the 2004 epoch, we combined ObsIDs 4770 and 4771 for a total exposure time of $\sim$114\,ks. See Table \ref{tab:data} for information on individual ObsIDs. In all cases, we confirm that the individual spectra do not change significantly over 3--6 month timescales, justifying their combination into the three epochs.


\subsection{\emph{XMM-Newton}}

To augment the \emph{Chandra} spectra, we incorporate observations from \emph{XMM-Newton} taken in 2001, 2013, 2014, 2016 and 2018. The \emph{XMM-Newton} spacecraft consists of three X-ray telescopes with identical mirror modules, each equipped with a CCD camera which together comprise the European Photon Imaging Camera \citep[EPIC;][]{Struder2001}. Two of the telescopes employ Metal Oxide Semi-conductor \citep[MOS;][]{Turner2001} CCD arrays, installed behind Reflection Grating Spectrometers \citep[RGS;][]{denHerder2001}; the MOS cameras only capture $\approx$44\% of the incident flux, after accounting for the $\approx$50\% diverted to the RGS detectors and structural obscuration. The third telescope focuses its unobstructed beam onto the \emph{pn} CCD camera. The EPIC cameras provide sensitive imaging over a $\approx$30$\arcmin$ field of view (FOV) in the $0.3$--$12.0$ keV energy range, with modest spectral ($E/\Delta E$ $\sim$20--50) and angular (PSF$\approx$6\farcs0 FWHM) resolutions. This spectral resolution equates to velocities of $\gtrsim$6000--15000 km s$^{-1}$, such that the EPIC cameras are only able to marginally constrain the largest velocities seen from SN\,1996cr \citep[e.g.,$\approx$4000--6700\,km\,s$^{-1}$;][]{Bauer2008a}. Thus, while the \emph{XMM-Newton} epochs have insufficient spectral resolution to constrain the velocity structure of the emission lines in the same way as the \emph{Chandra}-HETG spectra, they do provide useful constraints on the evolution of the continuum shape and line abundances of the SN. Table~\ref{tab:data} shows exposure times for the \emph{XMM-Newton} instruments at each epoch. Due to the poorer angular resolution of \emph{XMM-Newton} and the relative position of SN\,1996cr with respect to the bright AGN emission in the Circinus Galaxy, the spectra of SN\,1996cr suffer some contamination from the central AGN. Thus, particular care must be taken to select a region for appropriate and optimal background subtraction. 
In our analysis, we exclude the RGS data due to difficulties related to separating SN\,1996cr's emission from the bright extended emission associated with the AGN and circumnuclear star formation, coupled with the modest photon statistics obtained at all \emph{XMM-Newton} epochs \citep[see Figs. 1 and 2 of][]{Arevalo2014}.

Each epoch of \emph{XMM-Newton} data was reduced using \texttt{SAS} (v16.1.0) package. After standard processing and cleaning, we extracted MOS1, MOS2, and \emph{pn} spectra using a circular aperture of radius 8\farcs7 centered in the SNe using the \texttt{xmmextractor} script. To select a background region which removes the substantial radially symmetric contamination from the AGN \citep[e.g., due to the point spread function and Thomson scattered reflection continua and Fe K$\alpha$ line emission;][]{Arevalo2014}, we adopted a half-annulus centered on the AGN with inner and outer radii of 16\farcs2 and 33\farcs8 (i.e., at a radial offset comparable to that of SN\,1996cr from the nucleus), respectively, which excluded the extraction region of the SN itself and avoided the other bright off-nuclear source \citep{Bauer2001a} and ionization cone \citep{Arevalo2014}. Table~\ref{tab:data} provides the observation IDs, dates, useful exposure times and instruments used in this work. We note that there are known energy-dependent cross-calibration offsets between the various {\it XMM} and {\it Chandra} instruments, with the {\it XMM} \emph{pn} detector in particular yielding $\sim$10--20\% cooler temperatures or softer photon indices compared to {\it Chandra}'s ACIS detector for identical objects \citep[e.g.,][]{Nevalainen2010a, Madsen2017a}. We discuss such effects in $\S$\ref{sec:column_density}.

\begin{figure*}
    \centering
    \includegraphics[scale=0.65]{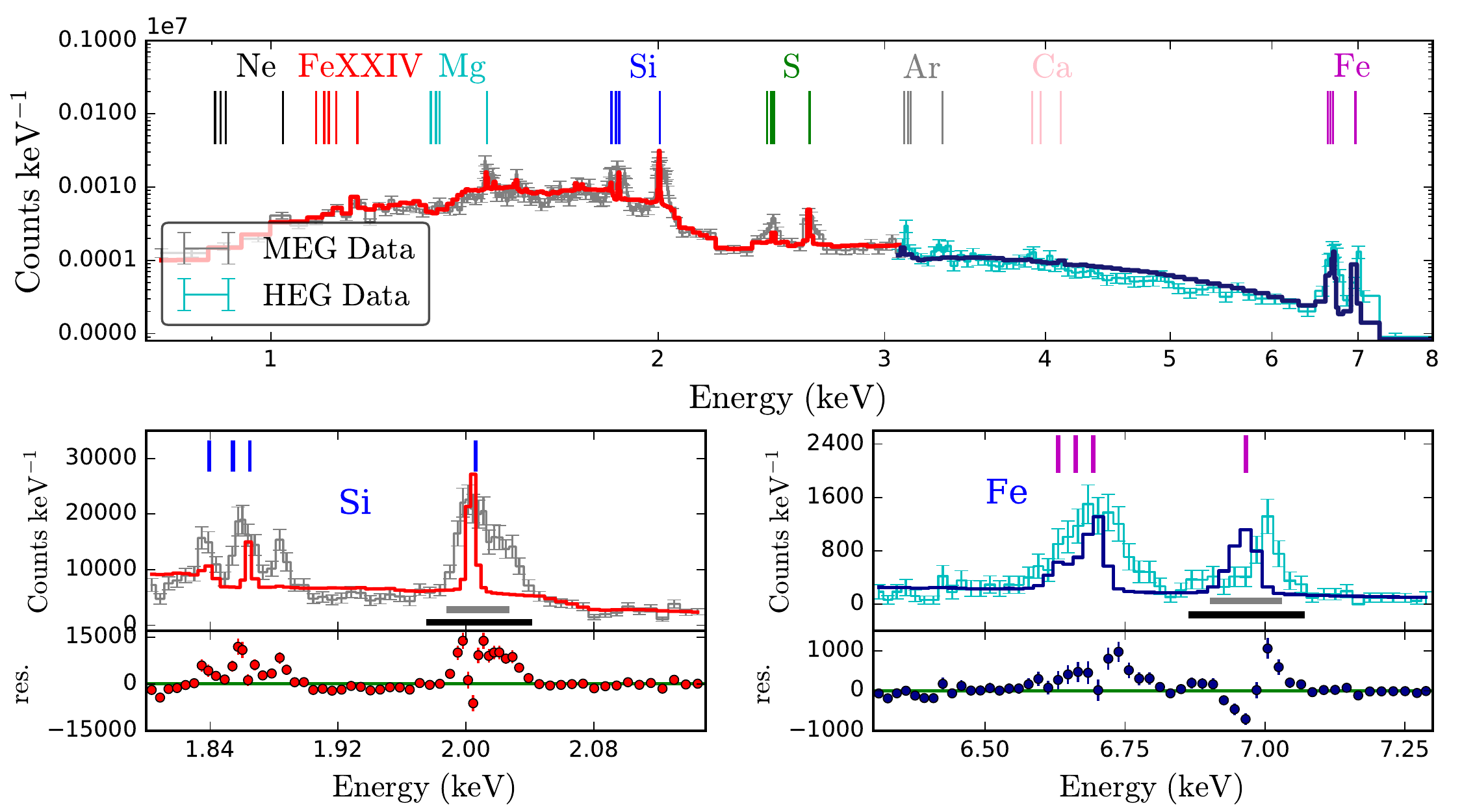}
    \caption{\emph{Top panel}: \emph{Chandra} HEG (\emph{cyan}) and MEG (\emph{grey}) X-ray spectra and 1-$\sigma$ errors from the 2009 epoch, as well as \emph{blue} and \emph{red} solid lines representing the fitted model M1 (single plasma, no velocity structure) for the HEG and MEG, respectively. For clarity, we only show MEG/HEG below/above 3.0\,keV. \emph{Bottom panels}: close-up spectra of the H-like and He-like emission complexes for Si and Fe (\emph{left} and \emph{right}, respectively, and their residuals). Vertical lines denote rest-frame energies. Horizontal \emph{grey} and \emph{black} bars represent line-widths of 3000 and 5000\,km\,$^{-1}$, respectively, with respect to the H-like lines. While the bulk of the Fe XXVI emission is seen with velocity width $\lesssim$3000\,km\,s$^{-1}$, other elements like Si show substantial emission up to velocity widths of $\sim$5000\,km\,s$^{-1}$, implying that Fe and Si are produced from distinct regions.} 
    \label{fig:Si_Fe_vpshock}
\end{figure*}

\section{Line structure, methodology and models}\label{sec:MethodModel}

In this section, we  explain the structure of emission lines observed in SN\,1996cr as compared to SN\,1987A (\S\ref{sec:line_structure}), describe the fitting process and  statistical methods adopted (\S\ref{sec:fit_stat}), apply a handful of models to the 2009 epoch (\S\ref{sec:2009}, including the development of a geometrical model called \texttt{shellblur} in \S\ref{sec:2009_geom}), and finally  extend the best-fit model to the other X-ray epochs (\S\ref{sec:other_epochs}).

\subsection{Line structure}\label{sec:line_structure}

As shown in Fig.~\ref{fig:Si_Fe_vpshock}, the intense, broad, asymmetric ionized emission lines in the 2009 epoch spectra of SN\,1996cr are indicative of a strong ejecta-CSM interaction. CSM geometries have been revealed/inferred for a number of SNe to date. Most notable is the remarkable SN\,1987A, for which a complex CSM ring was directly imaged \citep[e.g.,][]{Jakobsen1991,Panagia1991}. Others include, e.g., SNe\,1979C, 1986J, 1997eg, 1998S, 1993J, 2005kd, 2006jc, 2006gy, 2008iz, 2010jl, 2011dh, and 2014C \citep[e.g.,][]{Leonard2000a, Smith2007a, Foley2007a, Hoffman2008a, Bartel2008a, Chandra2009b, Bietenholz2010a, Martin-Vidal2011a, Bietenholz2012a, Katsuda2016a, Kimani2016a, Dwarkadas2016a, Bartel2017a, Bietenholz2018a}, many of which are classified as type IIn. 

Notably, SN\,1987A has remained bright enough, for long enough, to support several campaigns with X-ray observatories, and in some cases high resolution X-ray spectroscopy \citep{Burrows2000,Michael2002a, Zhekov2006a, Dewey2008a, Zhekov2009a, Sturm2010a, Frank2016a}. With thermal plasma temperatures in the range $T$ $\sim$0.5--2\,keV, SN\,1987A primarily exhibits ionized lines from Nitrogen (N), Magnesium (Mg), Oxygen (O), Neon (Ne) and Silicon (Si), but lacks higher ionization lines like Sulfur (S), Argon (Ar), Iron (Fe) and Nickel (Ni). Spatially resolved spectral analysis of these lines found that they have thermal widths of $\sim$60--300\,km\,s$^{-1}$ and Doppler broadening widths of $\sim$300--700\,km\,s$^{-1}$ that trace out the two-shock structure (forward and reverse) moving through the equatorial ring, the outer H{\sc ii} region, and the inner ejecta \citep[e.g.,][]{Michael2002a, Dewey2008a, Dewey2011a}. 

In our case, SN\,1996cr additionally presents strong lines associated with the He-like (Fe\,XXV K-$\alpha$; 6.7\,keV) and H-like (Fe\,XXVI Ly-$\alpha$; 6.9\,keV) ions of Iron, which are prominent in the X-ray spectra of other SNe such as SNe\,1986J \citep{Temple2005a}, 1998S \citep{Pooley2002a}, 2010jl  \citep{Chandra2012b}, 2006jd \citep{Chandra2012a}, 2009ip \citep{Smith2014b,Margutti2014a}, and 2014C \citep{Margutti2017a}, but only weakly detected in SN\,1987A \citep{Sturm2010a}. Such strongly ionized Fe lines in X-ray spectra are generally a sign of an exceptionally hot, multi-phased plasma ($T$ $\gtrsim$10\,keV), and possibly a strongly enriched medium with super solar abundances, associated with ejecta-CSM interaction \citep[e.g.,][]{Nymark2006a, Margutti2017a}.

As with SN\,1987A, our ultimate goal is to understand and interpret the geometrical and physical information that is encapsulated in the velocity profiles of the emission lines stemming from the ejecta-CSM interaction of SN\,1996cr. The observed line profiles shown in Fig.~\ref{fig:Si_Fe_vpshock} and \ref{fig:velocity_profile} are well-resolved compared to the native HETG resolution ($\sim$400-700 km\,s$^{-1}$ at 2 keV and $\sim$1500 km\,s$^{-1}$ at 6.0 keV) and show substantial broad, asymmetric structure up to $\sim$5000 km\,s$^{-1}$. 

We infer several things from the high signal-to-noise H-like and He-like Si and Fe profiles shown in Fig.~\ref{fig:velocity_profile}. First, both lines are well-resolved and asymmetric, demonstrating that they can provide critical insight on the kinematic sites of the ejecta-shock interaction(s). Such information was previously inferred from 1-D hydro-dynamical modeling of SN\,1996cr by \citetalias{Dwarkadas2010a}, but never measured directly. Second, the maximum velocities and shapes of the H-like Fe and Si profiles appear to be quite distinct. In Fig.~\ref{fig:velocity_profile}, the H-like Fe line exhibits maximum Doppler velocity offsets up to \hbox{$\sim$ $\pm$3000--4000\,km\,s$^{-1}$} from the systemic host velocity and is comprised of a strong, unresolved blueshifted peak at \hbox{$\sim$ $-$2000\,km\,s$^{-1}$} and a $\sim$4$\times$ weaker redshifted "peak" or "plateau". On the other hand, the H-like Si line is much more centrally peaked around the systemic velocity, but also shows a clear blueshifted asymmetric shoulder up to \hbox{$\sim$ $-$4000\,km\,s$^{-1}$}, with maximum Doppler velocity offsets approaching \hbox{$\sim$ $\pm$5000--6000\,km\,s$^{-1}$}. The other strong emission lines of Ne, Mg, and S generally all show profiles comparable to Si, while the Fe\,XXIV lines show characteristics of both Si and Fe. The maximum velocity and profile discrepancies between Fe and the rest of the elements suggest we are observing at least two kinematically and/or spatially distinct shocks.

To elucidate the nature of this structure, we investigate a few physically motivated models as described below.

\subsection{Fitting process and statistical methods}\label{sec:fit_stat}

\subsubsection{Fitting process}\label{sec:fitting}

To fit the spectra we utilized the X-ray software fitting package \texttt{XSPEC} v.12.8.2n \citep{Arnaud1996a,Arnaud1999a} using ATOMDB v.3.0.9 \citep{Smith2001a}\footnote{\url{http://www.atomdb.org/}}, which has been updated to include relevant inner-shell processes that can be important for X-ray plasma spectral modeling of SN\,1996cr, and \citet{Anders1989a} abundances.
The interaction of the SN blast wave with the CSM sets up forward and reverse shocks \citep[e.g.,][]{Chevalier1982a}, behind which one can find the shocked CSM and shocked ejecta, respectively, separated by a contact discontinuity, which is Rayleigh-Taylor unstable. The shocked plasma can generate copious thermal X-ray emission \citep{Chevalier1982b}, in proportion to the temperature, ionization state, and density of the plasma. As the blast wave is rapidly expanding, and the density of the plasma remains relatively low, the typical ionization equilibrium timescales are of order a few to thousands of years depending on the temperature, density and composition of the medium \citep[e.g.,][]{Smith2010b}, and hence the X-ray emission should be computed under non-equilibrium ionization (NEI) conditions as a precaution (e.g., \citealp{Borkowski2001a}; \citetalias{Dwarkadas2010a}).\footnote{Following \citet{Smith2010b}, metals such as Mg, S, Si, Fe, and Ni require $n_{e}t$ $>$ 10$^{12}$ s cm$^{-3}$ to be within 10\% of their equilibrium value at temperatures of $kT\approx$1--10 keV. Given that our model fits return values of $n_{e}t$ in the range (1--4)$\times$10$^{12}$ s cm$^{-3}$, which is close to the limit, we chose to adopt an NEI model to be conservative.} 
For this purpose, we employ the \texttt{XSPEC} NEI model \texttt{vpshock}, a plane-parallel shocked plasma model \citep{Borkowski2001a}. This model parametrizes the shock as a function of: electron temperature ($kT_e$); ionization time scale $\tau=n_e t$, where $n_e$ is the electron density and $t$ is the time since the plasma was shocked; individual atomic abundances for He, C, N, O, Ne, Mg, Si, S, Ar, Ca, Fe, Ni with respect to Solar; and normalization, which depends on the angular distance ($D_A$), redshift ($z$), and volume emission measure of plasma as $C(D_A,z)\int n_en_HdV$ \citep{Borkowski2001a}. Furthermore, we account for the small recessional velocity of the host galaxy \citep[434 km\,s$^{-1}$;][]{Koribalski2004}.

We begin by finding the best model to explain the high signal-to-noise X-ray grating spectra from the 2009 epoch, and then apply it to the other epochs to confirm this and investigate parameter evolution ($\S$\ref{sec:evol_interp}). In some epochs, due to the low signal-to-noise and/or spectral resolution, we are unable to constrain an elemental abundance robustly. In such cases, we either fix it: $i)$ to the best-fit value from nearest epoch where it was well-determined as a free parameter, $ii)$ to the Circinus Galaxy ionized gas-phase abundances, which lie in the range $\sim$0.3--1.0 $Z_{\odot}$ \citep[as determined by][]{Oliva1999a}, if associated with the CSM\footnote{This choice likely only provides a floor to the real CSM elemental abundances since the CSM around SN\,1996cr should be at least moderately further enriched with heavy elements due to stellar winds or mass-loss episodes prior to the explosion of the progenitor star, e.g., in line with the results of \citetalias{Dwarkadas2010a}.} or $iii)$ to solar values if associated with the ejecta\footnote{Again, this choice likely only provides a floor to the true ejecta elemental abundances, which should be super-solar at this early stage of the SN.}; $ii)$ and $iii)$ are particularly relevant for elements such as H, He, C, N, O, and Ni, since their contributions are poorly constrained by the fitting process in this energy range. Other SNe such as SN\,1987A \citep{Michael2002a, Zhekov2009a} or type IIn SN\,2010jl, SN\,2006jd, and SN\,2005kd \citep{Chandra2015a, Dwarkadas2016a, Katsuda2016a} present similar high abundance values.

\subsubsection{Statistical methods}\label{sec:statistic}

Due to the low number of counts per bin for high resolution X-ray spectroscopy, and to retain the highest spectral resolution with which to resolve emission lines, we adopt maximum likelihood statistics for a Poisson distribution, the so-called Cash-statistics \citep[C-stat, with \hbox{$C{=}{-}2\ln{L_{\rm Poisson}}{+}const$};][]{Cash1979a} to find the best-fit model. 

Although C-stat is not distributed like $\chi^2$, meaning that the standard goodness-of-fit is not applicable \citep{Kaastra2017a, Buchner2014}, $\Delta$C-stat is according to Wilk's theorem \citep{Wilks1938, Cash1979a}. Thus, to evaluate the statistical improvement between models, we use four alternative different methods. $i)$ We generate 1000 simulations in order to calibrate $\Delta$C-stat for application to the goodness-of-fit criteria \citep{Kaastra2017a}. $ii)$ The Bayesian Information Criteria \citep[BIC;][]{Schwarz1978}, whereby the lowest value of \hbox{BIC${=}C{-}m\ln{n}$} indicates which model is preferred, with $C$, $n$ and $m$ denoting the C-stat value, the number of spectral bins and number of free parameters of the model, respectively \citep{Buchner2014}. $iii)$ The Akaike Information Criteria \citep[AIC;][]{Akaike1974}, which quantifies the information loss by a specific model, whereby the preferred model is the one with the lowest \hbox{AIC${=}C{-}2m$}. For both BIC and AIC, models are penalized for increased numbers of free parameters, or model complexity, with BIC having a stronger penalty factor than AIC. $iv)$ The Bayesian X-ray Astronomy (BXA) package \citep{Buchner2014}, which joins the Monte Carlo nested sampling algorithm MultiNest \citep{Feroz2009} with the fitting environment of \texttt{XSPEC}. For model comparison, BXA computes the integrals over parameter space, called the evidence (Z), which is maximized for the best-fit model. For BXA, we assume uniform model priors over sensible upper/lower limits throughout, and, following \citet{Buchner2014}, we consider a difference of $\log{\rm Z_{1}} {-} \log{\rm Z_{2}} {>} \log{10}{=}1$ to denote a significant preference when comparing between models; typically the evidence is normalized for simplicity, such that the maximum value is 1. Given BXA's incorporation of all parameter space, we consider it to be the most robust indicator among the four methods.  In \S\ref{sec:2009_2compblur}, we employ these methods to discard models.

Unless stated otherwise, we consider typically a confidence interval of 1-$\sigma$ for the parameter errors. For each \emph{Chandra} HETG epoch, we fit simultaneously both the HEG and MEG first-order spectra to improve the statistics during the process of finding best-fit parameters. The HETG 0th order spectra were incorporated after arriving at a set of best fit values, to increase the number of photons and constrain the parameter errors better. For each \emph{XMM-Newton} epoch, we fit simultaneously the \emph{pn}, MOS1, and MOS2 spectra to arrive at a best fit. For lower signal-to-noise or lower spectral resolution epochs, we freeze some poorly constrained parameters to improve the stability of the fits.

\subsection{Epoch 2009}\label{sec:2009}

\subsubsection{Single plasma component (M1)}\label{sec:2009_1comp}

We begin the modeling process by fitting the 2009 epoch grating data with a single absorbed \texttt{vpshock} model at the systemic velocity (i.e., \texttt{TBabs*vpshock}; hereafter model M1). \texttt{TBabs} models the X-ray absorption due to the line-of-sight ISM \citep{Wilms2000a}, parametrized by the equivalent hydrogen column, $N_{\rm H}$; this model adopts typical Milky Way ISM abundances, and incorporates interstellar grains and $H_{2}$ molecules. For simplicity, we separate the line-of-sight absorption into Galactic ISM and Circinus Galaxy ISM + SN CSM components; in anticipation of our fitting results, we note that the Circinus Galaxy ISM absorption appears to be negligible \citep{Bauer2001a}. For M1, we model as free parameters $kT_{\rm e}$, $N_{\rm H}$, $\tau$ and the abundances of strong observed lines from Ne through Fe between $0.8$--$10.0$ keV. 

For model M1, we find best-fit parameters of $kT_{\rm e}$ $=$13.4$\pm$0.9 keV, $N_{\rm H}$ $=($3.9$\pm$0.2)$\times10^{21}$\,cm$^{-2}$, $\tau$ $=($8.1$\pm$1.3)$\times10^{12}$\,s\,cm$^{-3}$, and abundances ranging from 0.42--3.03 $Z_\odot$, with a C-stat of 10383.89 for 8545 degrees of freedom (DOF) (see Table \ref{tab:cstat}). As seen in Fig.~\ref{fig:Si_Fe_vpshock}, M1 provides a reasonable fit to the continuum (\emph{top panel}) and approximates the intensity of the emission lines, but fails to model the Doppler width ($\sim$3000--5000\,km\,s$^{-1}$) and line shapes (\emph{bottom panels}), leaving strong residuals around the H-like and He-like lines of Fe, S, Si, Mg, and Ne. We also note some residuals in the continuum fit between 4--6\,keV, with the model being too high.

\begin{figure}
    \centering
    \includegraphics[scale=0.5]{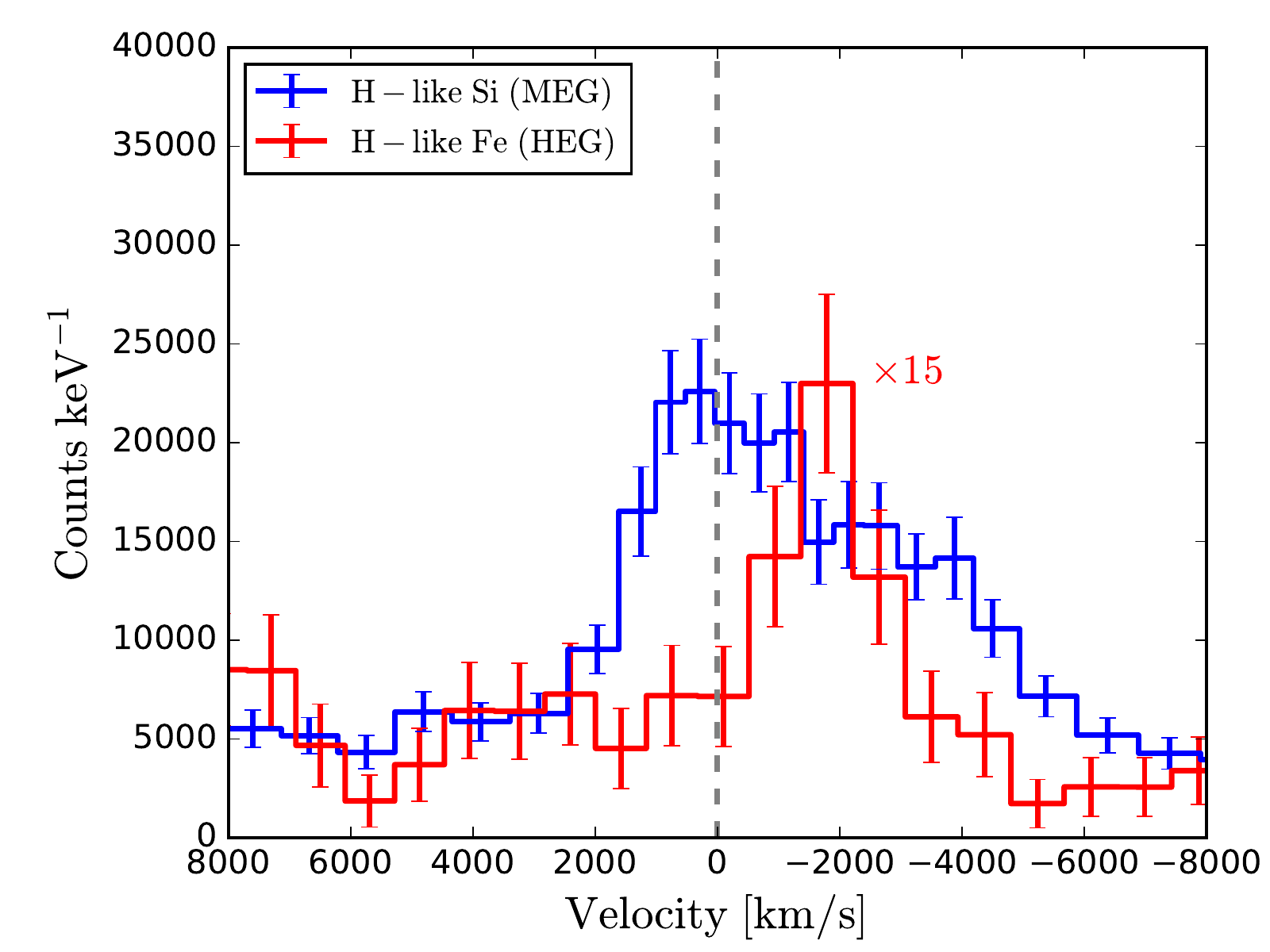}
    \caption{Comparison of the asymmetric velocity profiles of the H-like Si (\emph{blue} solid line) and Fe (\emph{red} solid line) emission lines from the 2009 epoch, as detected by \emph{Chandra}. The Fe counts have been multiplied by a factor of 15 for visualization purposes. The distinct profiles imply different physical/geometrical origins.} 
    \label{fig:velocity_profile}
\end{figure}

\subsubsection{Line Geometry}\label{sec:2009_geom}


\begin{figure}
    \includegraphics[scale=0.45]{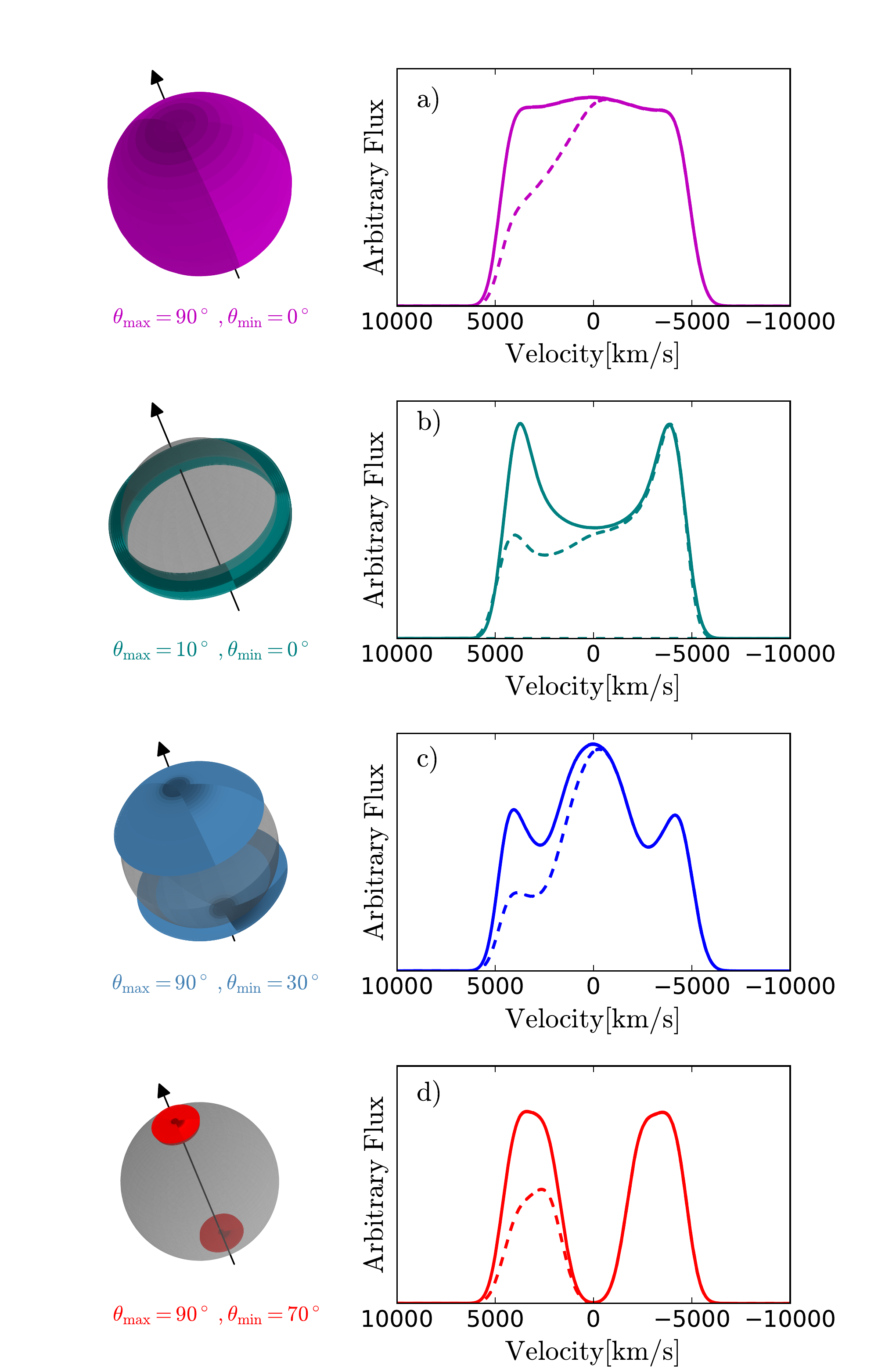}
    \caption{Examples of different expanding shock structure geometries, depicted by the colored regions in the \emph{left panels}: 
    \emph{(a)} spherically symmetric; 
    \emph{(b)} a $10^\circ$-wide equatorial belt;
    \emph{(c)} a $60^\circ$-wide polar cap;
    \emph{(d)} a $20^\circ$-wide polar cap interaction. 
    In all cases, we assume a maximum expansion velocity of $v_{\rm max}$$=$5000\,km\,s$^{-1}$, an axis of symmetry inclined by $55^\circ$ with respect to the line of sight, and that there exists a uniform-density ejecta core which provides a maximum obscuration of up to $N_{\rm ejecta}$ (measured at the diameter) to the farside of the shock. The \emph{right panels} show the resultant velocity profiles for each model assuming an input unresolved Gaussian emission-line centered at 6.0\,keV. Two line-profiles are shown, one assuming $N_{\rm ejecta}$$=$1$\times$10$^{20}$ cm$^{-2}$ (i.e.,  unobscured; \emph{solid curves}) and another $N_{\rm ejecta}$$=$2$\times$10$^{23}$ cm$^{-2}$ (i.e., obscured; \emph{dashed curves}), to demonstrate the variety of profiles that can be generated by a given combination of geometry and internal absorber.}
    \label{fig:estructure}
\end{figure}

\begin{figure}
\centering
   \includegraphics[scale=0.55]{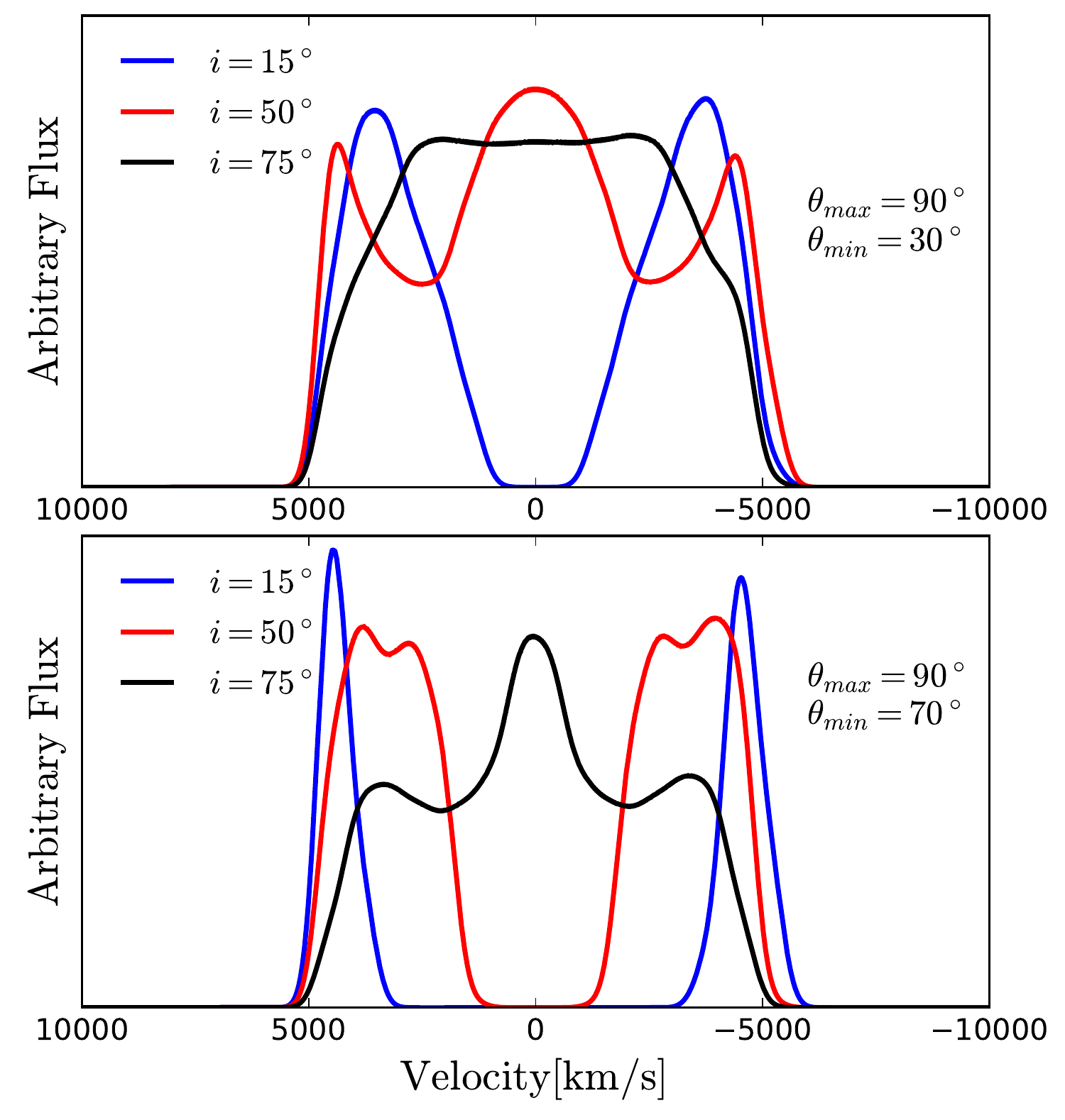}
    \vspace{-0.3cm}
    \caption{Comparison of velocity profiles as a function of line-of-sight inclination angles (15$^\circ$, 50$^\circ$, 75$^\circ$) for the latter two geometries [panels $(c)$ and $(d)$] in Fig.~\ref{fig:estructure}: wide polar angle ($\theta_{\text{min}}\sim30^\circ$) in the \emph{top panel} and narrow polar angle ($\theta_{\text{min}}\sim70^\circ$) in the \emph{bottom panel}. We convolved the above geometric models with an unresolved Gaussian emission-line centered at 6.0\, keV, a maximum expansion velocity of $v_{\rm max}$$=$5000\,km\,s$^{-1}$, and no internal obscuration.}
    \label{fig:line_profile}
\end{figure}

\begin{figure}
    \centering
    {\includegraphics[scale=0.68]{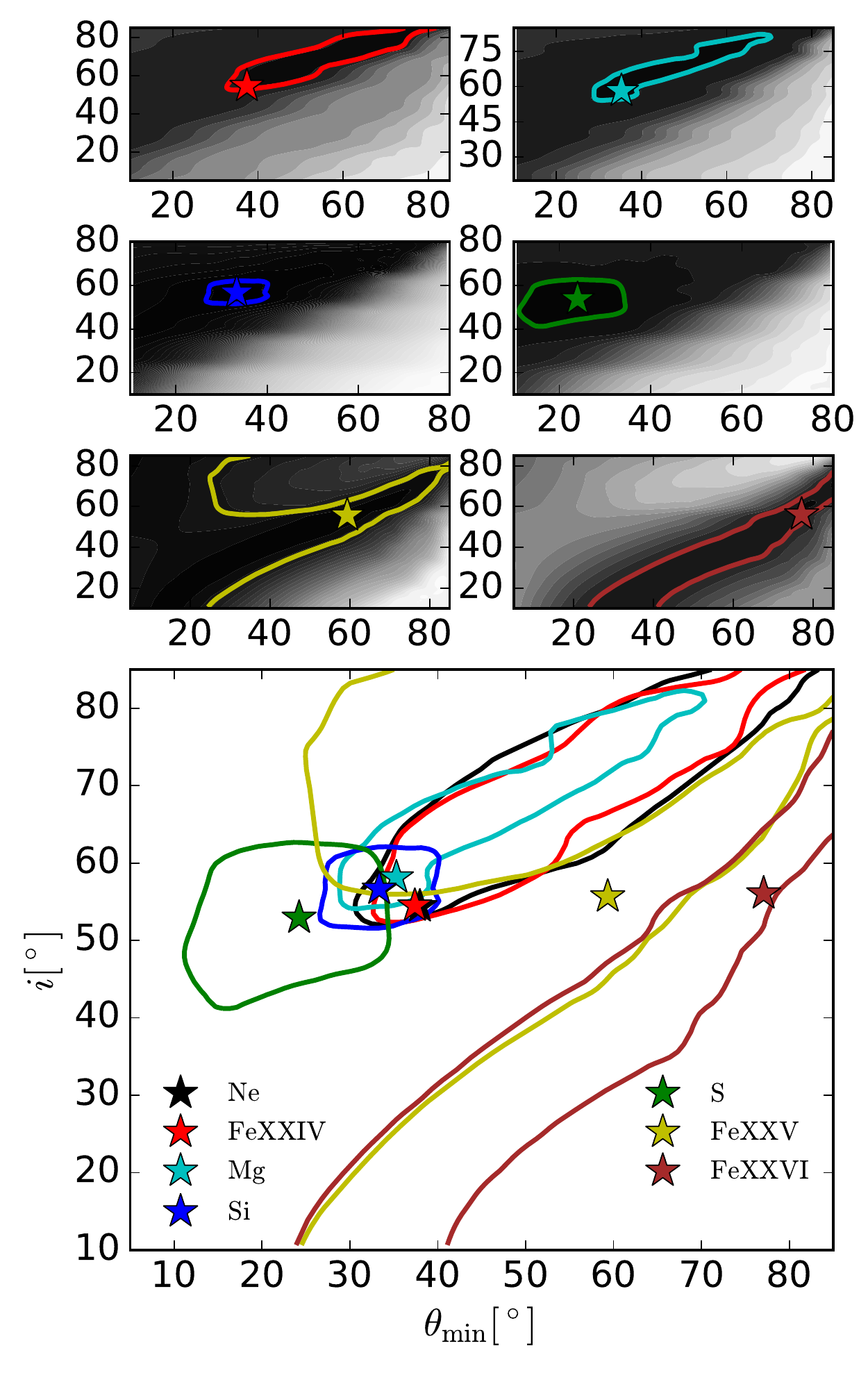}}
    \vspace{-0.8cm}
    \caption{Error confidence contour maps, comparing the minimum polar angle ($\theta_{\text{min}}$) and line-of-sight angle ($i$) from the polar emission line model, for 2009 {\it CXO} HEG/MEG spectra. \emph{Top panels} show confidence contours with greyscale shading for the most intense individual lines, with lighter and darker colors representing higher and lower C-stat values, respectively. The solid color curves denote the 1-$\sigma$ contours, while the stars are the best-fit values obtained. The \emph{bottom panel} compares all of the  1-$\sigma$ contours and best-fit values together.}
    \label{fig:contornos}
\end{figure}

The NEI-based model M1 in $\S$\ref{sec:2009_1comp} provides a reasonable fit to the overall continuum and intensity of lines, but fails to describe the asymmetric Doppler-broadened profiles, which presumably arise from rapidly expanding shocks, the velocities of which are completely unaccounted for in the model. As a sensible starting point, we assume that the density structure of both the expanding ejecta and CSM material have a spherical/conical shell-like symmetry. To incorporate the resulting velocity structure into the NEI plasma models, we develop an \texttt{XSPEC} convolution model called ``shellblur'',\footnote{"Shellblur" is available as a table model at https://www.dropbox.com/s/ts1jfrg68nx38fo/, while the full code can be found at https://github.com/jaquirola/shellblur-model} which adopts a spherical geometry parameterized by a maximum velocity ($v_{\rm max}$), an inclination angle with respect to the line-of-sight ($i$), minimum and maximum aperture angles ($\theta_{\rm min}$, $\theta_{\rm max}$),\footnote{When $\theta_{\rm max}$$=$90, we can consider $\theta_{\rm min}$ to be  the effective half-opening angle.} and an interior absorption term ($N_{\rm ejecta}$).

Interior to the reverse shock, we expect to find unshocked ejecta material, which, if sufficiently dense, will absorb the shock emission on the (redshifted) farside. Interior to the forward shock, we might expect mild additional contributions from the shocked ejecta and shocked CSM. For simplicity and coding efficiency, we naively assume a uniform, spherical density distribution with solar abundances; a radially decreasing profile would tend to shift the velocity dependence of the absorption as well as making it more severe and abrupt. The assumption of uniformity should be reasonable; while the ejecta density is very steep initially, after a few years (beyond day $\sim$2500) the density profile of the inner ejecta should become relatively flat (see Fig. 3 of \citetalias{Dwarkadas2010a}), although this could differ by factors of at least a few in practice given the polar geometries our models favor. The assumption of solar abundances is unlikely to be valid, given the expected ejecta composition. However, since high-Z elements dominate the total absorption cross section as a function of energy, even at solar abundances \citep[e.g.,][]{Kaastra2008a}, the impact on the ejecta column density estimate itself should be relatively minimal; a factor of $\sim$1.2--3 higher than for solar abundances, depending on exact composition and energy. Given this, we contend that our assumption of solar abundances is a simple and reasonable first approximation to model the internal absorption. Unfortunately, the quality of the spectra are not sufficient to determine relative abundances of the unshocked ejecta material directly, and thus must be based on theoretical arguments for heavy element yields \citep[e.g.,][]{Nomoto2006a}. Thus how we might relate the ejecta column density to the overall enclosed mass remains more uncertain. Nonetheless, we can still try to interpret our results to give some important insights. Finally, given the above uncertainties, we have chosen not to apply any absorption correction to the model components based on $N_{\rm ejecta}$); thus all quoted fluxes and abundances should be considered lower limits in this respect (with the potential upward correction of up to 2$\times$.

Importantly, the \emph{shellblur} convolution model allows us to infuse various geometrically motivated velocity profiles into our spectral fits. \citet{Maeda2008a} invoked a similar model to study asymmetric ejecta using nebular phase [O{\sc{i}}] emission-line profiles in Type Ib/c SNe. Fig.~\ref{fig:estructure} shows different geometrical interactions (\emph{left panels}) and the corresponding line-profiles assuming a 6\,keV emission line, a maximum expansion velocity of $v_{\rm max}$$=$5000\,km\,s$^{-1}$, an axis of symmetry inclined by $55^{\circ}$ with respect to the line of sight and ejecta column densities of $N_{\rm ejecta}$$=$10$^{20}$\,cm$^{-2}$ ('unobscured'; \emph{solid line}) and $N_{\rm ejecta}$$=$2$\times$10$^{23}$\,cm$^{-2}$ ('obscured'; \emph{dashed line}). In Fig.~\ref{fig:line_profile}, we provide an example of how the velocity-profile changes as a function of line-of-sight inclination (different colors) for the latter two geometries [panels $(c)$ and $(d)$] in Fig.~\ref{fig:estructure}. Here we convolve the geometric models with an unresolved Gaussian line centered at 6.0\,keV and assume no internal absorption.

Assuming the shock interaction is `uniform' and occurred in a geometrically thin, expanding shell \citep[e.g., as found for 1993J;][]{Fransson1998a, Martin-Vidal2011a}, we should observe a square velocity profile (`full shell' scenario), as depicted in \emph{panel a)} of Fig.~\ref{fig:estructure}, convolved with model M1. However, if the density of the unshocked ejecta region is high enough, then the receding side may be partially or fully obscured, effectively dampening the low-energy, redshifted portion of the profile. The observed Si and Fe both demonstrate this behavior, prompting us to also investigate a `full shell, obscured core' scenario. Intriguingly, we observe neither of these basic `full shell' scenarios, and rule them out at high confidence. Instead, we observe more complex profiles from both the Fe\,XXV/Fe\,XXVI and lower energy lines. In the `full shell' scenario convolved with M1, we obtained: $kT_{\rm e}$$\sim$12\,keV, $N_{\rm H}$$\sim$2.1$\times$10$^{21}$ cm$^{-2}$ with a C-value of 8970.61 for DOF 8542. Keeping with the theme of symmetry, we next investigate toroidal and polar geometries. For the former we fix $\theta_{\rm min}$$=$0$^{\circ}$, while for the latter we fix $\theta_{\rm max}$$=$90$^{\circ}$. 

With its resolved, pearl-necklace shock structure, SN\,1987A is the most famous case for a ring-like or equatorial-belt geometry. The velocity profile associated with such a morphology is a bullhorn shape, as depicted in \emph{panel b)} of Fig.~\ref{fig:estructure}. If the emission from the (redshifted) farside of the model is strongly obscured by the interior ejecta, the resulting profile resembles that of Fe\,XXVI (and less obviously the blended profile of Fe\,XXV) in Fig.~\ref{fig:velocity_profile}, although it remains difficult to fit the exact profiles of both Fe\,XXVI and Fe\,XXV with any combination of line-of-sight angle, torus height, and interior obscuration due to the relative ratio of the blue/red peaks and the strength of the emission at low / zero velocity in between. The other lines are all too centrally concentrated, and strongly rule out a simple ring/torus shape at high confidence. In the `equatorial belt' scenario convolved with M1, we obtained: $kT_{e}$$\sim$11.6 keV, $N_{\rm H}$$\sim$1.9$\times$10$^{21}$ cm$^{-2}$ with a C-value of 9120.9 for DOF 8542.

Another potential geometry for the shock interaction might be that of the polar cap of a sphere, for example, the Homunculus Nebula around the luminous blue variable star (LBV) $\eta$ Car \citep{Smith2013, Smith2006, Smith2007a, Davidson1997a}. If $\eta$ Car were to explode, the shock-interaction might develop primarily first along the equator and afterward along the polar axis due to the enhanced bipolar CSM density \citep{vanMarle2010}. Depending on the opening angle of this polar emission and line-of-sight orientation angle, we could observe it either as a centrally dominant line, as depicted in \emph{panel c)} of Fig.~\ref{fig:estructure}, or even a widely spaced double Gaussian shape, as depicted in \emph{panel d)} of Fig.~\ref{fig:estructure}. It is prudent to note here that clear degeneracies exist between $\theta_{\text{min}}$ and $i$, as shown in Fig.\ref{fig:line_profile}, such that similar line profiles can arise from different permutations of the two parameters.

Fig.~\ref{fig:contornos} shows confidence contour maps for the strongest individual lines (\emph{top panels}), considering polar geometry parameters $\theta_{\rm min}$ and $i$, and a comparison between them (\emph{bottom panel}). We achieve good fits to the Fe\,XXVI and Fe\,XXV lines with relatively narrow opening angles ($\theta_{\text{min}}$ $\approx$60--75$^{\circ}$), while the rest of the lines are well-fit with a wider polar angle ($\theta_{\text{min}}$ $\approx$25--35$^{\circ}$). Intriguingly, the inclination angle $i$ remains remarkably consistent across all individual line fits, at $\approx$55$^{\circ}$. This suggests that the morphological alignment of most elements around the polar axis are, to first order, the same. The internal absorption and maximum velocity terms required for the lower energy lines were typically $\approx$2$\times$ $10^{22}$\,cm$^{-2}$ and  $\approx$4600\,km\,s$^{-1}$, respectively, while for the Fe\,XXVI line we found better fits with values of $\approx$5$\times10^{23}$\,cm$^{-2}$ and $\approx$3000\,km\,s$^{-1}$, respectively. For a fixed inclination angle of 55$^{\circ}$, the Fe\,XXVI emission is more tightly concentrated around the polar regions, in agreement with the preliminary results from \citet{Dewey2011a}.

Finally, we note that the 1-$\sigma$ contours on the inclination angle $i$ and minimum opening angle $\theta_{\text{min}}$ highlight some interesting behavior. All of the error confidence maps show some degree of $i$-$\theta_{\text{min}}$ degeneracy traced by the lower (darker) C-stat values, but separate into 2--3 distinct regions. For instance, the contours for Si and S are robustly centred around their best-fit values, while the contours of H-like Fe (Fe XXVI) trace out a narrow band in $i$-$\theta_{\text{min}}$ parameter space. Given the degeneracy between $i$ and $\theta_{\rm min}$ for Fe\,XXVI in Fig.~\ref{fig:contornos}, an alternative physical scenario for this line might be with $i$ $\sim$80--90$^{\circ}$, $\theta_{\rm min}$ $\sim$90$^{\circ}$, and $v_{\rm max}$ $\sim$4600\,km\,s$^{-1}$, such that all of the lines shared a similar maximum velocity rather than a similar inclination angle. In addition, the Ne, Fe\,XXIV, and Mg  transitions, while having best-fit values close to Si, show skewed low-level contours toward higher $\theta_{\text{min}}$ and $i$ values. In contrast, the Fe\,XXV line exhibits best-fit contours that are sandwiched midway between the high (Fe\,XXIV) and low (Si and S) solutions, with a large degenerate range of $\theta_{\text{min}}$ and $i$ values. Taken together, the contours of the various lines appear to reinforce the notion that there are at least two distinct components.

\begin{figure}
    \centering
    \hglue-0.3cm{\includegraphics[scale=0.5]{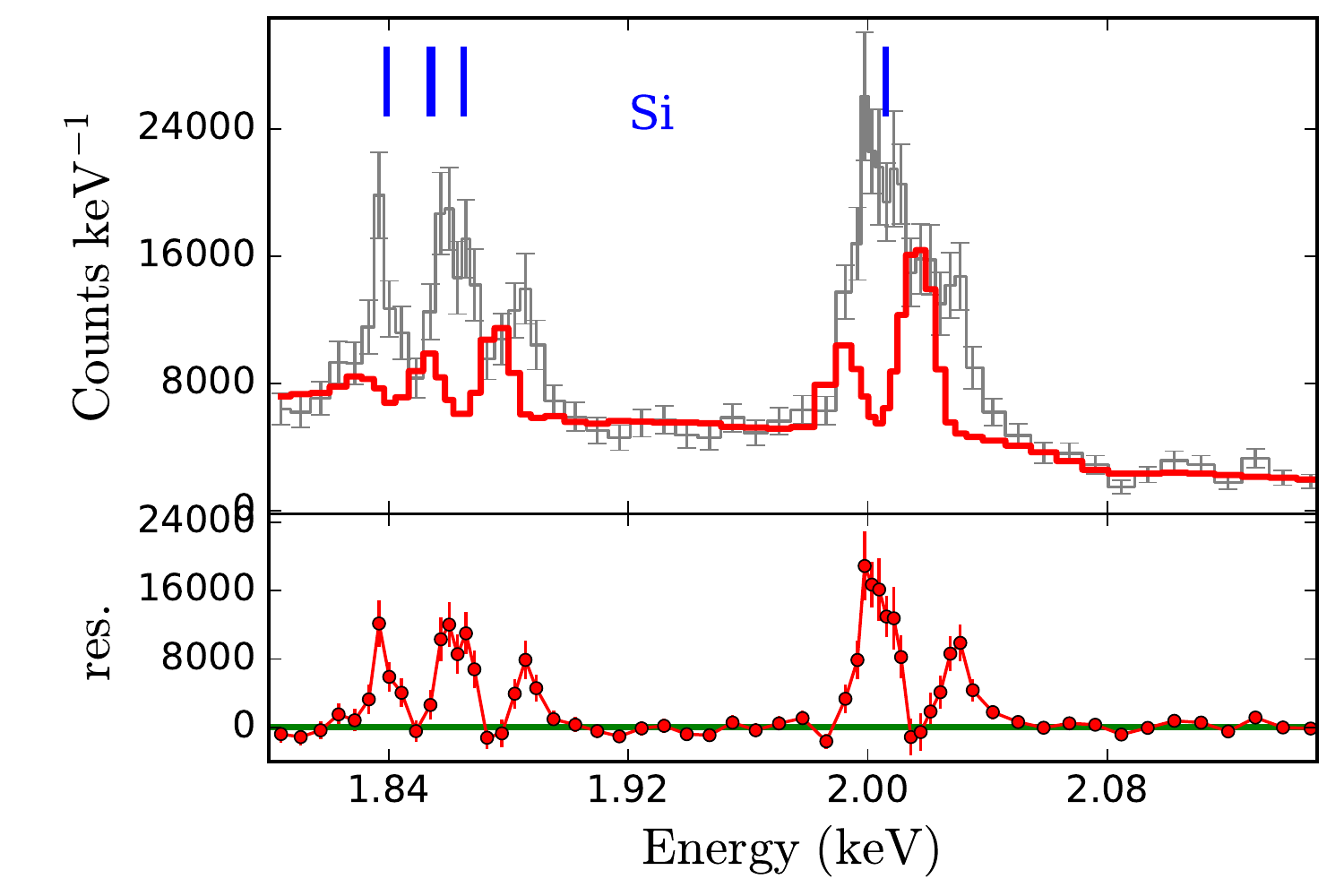}}
    \vspace{-0.3cm}
    \caption{Comparison between the best-fit narrow polar geometry from Fe\,XXVI and the Si line complex. Removal of this geometric component results in strong central (roughly Gaussian component with FWHM$\sim$1500\,km\,s$^{-1}$) and Doppler-shifted (unresolved component with $\sim$$+$5000\,km\,s$^{-1}$ offset) residuals.}
    \label{fig:Si-test}
\end{figure}

To further understand the potential degeneracy of the velocity profiles, we assessed how the contribution of the best-fit (\hbox{$v_{\rm max}$ $\sim3000$\,km\,s$^{-1}$}) Fe\,XXVI model appears for the observed lines below 4\,keV. As an example, Fig.~\ref{fig:Si-test} demonstrates how the Fe\,XXVI model, if normalized and removed from the Si line profile, leaves as residuals a large central Gaussian (FWHM$\sim$1500\,km\,s$^{-1}$) component and a smaller unresolved Gaussian offset by $\sim$$+$5000\,km\,s$^{-1}$.
Such unusual residuals are not easily accounted for by any single geometric model as described above and would require modeling as at least two additional {\it ad hoc} clumps.

Finally, we wish to highlight the presence of smaller residuals in the He-like lines of Si, Ar, and Fe, which remain that even after fitting the line emission with multiple components, as in $\S$\ref{sec:2009_2compblur}. These imply that further complexity (single or multiple components) likely exists. We revisit this theme later in this section.

\begin{figure*}
    \centering
    \includegraphics[scale=0.53]{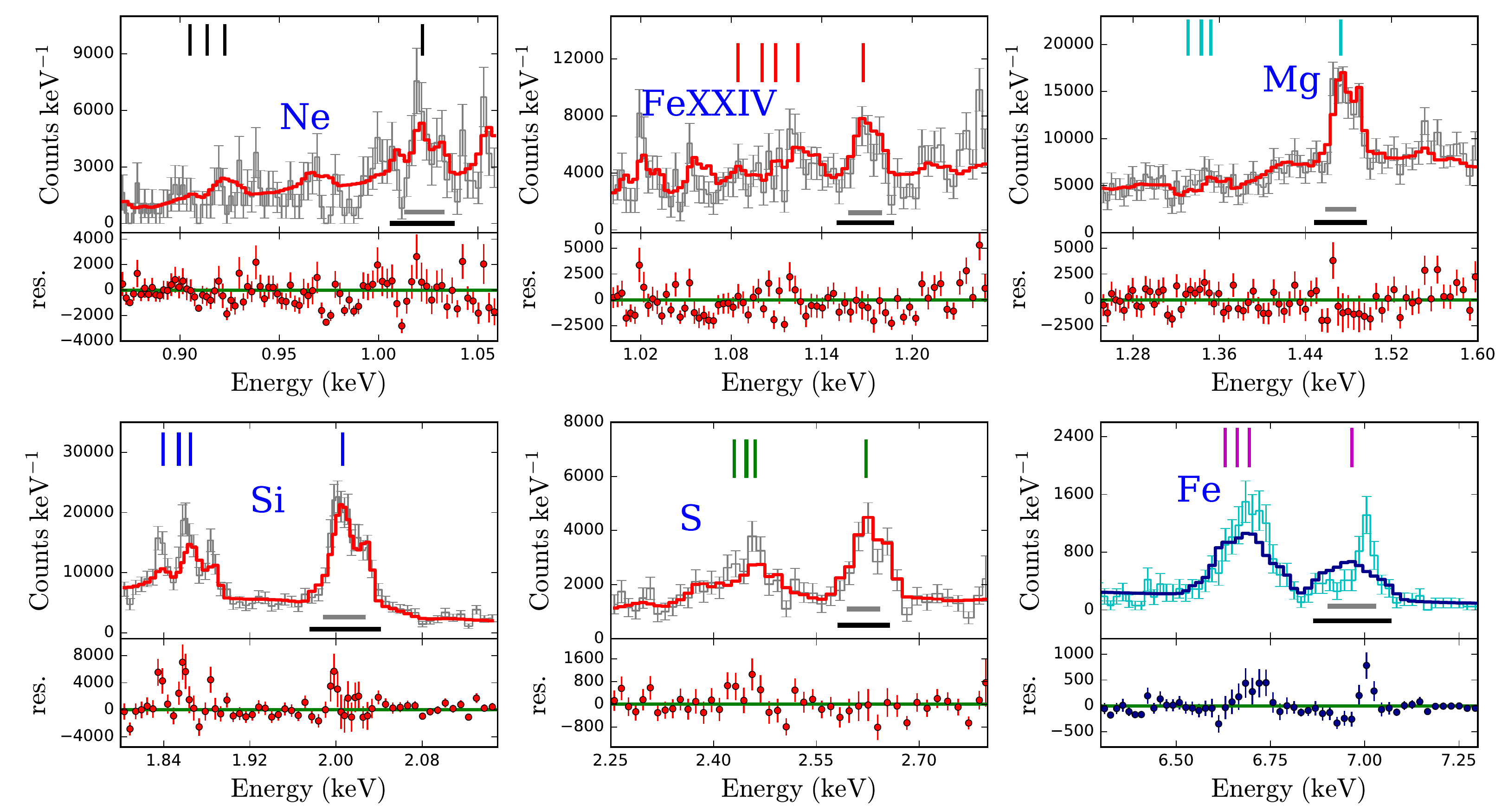}
    \vspace{-0.2cm}
    \caption{Close-up spectra of several well-detected H-like and He-like emission complexes from the 2009 \emph{Chandra} HEG/MEG spectra, overlaid with the fitted model M2 (single plasma, polar emission), and their residuals. Colors and lines are the same as those presented in Figure~\ref{fig:Si_Fe_vpshock}. Model M2 provides a reasonable fit to the low energy lines, but fails to match the Fe line profiles. As in Fig.~\ref{fig:Si_Fe_vpshock}, horizontal grey and black bars denote line-widths of 3000 and 5000\,km\,$^{-1}$, respectively, with respect to the H-like lines, highlighting the differences between the different lines.}
    \label{fig:Lineas_shell_vpshock}
\end{figure*}

\subsubsection{Single Plasma Component with Shellblur (M2)}\label{sec:2009_1comp_blur}

Given the overall success of the polar geometry in arriving at a single inclination angle for all lines and relatively consistent best-fit parameters for the Ne, Fe\,XXIV, Mg, Si, and S lines, we adopt it for the remainder of the analysis. We now return to the single temperature plasma model, convolving it with a single polar geometry [i.e., \texttt{TBabs*shellblur(vpshock)}; hereafter model M2], and attempt a global fit of the 2009 epoch grating data.

A best-fit is obtained with the following geometrical parameters:  $\theta_{\text{min}}$=$31\fdg2\pm4\fdg4$, $i$=$56\fdg\pm2\fdg8$,  $v_{\text{max}}$=4600$\pm71$\,km\,s$^{-1}$, \hbox{$N_{\text{ejecta}}$=$(1.9\pm0.3)\times10^{22}$\,cm$^{-2}$}, \hbox{$kT_{e}$=$12.1\pm0.8$\,keV}, \hbox{$N_{\text{H}}$=$(2.3\pm0.2)\times10^{21}$\,cm$^{-2}$}, \hbox{$\tau$=(4.4$\pm0.2$)$\times10^{12}$\,s\,cm$^{-3}$}, and abundances ranging from 1.0--7.7 $Z_\odot$, with a C-stat value of 8910.25 for 8541 DOF. Table~\ref{tab:cstat}
highlights the huge improvement between models M1 and M2 based on both BIC and AIC.

As seen in Fig.~\ref{fig:Lineas_shell_vpshock}, model M2 yields a reasonable match to the strong lines of Ne, Fe\,XXIV, Mg, Si and S, as well as the continuum (not shown in Fig.~\ref{fig:Lineas_shell_vpshock}). However, it fails to fit the lines of Fe\,XXVI ($\sim$6.9--7.0\,keV) and Fe\,XXV ($\sim$6.7\,keV), suggesting that this model remains incomplete and requires additional geometric/kinematic plasma components.


\begin{table*}
   \caption{Statistic values associated with each model for the 2000, 2004 and 2009 epochs (\emph{Chandra} observations). The first column denotes the model used. This is followed by three sets of four columns presenting the C-stat and degrees of freedom (DOF), AIC, BIC and log-evidence ($\log$~Z) values for each model and epoch.}
       \centering
       \scalebox{0.9}{
       \begin{tabular}{l|cccc|cccc|cccc}
       \hline
       Model & \multicolumn{4}{c}{2000} & \multicolumn{4}{c}{2004} & \multicolumn{4}{c}{2009} \\ \hline
       & C-stat (DOF) & AIC & BIC & $\log$~Z & C-stat (DOF) & AIC & BIC & $\log$~Z & C-stat (DOF) & AIC & BIC & $\log$~Z \\ \hline
       M1 & 4464.1 (8546) & 4448.1 & 4391.7 & -21.49 & 6543.4 (8545) & 6525.4 & 6461.9 & -120.05 & 10383.9 (8545) & 10365.9 & 10302.4 & -781.17 \\
       M2 & 4426.8 (8542) & 4402.8 & 4318.1 & 0.0 & 6301.0 (8542) & 6277.0 & 6192.3 & -1.36 &  8910.3 (8541) & 8884.3 & 8792.6 & -36.34 \\
       M3 & 4406.2 (8540) & 4378.2 & 4279.4 & -12.08 & 6290.7 (8539) & 6260.7 & 6154.9 & -11.25 &  8860.4 (8538) & 8828.4 & 8715.5 & -21.17 \\
       M4 & 4441.0 (8538) & 4409.0 & 4296.1 & -9.26 & 6332.0 (8536) & 6296.0 & 6169.0 & -14.55 &  8858.4 (8536) & 8822.4 & 8695.4 & -113.10 \\
       M5 & 4398.6 (8533) & 4356.6 & 4208.4 & -6.03 & 6264.3 (8533) & 6222.3 & 6074.2 & 0.0 &  8779.6 (8528) & 8727.6 & 8544.2 & 0.0 \\ \hline
       \end{tabular}
       }
       \label{tab:cstat}
\end{table*}

\subsubsection{Multiple Plasma Components with Shellblur (M3--M6)}\label{sec:2009_2compblur}

We therefore develop a few more complex model combinations. First, we consider two NEI models with different temperatures, modified by a single foreground absorption and \texttt{shellblur} term [\texttt{TBabs*shellblur(vpshock+vpshock)}, hereafter model M3]. The result is two independent best-fit temperatures of $kT_{e,1}$ $=$ 10.6$\pm0.4$\,keV and $kT_{e,2}$ $=$ 0.9$\pm0.1$\,keV and a column density of $N_H$ $=$ (0.19$\pm0.02$)$\times10^{22}$\,cm$^{-2}$, which improves the residuals around the 4--6\,keV continuum and lower energy lines somewhat compared to model M2, lowering the C-stat value to 8860.35 for 8538 DOF (see Table~\ref{tab:cstat}). 
In this case, the hotter component dominates the total line and continuum emission, with the cooler component contributing a modest amount to the continuum shape below $\sim$2\,keV. Unsurprisingly, we find that the Ne, Fe\,XXIV, Mg, Si, S lines are best-fit with geometric parameters similar to model M2, while the Fe\,XXVI and Fe\,XXV lines remain poorly fit. To limit the number of free parameters for model M3, the line-of-sight angle was fixed to the value of $55\fdg0$ obtained previously from M2. Moreover, the abundances of H, He, C, N, O, Ar, Ca, Ni were fixed to their gas-phase or solar values, while those of Ne, Mg, Si, S, Fe, as well as the ionization time scales and normalizations of both NEI models, were fit as free parameters. However, due to the overall dominance of the high-temperature component, only weak abundance constraints could be achieved in the low-temperature component. Given this, all of the abundance values between the low and high temperature components, except Fe, were tied together.

Next, we consider two NEI components with different temperatures and different geometric terms, all modified by a single foreground absorption term [\texttt{TBabs(shellblur*vpshock+
+shellblur*vpshock)}, hereafter model M4]. The first \texttt{shellblur*vpshock} term is associated with a narrow polar cap (i.e., tracking Fe\,XXVI), while the second \texttt{shellblur*vpshock} term is associated with a wider polar emitting region (i.e., tracking Ne, Fe\,XXIV, Mg, Si, S). Somewhat surprisingly, when the plasma temperatures are left free, both tend toward values of $kT{\sim}12$\,keV absorbed by $N_{\rm H}\sim1.6\times10^{21}$ cm$^{-2}$, resulting in a C-stat value of 8858.40 for 8536 DOF with no clear improvement over model M3 (see Table~\ref{tab:cstat}).
The wide polar angle component dominates the overall continuum fit, with very marginal contribution from the narrow polar cap component. As with model M3, to limit the number of free parameters in model M4, we fix the line-of-sight angle to $55\fdg0$, and the abundances of H, He, C, N, O, Ar, Ca, Ni to their gas-phase or solar values, and fit the abundances of Ne, Mg, Si, S, Fe as free parameters, with all parameters aside from Fe tied together between components.The best-fit values for this model fail to match the Fe\,XXVI and Fe\,XXV lines well. We note that increasing the abundances to $Z_{\rm Fe}{\sim}250Z_{\rm Fe\odot}$ in the narrow polar cap component (vs. $Z_{\rm Fe}{\sim}0.54Z_{\rm Fe\odot}$ in the wider angle component) can achieve good fits to the profiles of both high-energy lines without directly impacting the Mg, Si, and S profiles, but this subsequently produces strong Fe\,XXIV residuals at lower energies (0.8--1.5 keV) compared to what is observed. Furthermore, $Z_{\rm Fe}{\sim}250Z_{\rm Fe\odot}$ implies that there is essentially just Iron (e.g., no Hydrogen, Helium, or other heavy metals), which is probably not realistic for a CCSNe \citep[e.g.,][]{Thielemann1996, Hwang2003}. Such strong $Z_{\rm Fe}$ values also diverge from those obtained by \citetalias{Dwarkadas2010a}. Thus, although strongly inhomogeneous abundance distributions are not inconceivable, it does not appear viable for the observed spectrum.

\begin{figure*}
    \centering
    \includegraphics[scale=0.55]{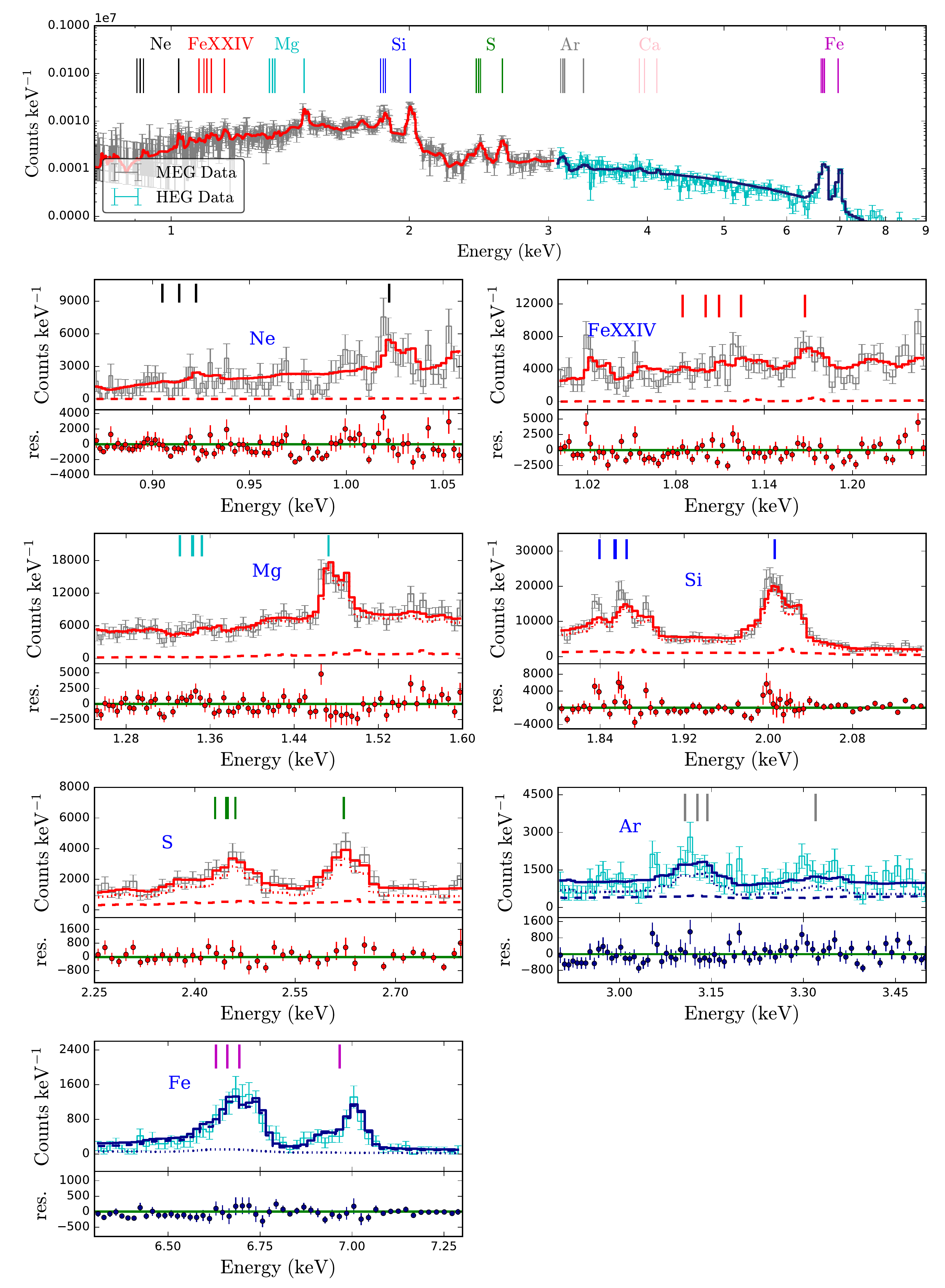}
    \caption{    
    \emph{Top panel}: Same as Fig.~\ref{fig:Si_Fe_vpshock} but with the best-fitting model M5 (two temperatures, two polar geometries, two absorptions) compared to the 2009 epoch spectra. \emph{Bottom panels}: Close-up spectra of all detected H-like and He-like emission complexes and their residuals. Colors and lines are the same as those presented in Fig.~\ref{fig:Si_Fe_vpshock}. MEG spectra are shown for Ne, Fe\,XXIV, Mg, Si, and S line complexes, while HEG spectra are shown for Ar and Fe. Model M5 provides a reasonable fit to all of the lines.}
    \label{fig:modM5_2009}
\end{figure*}

\begin{figure}
    \centering
    \hglue-0.3cm{\includegraphics[scale=0.4]{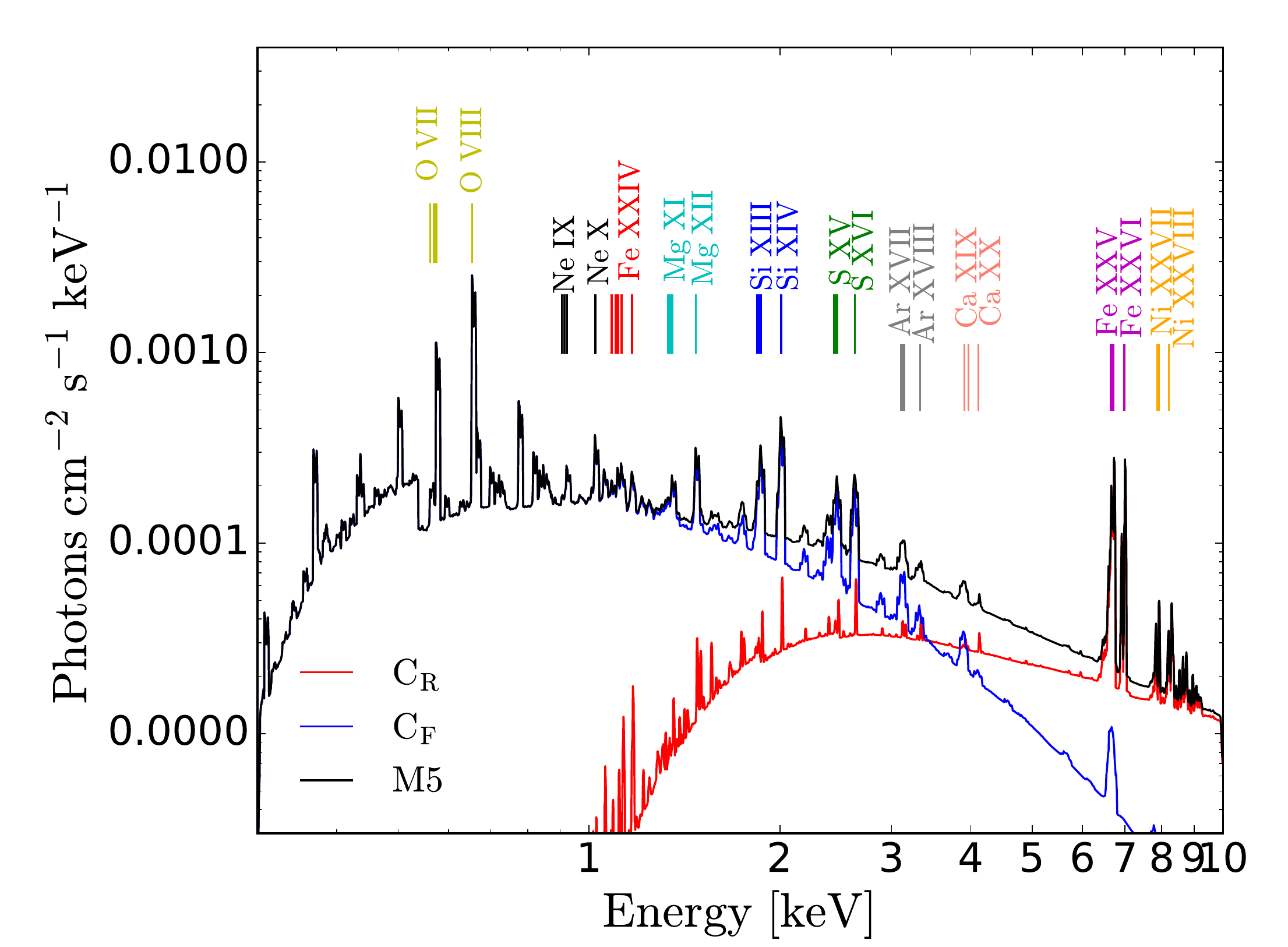}}
    \vspace{-0.5cm}
    \caption{Best-fit model M5 spectrum (\emph{black} line) between 0.5--10\,keV in units of Photons cm$^{-2}$ s$^{-1}$ keV$^{-1}$. The higher ($C_R$) and lower ($C_F$) temperature components are denoted in \emph{red} and \emph{blue}, respectively. The color vertical lines mark the most intense lines of the H-like and He-like ions of high-Z elements.}
    \label{fig:model_and_components}
\end{figure}

Thus far, models M1--M4 have failed to fit the velocity profiles of the Fe ions. We next consider two NEI components with distinct temperature, geometry, and foreground absorption terms [\texttt{TBabs(shellblur*vpshock)+ TBabs(shellblur*vpshock)}, hereafter model M5]. In the first \texttt{TBabs(shellblur*vpshock)} term, we fix the abundances for elements $Z{\leq}12$ to solar values, while for the second \texttt{TBabs(shellblur*vpshock)} term we fix the abundances for elements $Z{\leq}$9 to gas-phase abundance values, because of their role at different energies (see below). We further fix the line-of-sight to 55$^\circ$ for both terms. All other parameters in both components are left free. The fit resulted in: two absorption terms (lower and higher adsorptions), two different temperatures (a colder and hotter), and two interacting geometries (narrow and wide polar angles). The higher absorption term is coupled to the narrow polar cap plasma, allowing this component to contribute principally to the Fe\,XXVI and Fe\,XXV lines and the $>$4\,keV continuum, while minimizing its role at lower energies, which has already been shown to be problematic (e.g., Fig~\ref{fig:line_profile}). A best-fit is obtained with the parameters listed in Table~\ref{tab:parameters_error}, resulting in abundances ranging from 0.3--3.9 $Z_\odot$ in Table~\ref{tab:abundances} and a C-stat value of 8779.6 for 8528 DOF. Fig.~\ref{fig:modM5_2009} demonstrates that the best-fit M5 model results in reasonable fits to all of the strong lines in the 2009 epoch spectra. \emph{Dotted} and \emph{dashed} lines represent the  $C_F$ (low $kT$, low $N_{\rm H}$, wide polar angle) and $C_R$ (high $kT$, high $N_{\rm H}$, narrow polar angle) components, respectively; the subscripts 'F' and 'R' are in anticipation of our association of the components with the forward and reverse shocks in $\S$\ref{sec:evol_interp}. Table~\ref{tab:cstat} shows the C-stat values corresponding to each model and their DOFs, demonstrating that M5 yields the lowest C-stat (and BIC/AIC) value for the 2009 epoch. 


To confirm that model M5 is the best-fit model for the 2009 epoch, we consider the four different statistical assessment methods from \S\ref{sec:statistic}. First, we simulated 1000 spectra of the epoch 2009 using model M5 and computed best-fit  $\Delta$C-stat values, which should be distributed like $\chi^2$ \citep{Wilks1938}. Based on this distribution, model M5 provides a statistically better fit over models M1 and M2, at $>$95\% confidence among all realizations. This is reflected by the poor fit of M1 to the line profiles for the 2009 epoch (see Fig.~\ref{fig:Si_Fe_vpshock}) and the unsuccessful fit of the Fe lines for M2 for the 2009 epoch (see Fig.~\ref{fig:Lineas_shell_vpshock}). However, the fit distributions are unable to rule out models M3 and M4 for the 2009 epoch with high ($>$50\%) confidence.

We now turn to the AIC, BIC, and BXA methods. 
Table~\ref{tab:cstat} lists the AIC, BIC and BXA $\log$~Z values for models M1--M5 fit to the 2009 epoch spectra. The AIC and BIC show very similar behavior, highlighting a clear distinction between M1 and the rest, followed by modest decreases from models M2 to M3 to M4, and a subsequent drop for M5, which produces the lowest BIC and AIC values among the fits to the 2009 spectra, implying it should be considered the best-fit model. The BXA comparison method arrives at a similar conclusion, whereby model M1 produces the lowest evidence, followed by M4, M2, M3, and finally model M5 with the highest evidence value. Based on the criteria that $\log{\rm Z_1}{-}\log{\rm Z_2}{>}1$, model M5 is clearly preferred above all others. In summary, the AIC, BIC and BXA criteria all favor M5 to explain better the 2009 epoch spectra at high confidence.

Given that multiple temperature components are expected even in 1-dimensional shocks \citepalias[e.g.,][]{Dwarkadas2010a}, and that there appear to be two geometrically distinct shocks as traced by the line profiles, we are tempted to consider additional plasma components. To this end, we added an extra absorbed NEI model, both fixing its geometrical components to one of the previous polar scenarios (wide or narrow) as well as fitting the parameters freely. In none of these cases do we find a statistically significant improvement with respect to model M5, indicating that two dominant plasma components appear sufficient to explain the physical nature of the SN shock. 

Fig.~\ref{fig:model_and_components} shows the full theoretical model M5 (\emph{black} curve) along with the individual high ($C_R$) and low ($C_F$) temperature NEI plasma components (\emph{red} and \emph{blue} curves, respectively) and the most important emission lines (vertical colored lines). To limit the $C_R$ component from contributing to low-energy lines or dominating the continuum, it must be strongly absorbed ($2.47\times10^{22}$ cm$^{-2}$) compared to the $C_F$ component ($1.7\times10^{21}$ cm$^{-2}$). In this scenario, the higher temperature, narrow polar component (\emph{red}) contributes strongly to high-energy lines and underlying continuum of Fe, Ni, and modestly to mid-energy lines of S, Ar, and Ca, while the lower temperature, wider polar component (\emph{blue}) contributes strongly to the low-energy lines and continuum of Ne, Mg, Si, Fe\,XXIV, and modestly to mid-energy lines of S, Ar, and Ca. We note that the temperature, $kT_{e}$, in the hotter component is not well-constrained, owing to the HETG's limited energy range and decreasing effective area at high energies (see Table~\ref{tab:parameters_error}). While the best-fit value is $kT_{e}$ $=$33.5\,keV, we obtain a 3-$\sigma$ range spanning $\sim$10--80\,keV. The bright H-like and He-like Fe lines, which require a high degree of ionization, provide additional constraints on the temperature, although these are somewhat degenerate with abundance. Thus, it is crucial to have wide-band X-ray coverage to constrain the temperature of the SNe and disentangle instrumental and physical effects \citep[e.g., the case of SN\,2010jl in ][]{Chandra2015a,Chandra2018}. Although \emph{NuSTAR} observations covering the 3--79\,keV range exist for the Circinus galaxy, due to their coarse spatial resolution the emission from SN\,1996cr is severely contaminated by the much stronger AGN emission \citep[see Fig.~1 of][]{Arevalo2014}.

Given that the contours of the lower energy lines of Mg, Si, and Fe\,XXIV skew toward higher $\theta_{\rm min}$ in Fig.~\ref{fig:contornos}, these lines may indeed have a potential contribution from the narrow polar component which is not being modeled with M5. Therefore, we consider one final scenario, in which the hotter polar component is only partially absorbed [\texttt{TBabs(shellblur*vpshock)+TBpcf(shellblur*vpshock)}, hereafter model M6]. Model M6 introduces two additional parameters, the redshift $z$ and a partial covering fraction PCF, and allows us to evaluate whether the narrow polar component contributes to emission lines below $\sim$4\,keV. We obtained a best-fit with model M6 yielding a C-stat value of 8785.95 for 8527 DOF, which is slightly higher than the best-fit for model M5, resulting in higher AIC and BIC values, as well as lower evidence Z. In addition, the PCF parameter converged to a value of 1.0, suggesting that the narrow polar component does not contribute significantly to emission lines below $\sim$4\,keV. Given these results, we do not include model M6 in Table~\ref{tab:cstat} and consider model M5 to be the best and final model for the 2009 epoch (see Fig.~\ref{fig:modM5_2009}).


\subsection{Other epochs}\label{sec:other_epochs}

With a reasonable physical model for the 2009 epoch in hand, we review its applicability on the high-resolution 2000 and 2004 \emph{Chandra} HETG epochs, and then apply it to all of the epochs to explore how the ejecta-CSM interaction in SN\,1996cr evolved between years 5 and 21 post-explosion, thereby reconstructing its history. 

\begin{figure*}
    \centering
    \includegraphics[scale=0.7]{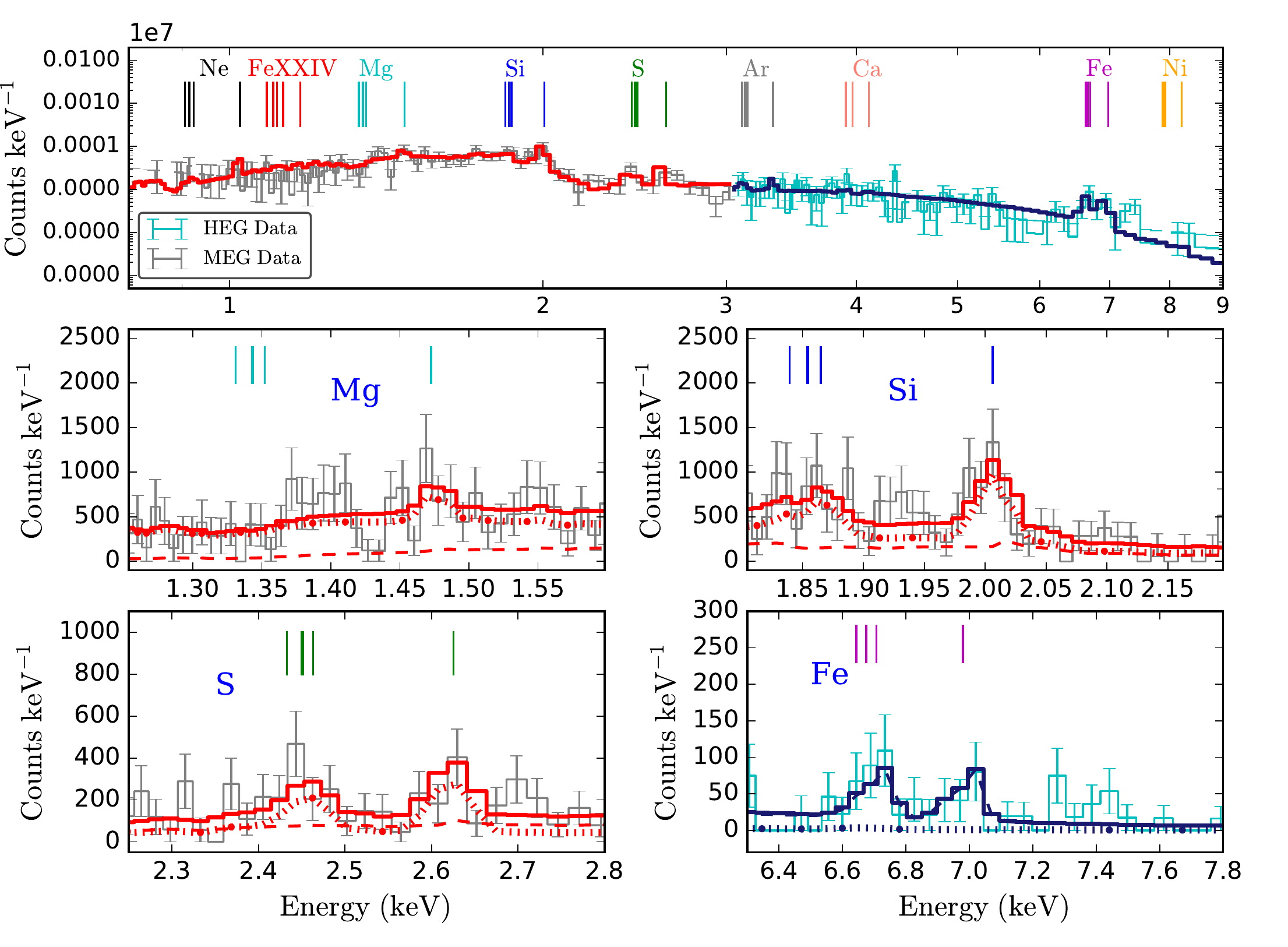}
    \caption{Same as Fig.~\ref{fig:modM5_2009} but for the 2000 epoch \emph{Chandra} HEG/MEG spectra.}
    \label{fig:modM5_2000}
\end{figure*}

\begin{figure*}
    \centering
    \includegraphics[scale=0.7]{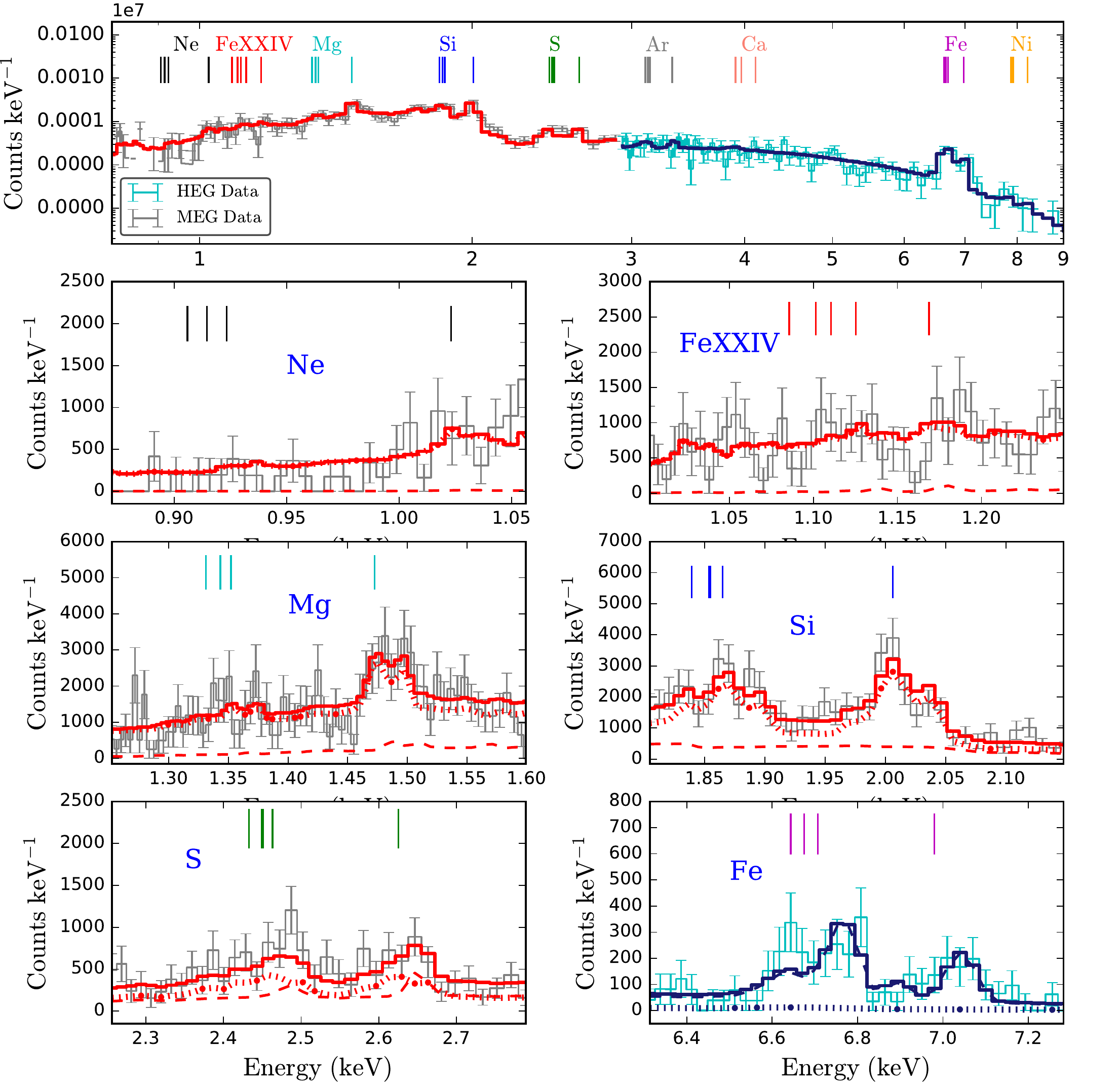}
    \caption{Same as Fig.~\ref{fig:modM5_2009} but for the 2004 epoch \emph{Chandra} HEG/MEG spectra.
    }
    \label{fig:modM5_2004}
\end{figure*}

We applied models M1 through M5 to the 2000 and 2004 \emph{Chandra} HETG epochs, fitting the geometry, temperature, absorption and abundance parameters as for the 2009 epoch, to confirm that model M5 remains the best one. In Table~\ref{tab:cstat}, we see that M5 remains the best-fitting model for the 2004 epoch, based on the C-stat, AIC, BIC and BXA $\log$~Z values, while for the 2000 epoch the results are mixed, with the C-stat, AIC and BIC values all supporting model M5 but the BXA $\log$~Z values favoring model M2. The mixed results for the 2000 epoch stem from the poor photon statistics (e.g., larger parameter errors, failure to detect some line complexes), as well as possible degeneracies in the models at early times. Given the consistency between the 2004 and 2009 results, the two highest signal-to-noise ratio HETG epochs, we adopt M5 as our fiducial best-fit model for all that follows. Figs.~\ref{fig:modM5_2000} and \ref{fig:modM5_2004} show the best-fitting model M5 for the 2000 and 2004 \emph{Chandra} epochs, respectively.

For the 2001, 2013, 2014, 2016, and 2018 \emph{XMM-Newton} epochs, the low spectral resolution of the \emph{pn} and MOS1/2 cameras results in geometrical degeneracies with the \texttt{shellblur model}, and thus we fix all of the geometrical parameters except the velocity expansion ($v_{\rm max}$) and the column density of the core ejecta ($N_{\rm ejecta}$) in component $C_R$ (since the spectral resolution of \emph{XMM-Newton} at $>$4\,keV is sufficient to estimate the width of the Fe\,XXVI and Fe\,XXV lines). 
Fitting model M5 to these epochs, we constrain the temperature ($T_e$), absorption ($N_H$) and abundance parameters of components $C_R$ and $C_F$, as well as $v_{\rm max}$ and $N_{\rm ejecta}$ for $C_R$. The spectral fits for the \emph{XMM-Newton} epochs using model M5 are shown in Figs. \ref{fig:modM5_2001}--\ref{fig:modM5_2018}.

We highlight an interesting feature between $\sim$7.3--7.5\,keV in the 2000 epoch spectrum (see Fig.~\ref{fig:modM5_2000}) that model M5 fails to fit. This feature is comprised of 11 counts, well above the expected continuum signal and unexpected given \emph{Chandra's} strongly decreasing effective area here. We verified that this emission at $\sim$7.4\,keV in the 2000 epoch does not come from contamination of other sources (AGN or off-nuclear point-sources) in the HETG dispersed spectra. The feature does not obviously correspond to any previously modeled element (e.g., H-like or He-like Fe, as indicated in Fig.~\ref{fig:modM5_2000} or Ni XXVII and XXVIII at $\sim$7.8\,keV and 8.1\,keV, respectively). We do not see similar velocity components from other elements. 

\begin{figure}
    \centering
    \hglue-0.5cm{\includegraphics[scale=0.7]{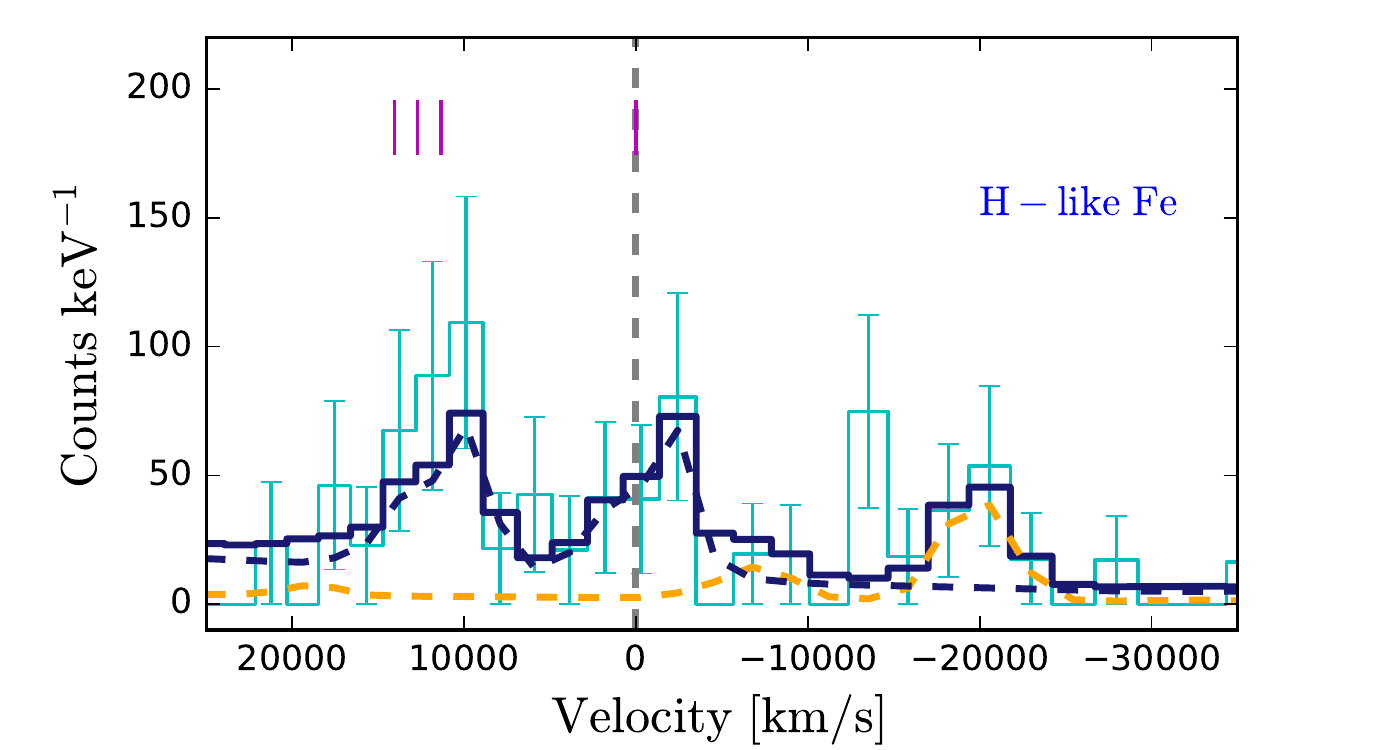}}
    \caption{The Fe emission complex of the 2000 epoch HEG-\emph{Chandra} data (\emph{cyan} histogram), centered on the H-like line in velocity space (vertical \emph{grey dashed} line). In addition to the hotter $C_R$ component (\emph{dashed blue} curve) obtained from model M5, we show a possible third NEI plasma component (\emph{yellow} dashed line) with a shell expansion velocity of $\sim23000$ km/s; the total model is shown as the \emph{dark, blue} histogram. The \emph{magenta vertical lines} denote the location in velocity space of the  H-like and He-like Fe lines.}
    \label{fig:Fe-bump}
\end{figure}

We consider briefly that this line complex arises from possible He-like and/or H-like Fe emission associated with "bullet"-like Fe ejecta \citep[e.g., similar to Cas A;][]{Willingale2002}, and model it with a third NEI plasma component convolved with a highly polar ($\theta_{\rm min}$$\sim$85$^{\circ}$, $i$$=$90$^{\circ}$) geometry and an exceptionally high expansion velocity ($v_{\rm max}$$\sim$23000\,km\,s$^{-1}$). The result provides a reasonable match to the data, as seen in Fig.~\ref{fig:Fe-bump}. The C-stat value for the epoch 2000 modestly improves by  $\Delta$C-stat$=$2.13 compared to the nominal M5 model. This Fe complex also appears weakly as a residual in the 2001 spectrum (see Fig.~\ref{fig:modM5_2001}), but not in the following epochs. For the 2001 epoch, adding such a "bullet"-like Fe plasma structure improves somewhat the fit to the \emph{XMM-Newton} data at $\sim$7.4\,keV (see Figure~\ref{fig:modM5_2001}). Alternatively, the line could be associated with a highly redshifted Ni~XXVII ($\sim$7.8\,keV) or Ni~XXVIII (8.1\,keV) "bullet"-like structure ($v_{\rm max}$$\sim$ 15000--26000\,km\,s$^{-1}$, $i$$=$90$^{\circ}$). We do not consider this possibility as viable as Fe, however, because the flux of this component would be roughly equal to what we estimate for all of the lower velocity Ni, even before we correct for any potentially high $N_{\rm ejecta}$, as found for Fe. Finally, another possible identification could be the 7.47\,keV Ni K$\alpha$ fluorescent line, but this would be quite unexpected as it requires cold reflection \citep[e.g.,][]{Yaqoob2011} and we do not see the correspondingly stronger 6.4\,keV Fe K$\alpha$ in the 2000 epoch spectra. 

We also find discrepancies between the model and the He-like S and Fe lines in the 2004 epoch spectrum (see Fig.~\ref{fig:modM5_2004}). We do not attempt further fine-tuning, as the differences do not appear internally consistent. That is, the bright unmodeled peaks in the He-like S or Fe lines do not occur in the same portion of the velocity profile. It is possible that the peak at $\sim$6.65\,keV comes from a different non-polar geometrical origin or clumped material, but we do not explore these possibilities. We simply note that abundance inhomogeneities and asymmetries may exist.

\begin{figure*}
    \centering
    \includegraphics[scale=0.73]{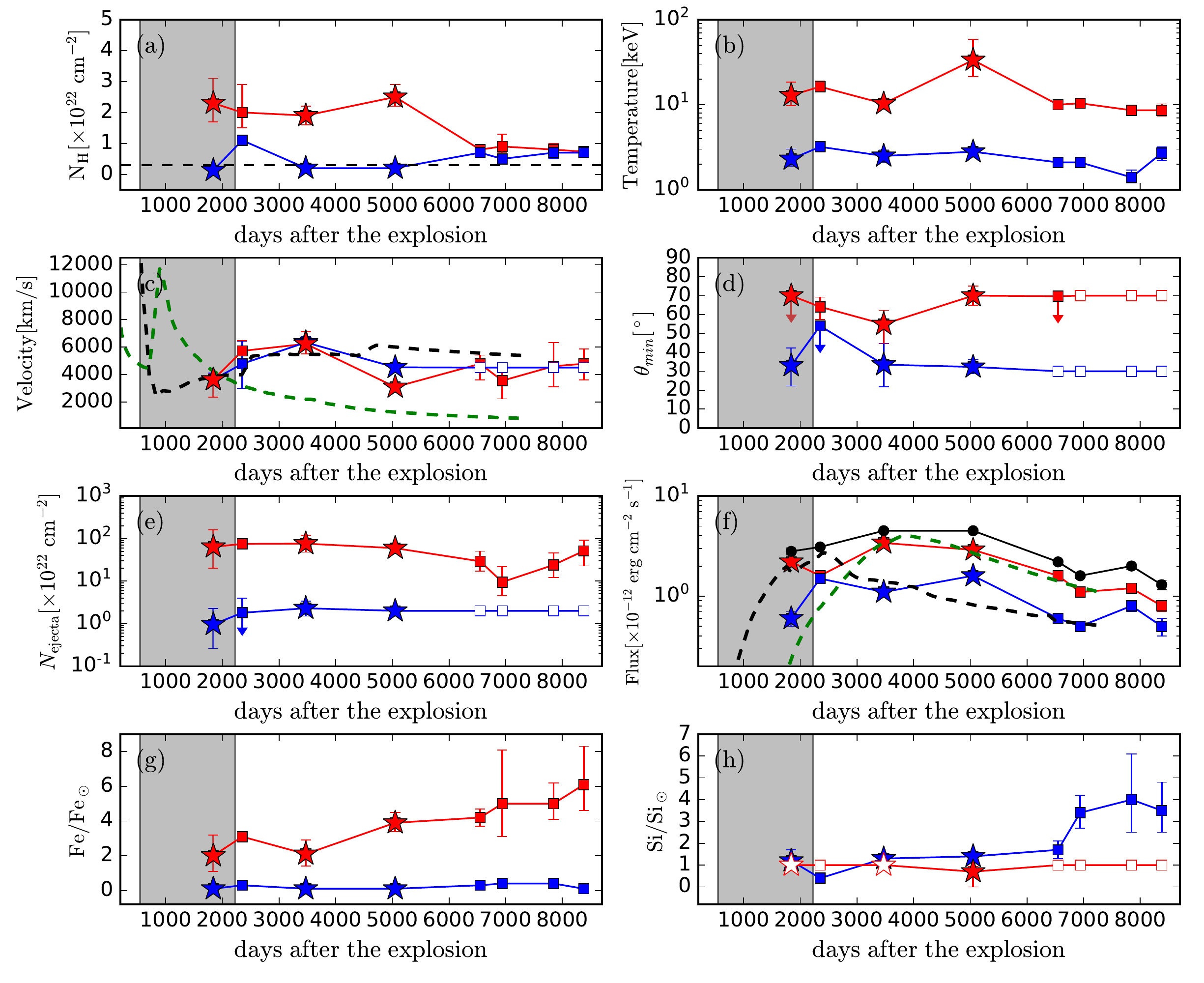}
    \vspace{-0.5cm}
    \caption{Evolution of parameters in different epochs. \emph{Red symbols} and \emph{blue symbols} denote parameters associated with the $C_R$ (high $kT$, high $N_{\rm H}$, narrow polar angle) and $C_F$ (low $kT$, low $N_{\rm H}$, wider polar angle) components, respectively. \emph{Open symbols} denote parameters that were fixed to previous best-fit values in the fitting analysis. The \emph{gray region} in each panel denotes the epoch during which the forward shock interacted with the dense CSM shell, according to the model of \citetalias{Dwarkadas2010a}. \emph{Stars} and \emph{squares} represent \emph{Chandra} and \emph{XMM-Newton} data, respectively. Panels $(a)$ and $(b)$ show the foreground absorption ($N_{\rm H}$) and NEI temperatures of components $C_R$ and $C_F$. The \emph{dashed black} line in panel $(a)$ represents the estimated Galactic column density (3$\times$10$^{21}$ cm$^{-2}$); values significantly above this imply extra internal absorption either from the disk of the Circinus Galaxy or the immediate vicinity of SN\,1996cr \citep{Bauer2008a}. Panels $(c)$, $(d)$ and $(e)$ show the geometrical model velocity expansion, opening-angle, and absorption from inner ejecta, respectively. The \emph{dashed black} and \emph{green} lines in panel $(c)$ represent the forward and reverse velocities obtained by \citetalias{Dwarkadas2010a}. Panel $(f)$ shows the absorption-corrected 0.5--8.0\,keV fluxes from each component separately and the total flux (\emph{black} circles), as well as the individual shocked ejecta and CSM contributions from the simulations of \citetalias{Dwarkadas2010a}. Finally, panels $(g)$ and $(h)$ show the Fe and Si abundances, respectively.}
    \label{fig:parameters_vs_time}
\end{figure*}

In Fig.~\ref{fig:parameters_vs_time}, we can see how various parameters evolve with time since the explosion of the SNe; we adopt an explosion date of 1995.4. The \emph{grey} region denotes the time during which the SN forward shock interacted with a dense shell, based on 1-D hydro-dynamical simulations \citepalias{Dwarkadas2010a}. Table~\ref{tab:parameters_error} gives the errors from M5 for the 2000, 2001, 2004, 2009, 2013, 2014, 2016 and 2018 epochs, while Table~\ref{tab:abundances} shows the abundances obtained in these epochs. In the next section, we search for plausible physical explanations for model M5 and its parameter evolution, and try to discuss the CSM geometry of SN\,1996cr.

\begin{table*}
   \caption{Spectral parameters for model M5 at different epochs. \emph{Col. 1}: Epoch of combined X-ray observations. \emph{Col. 2:} satellite and instrument. \emph{Column 3:} spectral model component. \emph{Col. 4:} column density $N_{\rm H}$ in units of $\times10^{22}$ cm$^{-2}$. \emph{Col. 5:} temperature in keV. \emph{Col. 6:} opening-angle measure in degrees ($^\circ$). \emph{Col. 7:} inclination angle respect to the line-of-sight in degrees ($^\circ$). \emph{Col. 8:} maximum expansion of the emission region $v_{\rm max}$ in units of km\,s$^{-1}$. \emph{Col. 9:} column density $N_{\text{ejecta}}$ of ejecta interior to reverse shock in units of $\times10^{22}$ cm$^{-2}$. \emph{Col. 10} unabsorbed $0.5$--$10.0$\,keV flux in units of $\times10^{-12}$ erg cm$^{-2}$ s$^{-1}$. \emph{Col. 11:} C-stat value and degree of freedom of the model. Upper limits are defined with 3-$\sigma$ confidence level.} 
       \scalebox{0.8}{
       \begin{tabular}{lccccccccccc}
       \hline
       Epoch & Telescope & Model & $N_{\rm H}$ & $kT$ & $\theta_{\text{min}}$  & $i$ & $v_{\text{max}}$ & $N_{\text{ejecta}}$ & Unabs. Flux & C-stat (DOF) \\ 
       (1) & (2) & (3) & (4) & (5) & (6) & (7) & (8) &
       (9) & (10) & (11)\\ \hline\hline
       
       2000 & \emph{Chandra}-HETG & \texttt{TBabs$_{C_R}$(shellblur$_{C_R}$*vpshock$_{C_R}$)+} & $2.3_{-0.6}^{+0.8}$ & $12.9_{-3.2}^{+5.5}$ & $<70\fdg0$ & $55\fdg0$(fix) & $3618.3_{-1279.3}^{+410.3}$ & $63.8_{-43.5}^{+97.1}$ & $2.2_{-0.2}^{+0.2}$ & \multirow{2}{*}{4398.6(8533)}\\
       \vspace{0.1cm}
            &  & \texttt{TBabs$_{C_F}$(shellblur$_{C_F}$*vpshock$_{C_F}$)} & $0.13_{-0.1}^{+0.1}$ & $2.3_{-0.4}^{+0.7}$ & $33\fdg0_{-10.8}^{+9.4}$ & $55\fdg0$(fix) & $3619.2_{-288.9}^{+482.6}$ & $0.96_{-0.7}^{+1.3}$ & $0.6_{-0.1}^{+0.1}$\\

      2001 & \emph{XMM-Newton} & \texttt{TBabs$_{C_R}$(shellblur$_{C_R}$*vpshock$_{C_R}$)+} & $2.0_{-0.5}^{+0.9}$ & $16.3_{-2.2}^{+2.6}$ & $64\fdg1_{-6.7}^{+5.1}$ & $55\fdg0$(fix) & $5717.3_{-878.8}^{+760.7}$ & $74.9_{-18.1}^{+23.9}$ & $1.6_{-0.1}^{+0.1}$ & \multirow{2}{*}{3054.4(3063)}  \\
       \vspace{0.1cm}
            &  & \texttt{TBabs$_{C_F}$(shellblur$_{C_F}$*vpshock$_{C_F}$)} & $1.1_{-0.1}^{+0.1}$ & $3.2_{-0.3}^{+0.3}$ & $<54\fdg0$ & $55\fdg0$(fix) &  $4780.0_{-1778.8}^{+1633.6}$ & $1.8_{-1.8}^{+2.2}$ & $1.5_{-0.1}^{+0.1}$ \\
            
      2004 & \emph{Chandra}-HETG & \texttt{TBabs$_{C_R}$(shellblur$_{C_R}$*vpshock$_{C_R}$)+} & $1.9_{-0.3}^{+0.4}$ & $10.3_{-1.4}^{+1.6}$ & $54\fdg9_{-10.1}^{+7.3}$ & $55\fdg0$(fix) & $6232.8_{-733.7}^{+884.5}$ & $76.1_{-28.4}^{+43.4}$ & $3.4_{-0.2}^{+0.2}$ & \multirow{2}{*}{6264.3(8533)} \\
       \vspace{0.1cm}
            &  & \texttt{TBabs$_{C_F}$(shellblur$_{C_F}$*vpshock$_{C_F}$)} & $0.2_{-0.1}^{+0.1}$ & $2.5_{-0.4}^{+0.5}$ & $33\fdg5_{-11.8}^{+11.0}$ & $55\fdg0$(fix) & $6333.2_{-203.9}^{+415.2}$ & $2.3_{-0.8}^{+1.1}$ & $1.1_{-0.1}^{+0.1}$ \\
       
      2009 & \emph{Chandra}-HETG & \texttt{TBabs$_{C_R}$(shellblur$_{C_R}$*vpshock$_{C_R}$)+} & $2.5_{-0.3}^{+0.4}$ & $33.5_{-12.1}^{+25.3}$ & $70\fdg0_{-5.1}^{+5.1}$ & $55\fdg0$(fix) & $3085.9_{-160.6}^{+353.7}$ & $59.5_{-11.6}^{+13.9}$ & $2.9_{-0.1}^{+0.1}$ & \multirow{2}{*}{8779.6(8528)} \\
       \vspace{0.1cm}
            &  & \texttt{TBabs$_{C_F}$(shellblur$_{C_F}$*vpshock$_{C_F}$)} & $0.2_{-0.03}^{+0.03}$ & $2.8_{-0.1}^{+0.2}$  & $32\fdg3_{-3.6}^{+3.7}$  & $55\fdg0$(fix) & $4522.1_{-108.1}^{+166.7}$ & $2.0_{-0.3}^{+0.4}$ & $1.6_{-0.03}^{+0.02}$ \\     

      2013 & \emph{XMM-Newton} & \texttt{TBabs$_{C_R}$(shellblur$_{C_R}$*vpshock$_{C_R}$)+} & $0.8_{-0.1}^{+0.1}$ & $10.0_{-0.7}^{+0.9}$ & $<69\fdg7$ & $55\fdg0$(fix) & $4777.0_{-1174.0}^{+636.4}$ & $29.2_{-12.2}^{+21.2}$ & $1.6_{-0.1}^{+0.1}$ & \multirow{2}{*}{563.9(534)} \\
       \vspace{0.1cm}
            &  & \texttt{TBabs$_{C_F}$(shellblur$_{C_F}$*vpshock$_{C_F}$)} & $0.7_{-0.1}^{+0.1}$ & $2.1_{-0.2}^{+0.3}$ & $30\fdg0$(fix) & $55\fdg0$(fix) & 
            $4500.0$(fix) & 2.0(fix) & $0.6_{-0.04}^{+0.04}$ \\
            
      2014 & \emph{XMM-Newton} & \texttt{TBabs$_{C_R}$(shellblur$_{C_R}$*vpshock$_{C_R}$)+} & $0.9_{-0.2}^{+0.4}$ & $10.4_{-1.1}^{+1.6}$ & $70\fdg0$(fix) & $55\fdg0$(fix) & $3546.5_{-1321.4}^{+783.7}$ & $9.5_{-5.0}^{+12.2}$ & $1.1_{-0.1}^{+0.1}$ & \multirow{2}{*}{180.31(503)}\\ 
       \vspace{0.1cm}
            &  & \texttt{TBabs$_{C_F}$(shellblur$_{C_F}$*vpshock$_{C_F}$)} & $0.5_{-0.1}^{+0.1}$ & $2.1_{-0.2}^{+0.3}$ & $30\fdg0$(fix) & $55\fdg0$(fix) & 4500.0(fix) & 2.0(fix) & $0.5_{-0.04}^{+0.04}$ \\
      
      2016 & \emph{XMM-Newton} & \texttt{TBabs$_{C_R}$(shellblur$_{C_R}$*vpshock$_{C_R}$)+} & $0.8_{-0.1}^{+0.2}$ & $8.6_{-1.0}^{+1.1}$ & $70\fdg0$(fix) & $55\fdg0$(fix) & $4602.4_{-1502.5}^{+1721.5}$ & $24.0_{-11.8}^{+22.0}$ & $1.2_{-0.1}^{+0.1}$ & \multirow{2}{*}{195(372)}\\
       \vspace{0.1cm}
            &  & \texttt{TBabs$_{C_F}$(shellblur$_{C_F}$*vpshock$_{C_F}$)} &  $0.7_{-0.2}^{+0.2}$ & $1.4_{-0.2}^{+0.3}$ & $30\fdg0$(fix) & $55\fdg0$(fix) & 4500.0(fix) & 2.0(fix) & $0.8_{-0.1}^{+0.1}$ \\
      
      2018 & \emph{XMM-Newton} & \texttt{TBabs$_{C_R}$(shellblur$_{C_R}$*vpshock$_{C_R}$)+} &  $0.73_{-0.1}^{+0.1}$ & $8.6_{-1.2}^{+1.6}$ & $70\fdg0$(fix) & $55\fdg0$(fix) & $4781.3_{-1180.8}^{+1045.3}$ & $51.3_{-28.6}^{+41.1}$ & $0.8_{-0.1}^{+0.1}$ & \multirow{2}{*}{2007.7(2045)}\\
            &  & \texttt{TBabs$_{C_F}$(shellblur$_{C_F}$*vpshock$_{C_F}$)} & $0.7_{-0.1}^{+0.1}$ & $2.7_{-0.5}^{+0.5}$ & $30\fdg0$(fix) & $55\fdg0$(fix) & 4500.0(fix) & 2.0(fix) & $0.5_{-0.1}^{+0.1}$ \\
\hline
       \end{tabular}
       }
       \label{tab:parameters_error}
\end{table*}

\vspace{-0.3cm}
\section{Results and discussion}\label{sec:Results}

In $\S$\ref{sec:MethodModel}, we derived a best-fit model to match the 2009 epoch X-ray spectra of SN\,1996cr and extended it to other epochs spanning $\pm$9 years. The spectra are successfully explained with only two distinct NEI components: a hot, heavily absorbed, high-latitude polar shock ($C_R$) and a cooler, moderately absorbed, wider polar shock ($C_F$). Here we explore the physical nature of each component and how the parameters evolve over time.

\vspace{-0.3cm}
\subsection{Interpretation of Model M5 at 2009 epoch}\label{sec:2009_interp}

The shock interaction in SNe can be quite complex \citep{Chevalier1992a, Michael2002a, Dwarkadas2007a, Orlando2019}, since it depends on the 3-D density distributions of the expanding ejecta and pre-existing CSM \citep[e.g.][]{DeLaney2010, Milisavljevic2013, Orlando2015, Orlando2016}. The canonical self-similar description of a spherically symmetric shock traveling into a spherically symmetric power-law medium produces a double-shock structure, consisting of a blast wave (forward shock) that travels outwards into the CSM and a reverse shock that travels back (in a Lagrangian sense) into the SN ejecta \citep{Chevalier1982a}. Between the forward and reverse shocks, there should exist shocked CSM and ejecta material separated by a contact discontinuity. Shock expansion typically results in X-ray and radio emission associated with the two shocks \citep{Chevalier1982b}, with the forward/reverse shocks thought to dominate at hard/soft X-rays energies ($\gtrsim$2\,keV / $\lesssim$2\,keV), respectively.

Core-collapse SNe evolve into the wind-driven regions created by prior mass-loss from their progenitor stars, and it is these regions that subsequently define the SN expansion, dynamics and kinematics. If the wind parameters remain constant, as is expected for most SNe, then the wind region should have a density profile which goes as $\rho_{\rm CSM}$$\propto r^{-2}$ \citep{Chevalier1982c,Chevalier1982b}. In the particular case of SN\,1996cr, \citetalias{Dwarkadas2010a} previously demonstrated that the wind parameters changed hundreds to thousands of years before explosion, resulting in a fast-to-slow wind collision and the formation of a dense shell of swept-up CSM. Thus, SN\,1996cr is somewhat different from the canonical picture, due to the nature of the medium surrounding the SN. A comparison between several epochs of X-ray spectra and \hbox{1-D} hydrodynamical simulations led \citetalias{Dwarkadas2010a} to propose that SN\,1996cr initially exploded into a low-density medium and, after $\sim$1.5\,yrs, the blast wave encountered a dense shell of material, at $\approx$0.03\,pc \citep[three times smaller than SN\,1987A,][]{Dewey2012a} from the progenitor star. The estimated width and density of this shell was found to be consistent with expectations from a WR or BSG wind having swept up a previously existing RSG wind. In this model, the bulk of the X-ray emission during the first 7 years arose from the forward shock (shocked CSM), while the reverse shock emission (shocked ejecta material) was dominant thereafter. The \citetalias{Dwarkadas2010a} model offered up a successful physical framework that fit the continuum shapes and emission-line strengths reasonably well, although that work made no attempt to model the line profiles (shape or velocity) as we have done here. Of particular importance, \citetalias{Dwarkadas2010a} assumed spherical symmetry and found that the observed temperature stratification, as evidenced by the flat continuum and line strengths of the SN\,1996cr spectra, could be naturally explained as the sum of the different radial components of the shock. \citet{Dewey2011a} built upon the 1-D model of \citetalias{Dwarkadas2010a} using a 3-D convolution technique \citep[based on][]{Dewey2009a} to fit the velocity profiles. \citet{Dewey2011a} principally reported on an analysis of the Si and Fe lines profiles for the 2009 epoch, which implied a non-spherical ejecta--CSM interaction geometry, but did not investigate the evolution as we do here.


\subsubsection{Two distinct shocks}\label{sec:2009geometry}
From our analysis of the emission line shapes in $\S$\ref{sec:2009_geom}, we are able to reject a spherically symmetric $4\pi$ emission geometry at high confidence. We can likewise reject a ring-like emission geometry (e.g., similar to SN\,1987A) with high confidence. Instead, we find that the S, Si, and Mg lines are best-fit by an inclined, wide-angle polar geometry ($C_F$) covering ${\approx}2{\pi}$ solid angle on the sky, while the highest ionization Fe XXVI line is best fit by a similarly inclined narrow-angle polar geometry. Intriguingly, the more modest ionization Fe XXIV and XXV lines have best-fit values intermediate between the wide and narrow components, suggesting potential contributions from both, although the uncertainties that remain are large. Substantial degeneracy appears to exist between the opening and inclination angles for the narrow-angle component, as seen in Fig.~\ref{fig:contornos}. However, it is reassuring that the best-fit inclination angle for both the $C_R$ and $C_F$ components appears to be ${\sim}55^\circ$. 

More complex geometries (e.g., a partial or elliptical ring, clumps) may provide acceptable fits to the line profiles --- e.g., emission from two points opposite each other on a ring will be highly degenerate with the narrow-angle polar emission we currently observe --- however, we feel that these need substantial further observational or theoretical justification to consider them. Unfortunately, the X-ray observations are unresolved and VLBI observations have not yet managed to define the structure in SN\,1996cr \citep{Bietenholz2014a}.

Our geometrical constraints do not necessarily invalidate the work of \citetalias{Dwarkadas2010a}, which appears to effectively capture the broad characteristics of the shock interaction and explain several key observational signatures (e.g., the CSM density profile and shock energetics leading to the X-ray light curve, and elemental abundances). We already noted the impressive self-consistency between the \citetalias{Dwarkadas2010a} inner density and the best-fit $N_{\rm ejecta}$ for component $C_F$. Notably, \citetalias{Dwarkadas2010a} assumed spherical symmetry and convolved their spectra with a simple doppler broadening, but never incorporated the actual HETG emission line profiles into their model, so it is perhaps not surprising that we find discrepancies by factors of 2--5 between our $C_R$ and $C_F$ component velocities and the forward and reverse shock velocities derived by \citetalias{Dwarkadas2010a} from the 2009 epoch (see panel $(c)$ of Fig.~\ref{fig:parameters_vs_time}). Our results are generally consistent with those of \citet{Dewey2011a}, who incorporated 3-D convolution models with velocity effects to fit the 2009 epoch HETG continuum and emission line spectra. Furthermore, \citet{Dewey2011a} found that the temperature-dependent line profiles implied that the progenitor CSM around SN\,1996cr was most likely denser at the poles.

Based on our best-fit polar geometry, our results imply that the solid area covered by the shock must be proportionally smaller than 4$\pi$ by factors of $\approx$2 for component $C_F$ and $\approx$15--30 for component $C_R$. For the $C_F$ component, this naively implies only minor adjustments to the CSM density, radius, or ejecta energetics, while for $C_R$, a more dramamtic adjustment will be required. A more complex issue is how the introduction of two spatially distinct shocks, components $C_R$ and $C_F$, will affect the layered temperature stratification and small-scale clumping introduced in the \citetalias{Dwarkadas2010a} model, which ultimately contribute to the continuum shape and emission-line strengths.

The potential polar geometry of the ejecta-CSM interaction of SN\,1996cr could result from either bullet-like ejecta \citep[as in Cas A;][]{Orlando2016} or previous mass-loss phases of a massive progenitor. In the case of the former, we might expect higher density ejecta at higher velocities. Regarding the latter, one possible channel to sculpt such a CSM feature is from an eccentric binary system undergoing eruptive mass loss. We directly observe similar dense bipolar CSM regions in evolved stars like $\eta$ Carinae \citep{Davidson1997a, Smith2007a, Smith2018a} and Betelgeuse \citep{Kervella2018a}, as well as indirectly in SN imposters like UGC2773-OT and SN\,2009ip \citep[e.g.,][]{Smith2010a, Mauerhan2014a, Reilly2017a}, or type IIn SNe such as SN\,2012ab \citep{Bilinski2018a}, and of course SN\,1987A, which is the clearest example of a SN evolving into a bipolar bubble. We return to this point in $\S$\ref{sec:evol_interp} and \S\ref{sec:scenarios}

The thermal X-ray emission from CCSNe typically comes from the higher density of the reverse shock. Nevertheless, in the case of SNe IIn, the X-ray plasma temperatures are generally higher, more characteristic of high velocity expansion into lower density material, and thus implies that the X-ray emission arises from the forward shock \citep[e.g., SNs IIn 2005ip, 2005kd, 2006jd, 2010jl][]{Chandra2012a, Chandra2012b, Katsuda2014a, Dwarkadas2016a}. More generally, authors adopt two plasma components to fit the SNe X-ray spectra, where typically one component is associated with the forward shock emission region while the other is related to the reverse shock emission region \citep[e.g.,][among others]{Yamaguchi2008a, Yamaguchi2011a, Schlegel2004a, Chandra2009a, Nymark2009a}. In some cases, both components are argued to arise from the shocked CSM \citep[e.g.,][]{Katsuda2016a}, while more rarely additional non-thermal components are introduced to understand the role of synchrotron or inverse Compton processes \citep{Tsubone2017a}.

Notably, the two distinct shock components here draw some parallels to the two shock components seen in SN\,1987A. For example, \citet[][ see also Orlando et al. 2015]{Dewey2012a} were able to successfully model the X-ray spectra of SN\,1987A as the weighted sum of two NEI components from two simple 1-D hydrodynamic simulations: a $\approx$0.5\,keV component associated with the interaction of a dense equatorial ring ($\sim$2$^{\circ}$ width) and a $\sim$2--4\,keV component associated with the interaction of a sparser surrounding H{\sc ii} region ($\sim$30$^{\circ}$ width) to produce very-broad emission lines. The shock going into the ring is slower and cooler, while the shock above and below the ring (into less dense material) is faster and hotter. While SN\,1996cr does not appear to have an equatorial ring geometry, the concept of two shocks propagating into dense and less dense CSM still may apply. Ultimately, our favored interpretation is somewhat different, as we discuss in $\S$\ref{sec:evol_interp}.

\subsubsection{Shock symmetry and internal ejecta obscuration ($N_{\rm ejecta}$)}\label{sec:2009-NH_ejecta}

One novelty of spectral models M2--M6 is that, under the assumption of symmetry, they place constraints on the overall inner ejecta column density, $N_{\rm ejecta}$. We found in $\S$\ref{sec:2009_2compblur} that the farside of the $C_R$ velocity profile must be absorbed by an ejecta neutral hydrogen column density of $N_{\rm ejecta}$=59.5$^{+13.9}_{-11.6}\times10^{22}$\,cm$^{-2}$, while the $C_F$ velocity profile only requires $N_{\rm ejecta}$=2.0$^{+0.4}_{-0.3}\times10^{22}$\,cm$^{-2}$. Assuming a spherical shock radius of $\sim$0.065\,pc at the 2009 epoch based on the \citetalias{Dwarkadas2010a} model, these values translate to estimated angle-averaged densities of 1.7$\times$10$^{6}$ and 2.6$\times$10$^{5}$\,amu\,cm$^{-3}$, respectively, for an inclination angle of 55$^{\circ}$ and adopting mean ion masses of 3.6 and 1.7\,amu for the ejecta and CSM, respectively, following \citetalias{Dwarkadas2010a}. The latter is in relatively good agreement with the 1-D model of \citetalias{Dwarkadas2010a} (see their Fig. 3), demonstrating the rough validity of that model. The former, however, is higher by a factor of $\sim$7, indicating a much higher concentration of high-Z material along this line-of-sight. The large disparity between the $N_{\rm ejecta}$ values, which in theory should probe roughly comparable inner ejecta densities, implies either a strongly inhomogeneous ejecta structure, with  substantially denser material associated with the narrow-angle $C_R$ component,\footnote{In principle, drastically higher ionization levels associated with the ejecta in between component $C_F$ could also bring $C_R$ and $C_F$ closer, although this seems unlikely given that the ionization would likewise have to be patchy and there is no obvious source for the necessary strong ionizing radiation.} or our assumption of symmetry breaks down with component $C_R$ being 2--3 times stronger on the front side.

\subsubsection{Fe K-shell in broader context}\label{sec:FeKLum}

Finally, we note that the Fe K-shell line luminosities and energy centroids observed in nearby SNRs have been found to exhibit clear distinctions based on Ia and CCSNe progenitor types, explosion energy, ejecta mass, and circumstellar environment \citep[e.g.,][]{Yamaguchi2014a, Patnaude2015}. Typically, the Fe K-shell centroid energies from type Ia SNe have $<$6.55\,keV, implying lower ionization than those from CCSNe with $\geq$6.55\,keV. SN\,1996cr follows this general scheme, with a centroid energy of 6684.4$_{-7.6}^{+9.6}$\,eV and total Fe K-shell flux and luminosity of (2.4$\pm$0.3$)\times10^{-5}$ photons\,cm$^{-2}$\,s$^{-1}$ and (3.9$\pm$0.5$)\times10^{46}$ photons\,s$^{-1}$ for the 2009 epoch, respectively. As such, SN\,1996cr lies in the extreme upper right corner of the ``Yamaguchi plot'' \citep[see Fig.~1 of][]{Yamaguchi2014a}, implying that Fe-rich ejecta reach the shock interaction region relatively quickly during the explosion. Notably, SN\,1996cr's values lie well outside of the theoretical models, consistent with it being a relatively extreme SN.

\begin{figure*}
    \begin{center}
    $
    \begin{array}{lccr}
    (a) \hspace{3.5cm} & (b) \hspace{3.5cm} & (c) \hspace{1.5cm} & \hspace{2.0cm} (d)   
    \end{array}$
    $
    \begin{array}{llll}
    \includegraphics[scale=0.16]{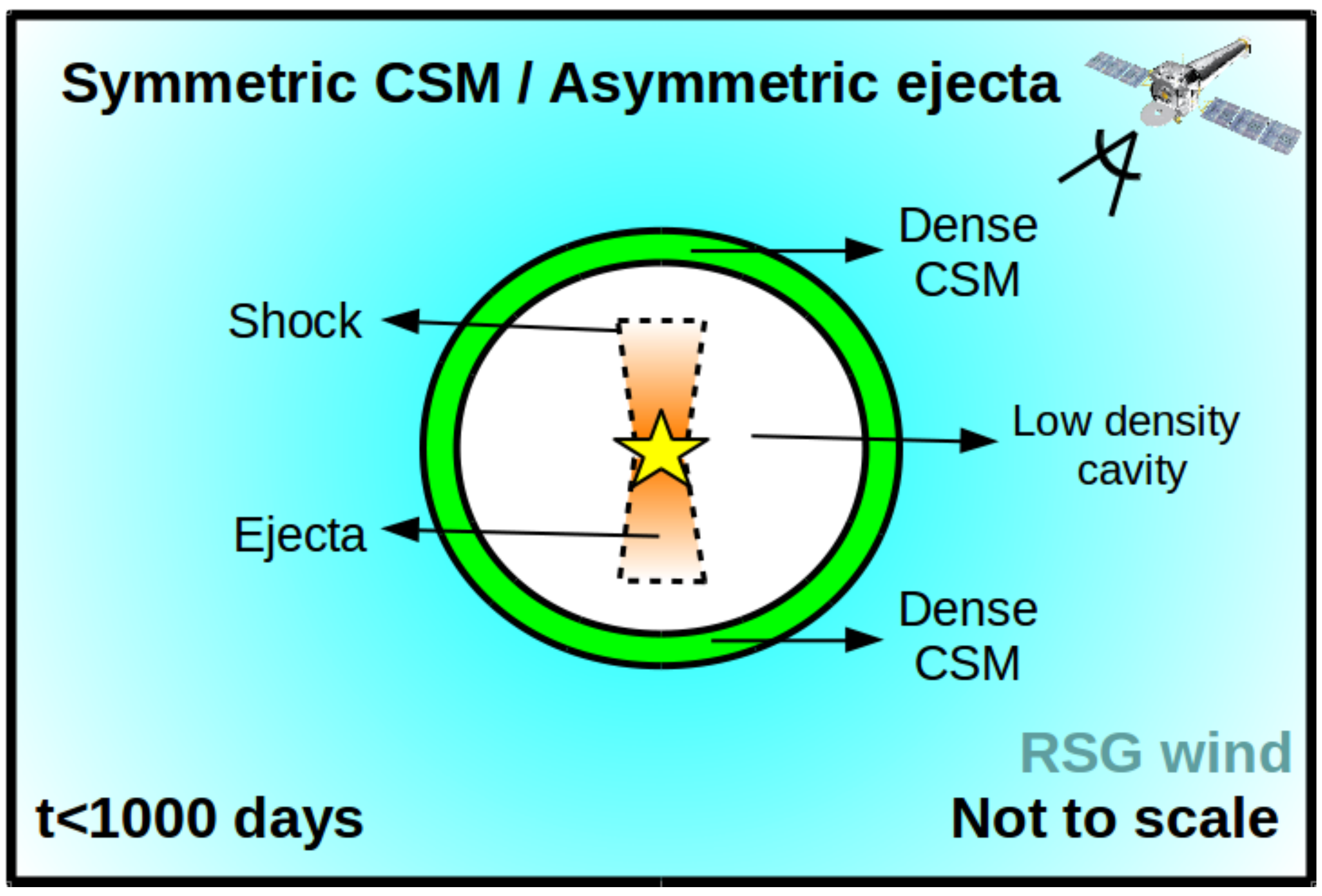} &
    \includegraphics[scale=0.16]{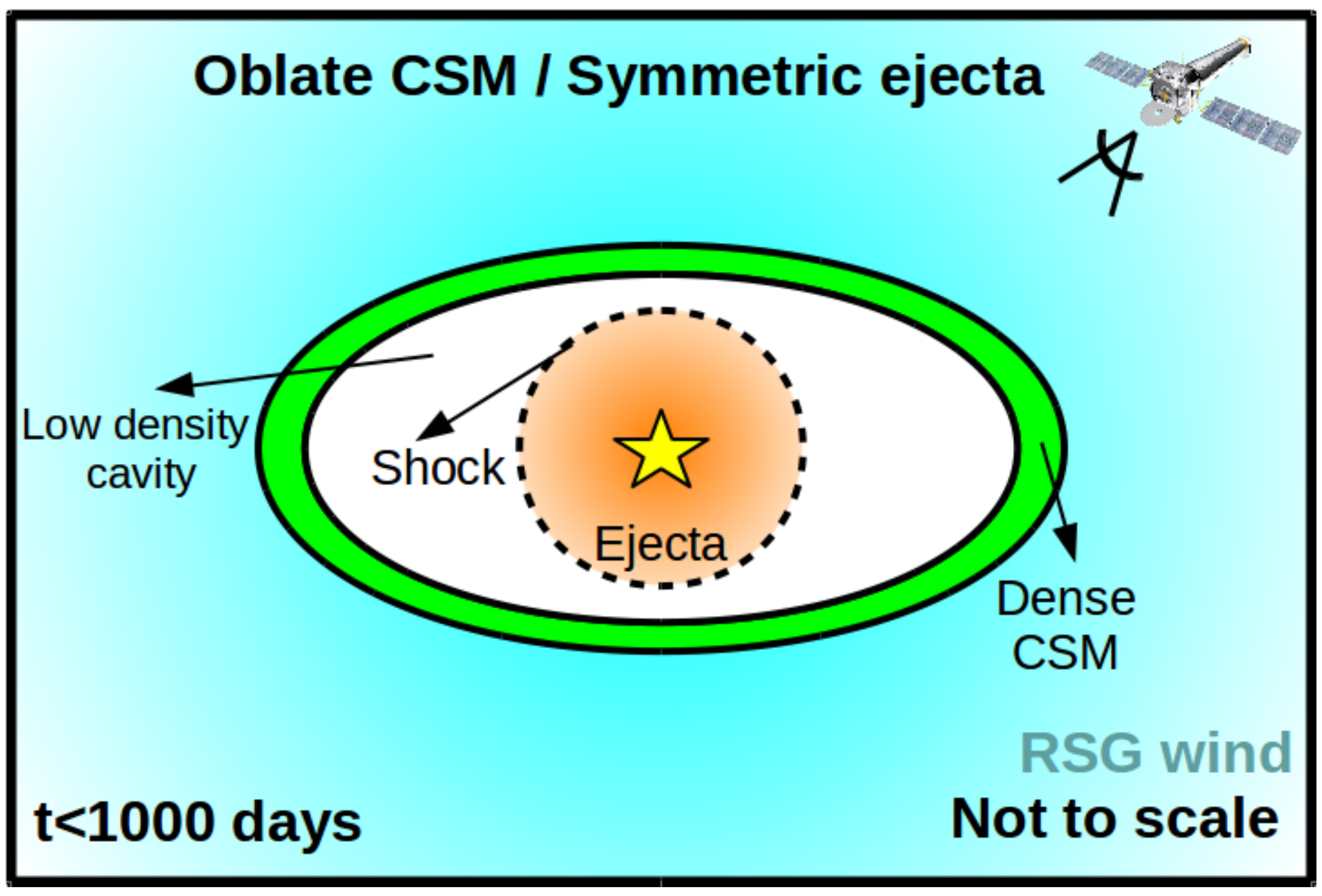} &
    \includegraphics[scale=0.16]{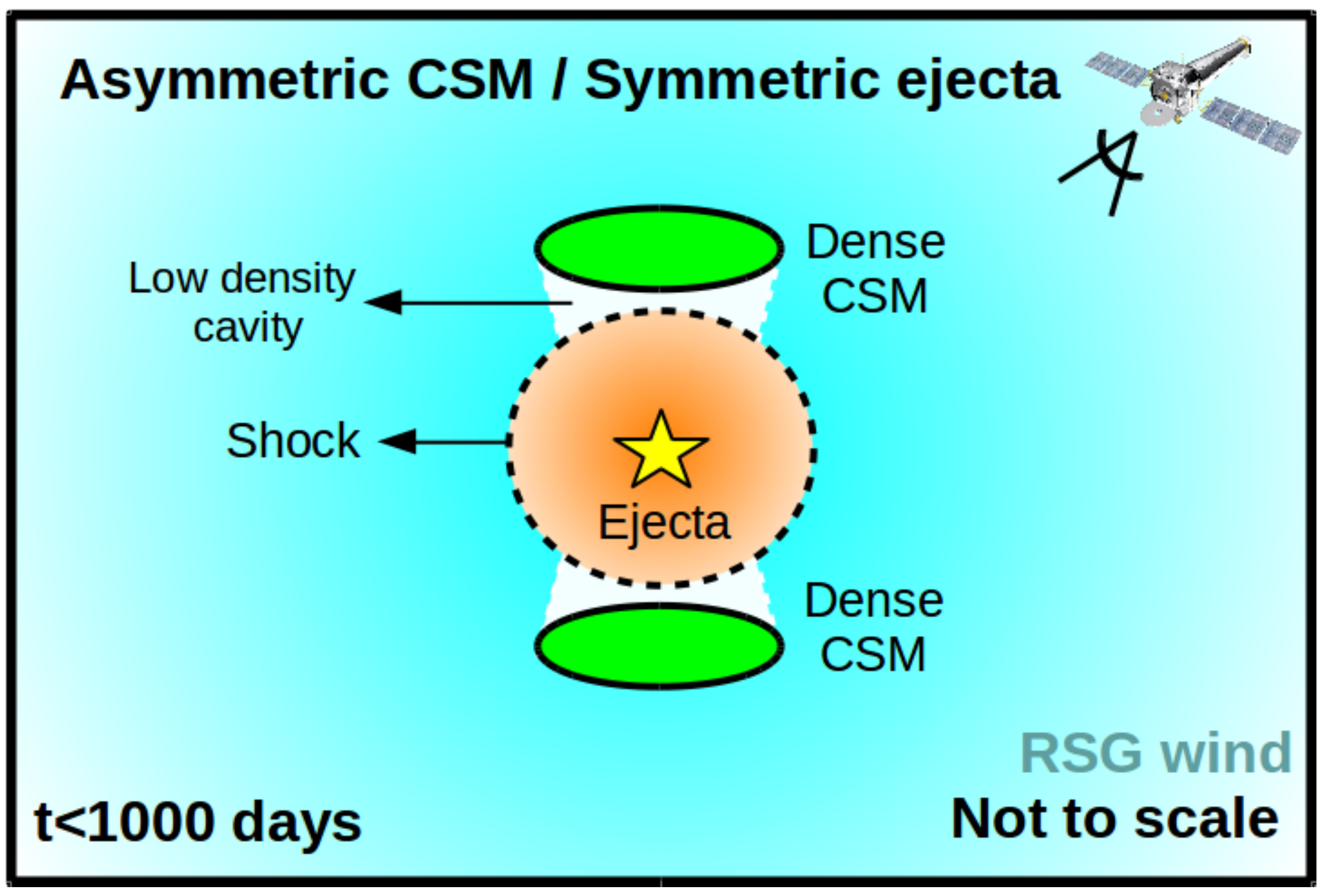} &
    \includegraphics[scale=0.16]{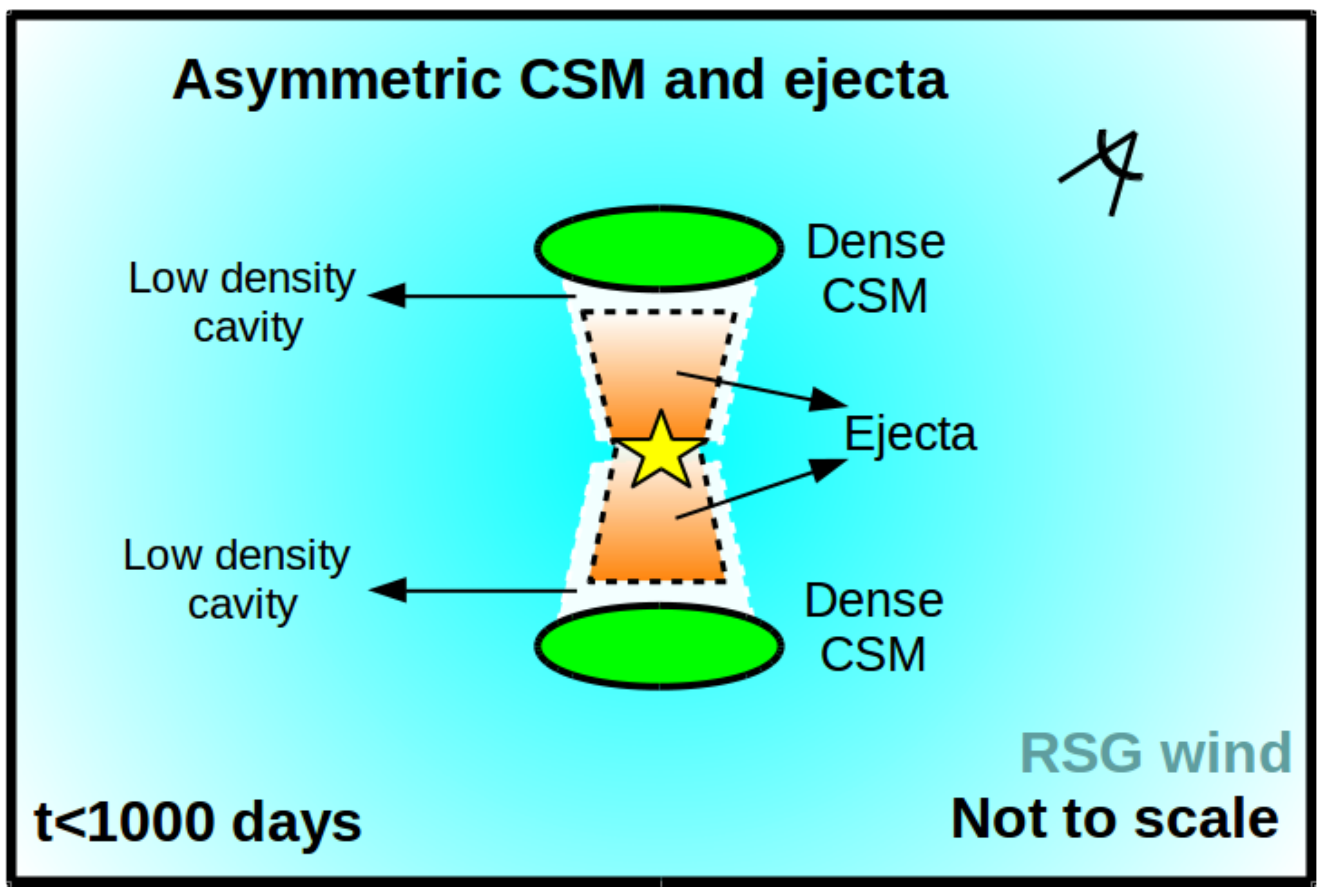} \\
    \includegraphics[scale=0.16]{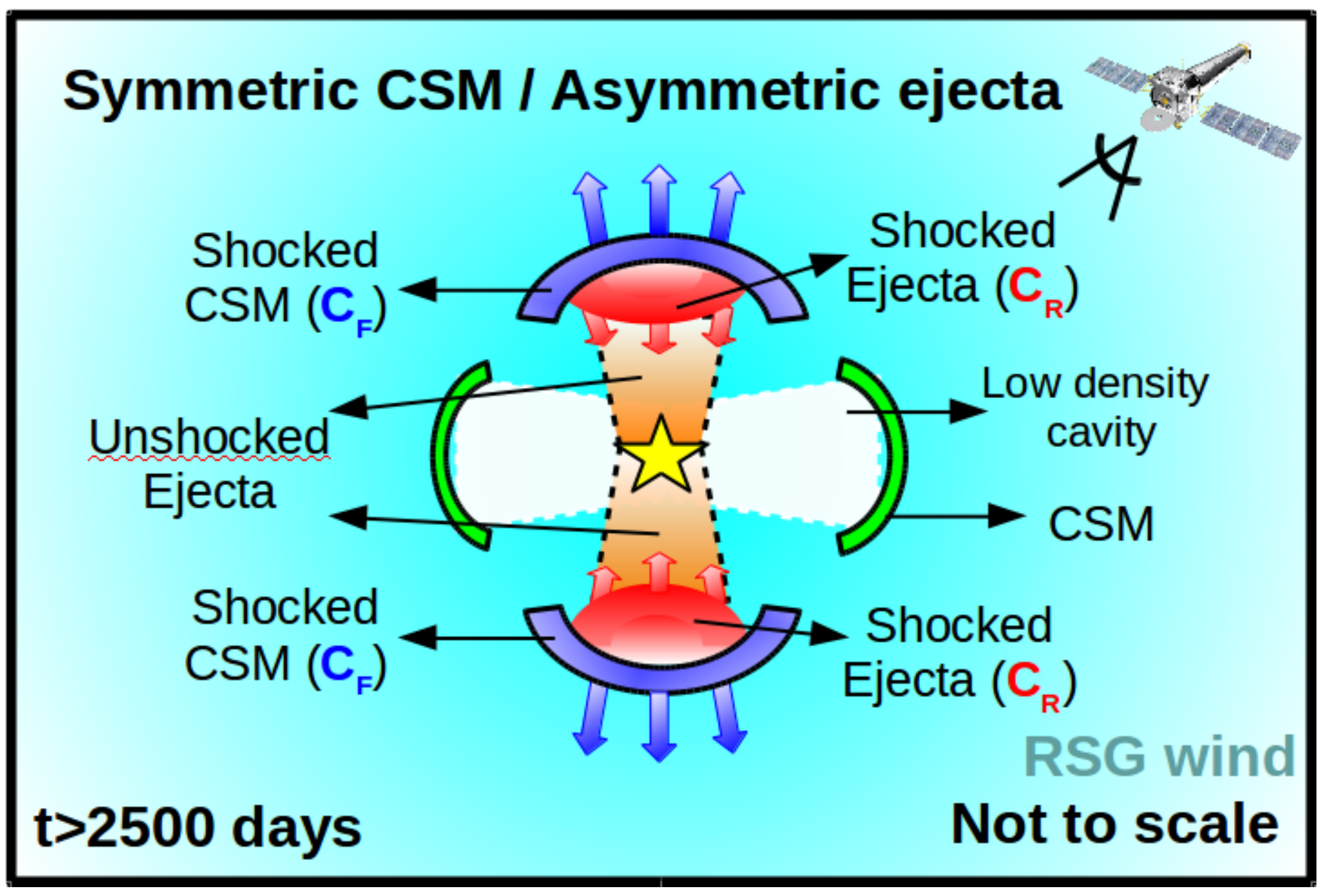} &
    \includegraphics[scale=0.16]{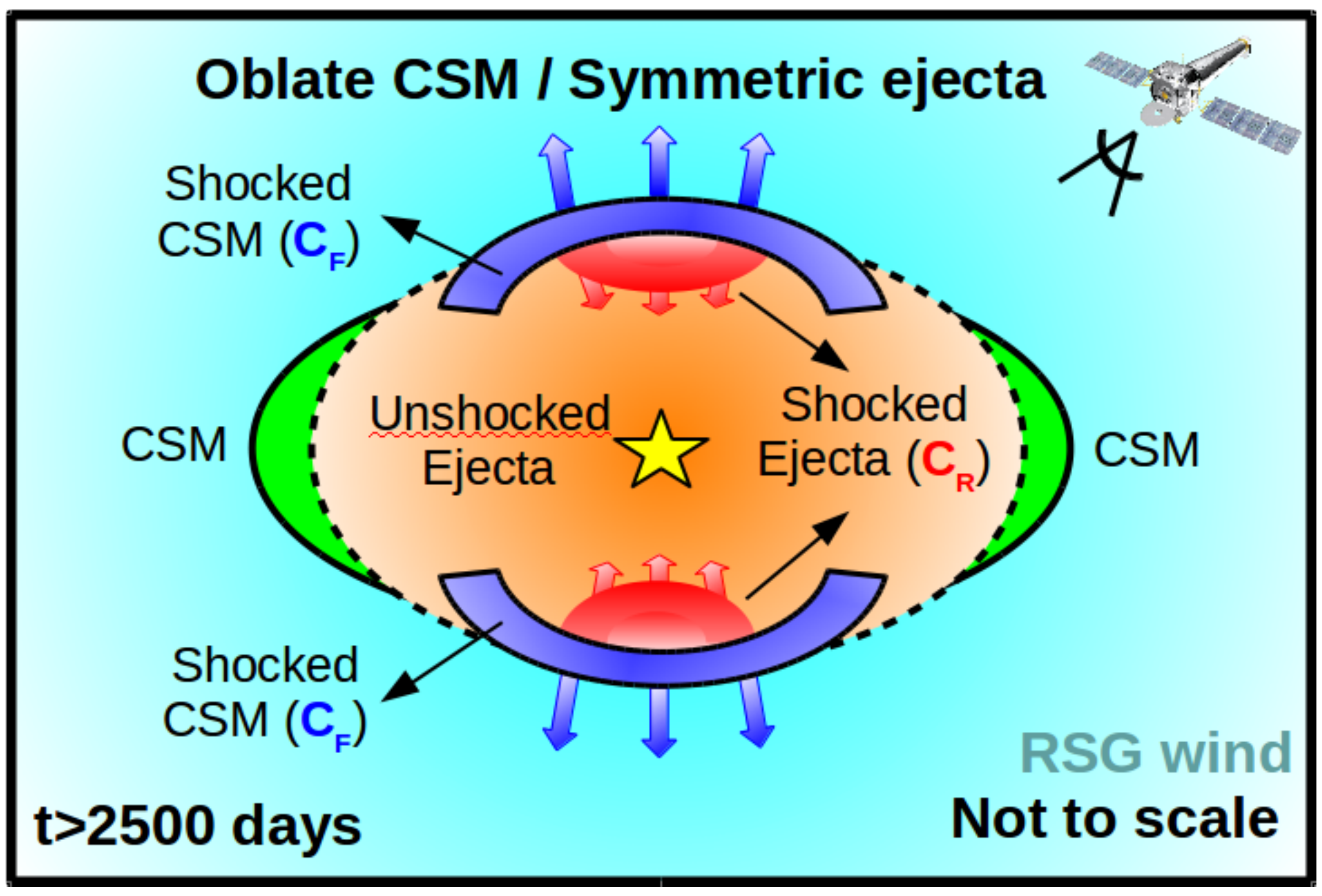} &
    \includegraphics[scale=0.16]{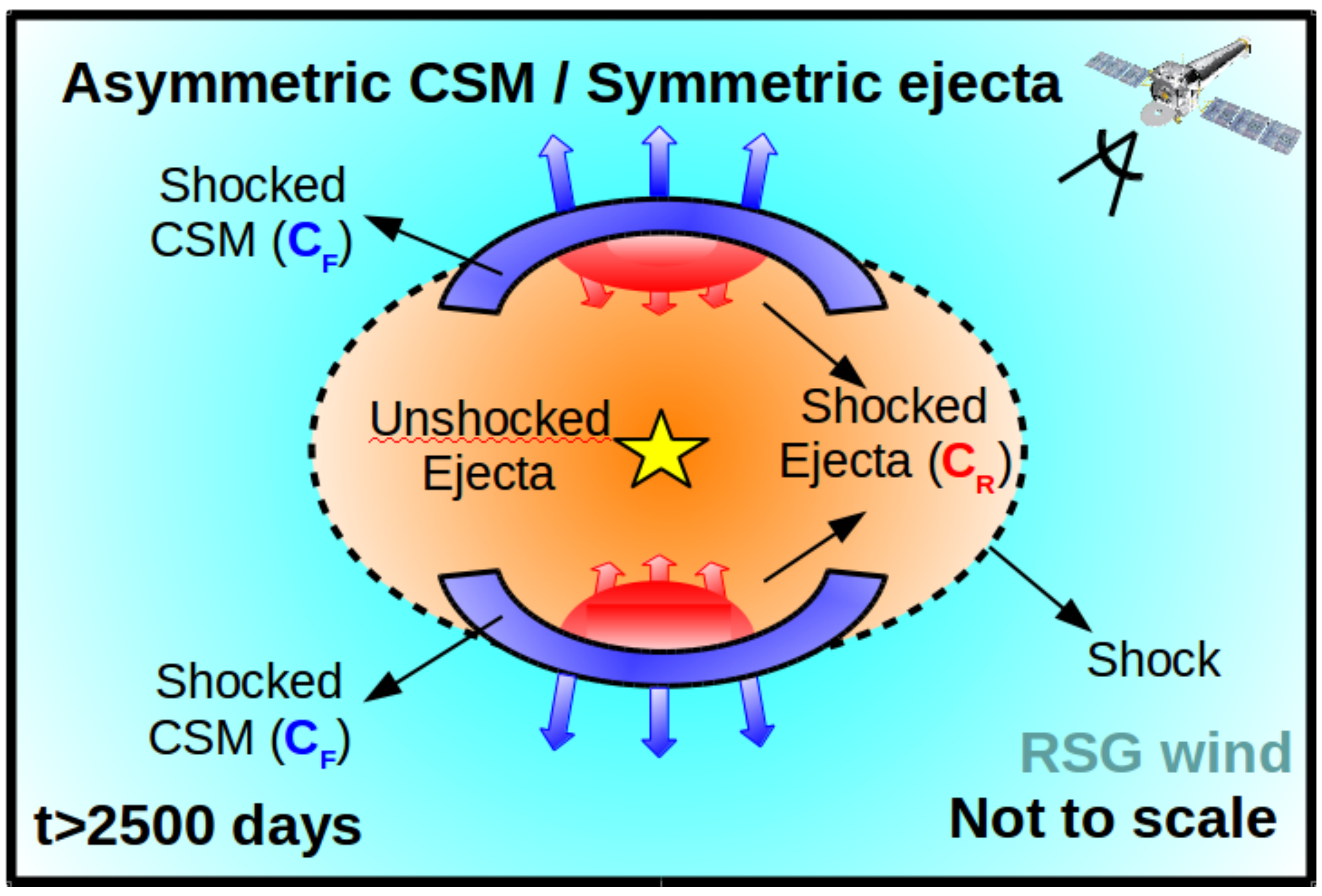} &
    \includegraphics[scale=0.16]{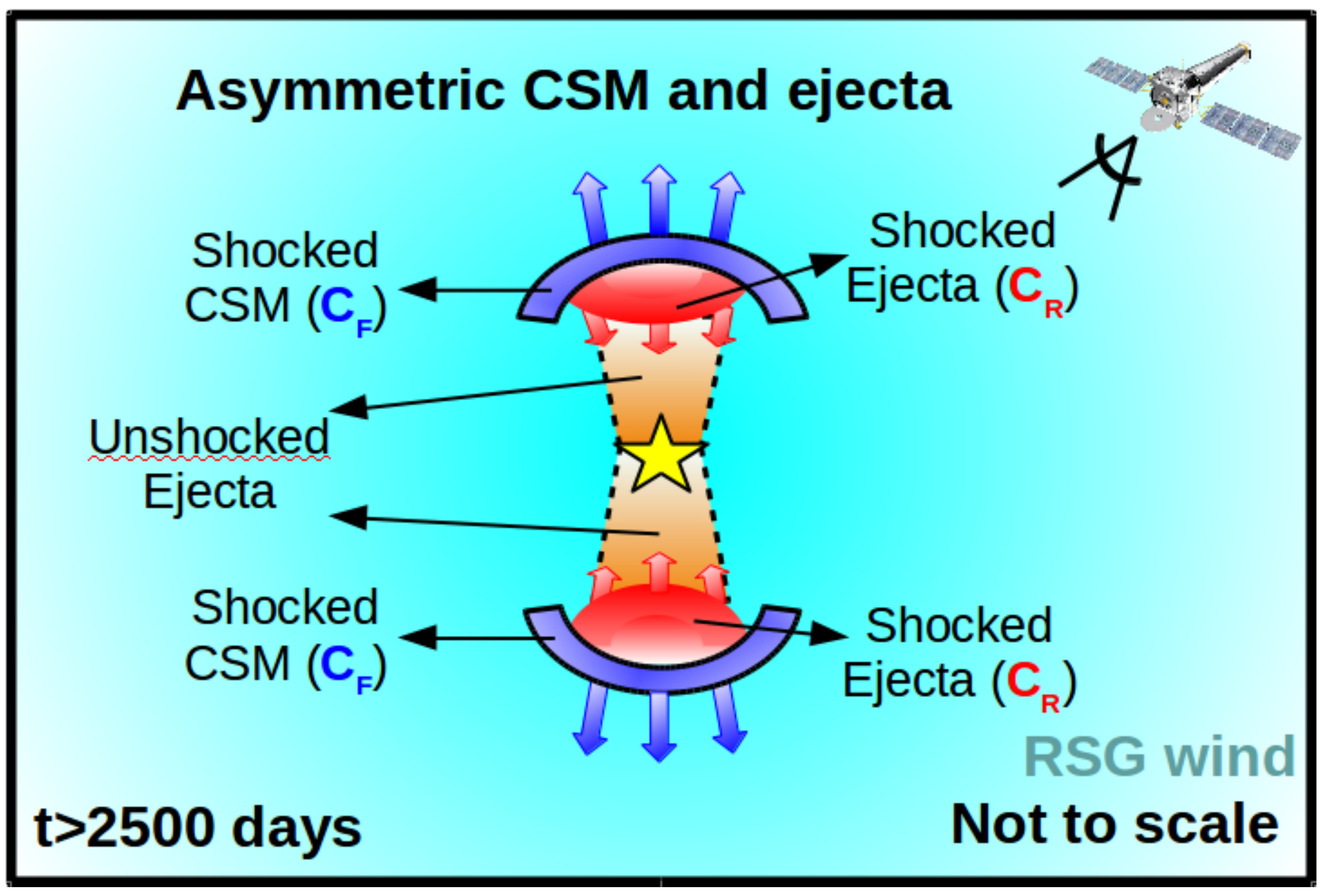}
    \end{array}$
    \end{center}
    \caption{Cartoon depicting four possible geometrical scenarios to generate the X-ray emission of SN\,1996cr. \emph{Top} and \emph{bottom} panels show the SN explosion before (${<}$1000 days) and after (${>}$2500 days) the interaction with the dense CSM, respectively. 
    \emph{Panels (a):} highly collimated ejecta into spherically symmetric CSM; \emph{Panels (b):} spherically symmetric ejecta into dense, oblate CSM; \emph{Panels (c):} spherically symmetric ejecta into dense polar CSM; \emph{Panels (d):} highly collimated ejecta into dense polar CSM. Components $C_R$ and $C_F$ described throughout the manuscript correspond to the \emph{red} and \emph{blue} regions in the \emph{bottom panels}, respectively. The colored arrows represent the direction of movement of the shocks. The \emph{cyan} gradient represents the outer RSG wind, the \emph{green} regions denote the dense shell, and the \emph{white} regions indicate the evacuated WR bubble according to \citetalias{Dwarkadas2010a}.}
    \label{fig:geometry}
\end{figure*}

\subsection{Time Evolution of Parameters}\label{sec:evol_interp}

Following on from our interpretation of the 2009 epoch, we investigate the evolution of the spectral parameters with time in Fig.~\ref{fig:parameters_vs_time}. It is important here to make the distinction between forward, reverse, and reflected shocks. The interaction of the SN ejecta with the CSM around the SN generates the canonical two structures confined by a forward shock (moving inside the CSM) and a reverse shock (traveling back into the expanding ejecta) \citep{Chevalier1994a, Zhekov2010, Bauer2008a}. In the self-similar solution, these shocks are separated by a contact discontinuity. The interaction of the forward shock with a dense CSM will additionally produce a reflected shock, which propagates back through previously shocked ejecta, further heating and compressing it \citep{Levenson2002}. This additional compression from the reflected shock can enhance the X-ray emission \citep{Hester1986}.

Taking cues from \citetalias{Dwarkadas2010a}, we tentatively identify component $C_F$ with forward-shocked CSM and component $C_R$ with either reverse-shocked or reflected-shocked ejecta \citepalias[as considered  following][]{Dwarkadas2010a}. A key issue to reconcile, however, is why $C_R$ and $C_F$ have distinct velocity (line) profiles and evolution, which we infer to demark separate spatial origins. The difference is clearest for 2009 epoch, where we have sufficient spectral resolution and photon statistics to measure clear differences between $C_R$ and $C_F$ in terms of expansion velocity and opening-angle of $\Delta v_{\rm max}{=}1436^{+328}_{-462}$\,km\,s$^{-1}$ and $\Delta\theta_{\rm min}{=}37\fdg3^{+8.8}_{-8.7}$, respectively; for the other epochs, the poor photon statistics and/or lack of high spectral resolution do not allow us to observe such strong distinctions, with 1-$\sigma$ differential errors of $\sim$700--1700\,km\,s$^{-1}$ and $\sim$18--21$^{\circ}$ (effectively encompassing the epoch 2009 differences).

There are a few plausible origins for the differences, which we depict in Fig.~\ref{fig:geometry}: 
(\emph{a}) asymmetric ejecta at the onset of the explosion  \citep[e.g., as has been argued for SN\,1987A;][]{Larsson2016}; 
(\emph{b}) mildly asymmetric CSM due to an oblate wind-blown shell, whereby the ejecta impacts one portion of the shell before another;
(\emph{c}) highly asymmetric CSM, similar to the aspherical, often bipolar or toroidal, structures surrounding many massive Milky Way and Magellanic Cloud stars  \citep[e.g.,][]{vanMarle2010};
(\emph{d}) strong asymmetries in both the ejecta and CSM.
All of these would naturally produce asymmetric shocks, and could lead to differences in forward and reverse shocks if the reverse/reflected shock were somehow enhanced and focused inward over a much smaller opening angle, regardless the shape or symmetry of the CSM. Finally, even if the explosion began with spherically symmetric ejecta, realistic multi-dimensional simulations suggest that relatively small instabilities or variations within the CSM or the wind-blown shell \citep[e.g.,][]{Dwarkadas2008} could lead to wrinkled or corrugated SN shocks \citep{Dwarkadas2007b}. The interaction of such a shock wave with even a spherical circumstellar shell will occur at somewhat different times along the length of the shell, resulting in a potentially asymmetric reverse shock. Thus, we should not be too surprised to see aspherical reflected shocks \citep{Dwarkadas2007b} as a result of any combination of these effects (see Fig.~\ref{fig:geometry}). Moreover, because of the high density of the ejecta, we might expect the reflected shock to dominate the overall emission at late times, similar to what is seen in SN\,1987A \citep{Dewey2012a,Orlando2015,Orlando2019}.

Finally, a word of caution regarding subsequent interpretations of model M5 and comparisons to the simulations of \citetalias{Dwarkadas2010a}. The NEI plasma model implicitly assumes a constant density over the emission region, while the ejecta profile in particular is expected to retain a strong power law ($\propto$ $r^{-9}$) radial dependence at early times \citepalias[see Fig. 3 of][]{Dwarkadas2010a}. This difference could subtlety bias the flux, velocity, temperature, line-of-sight column density and abundance estimates in low-quality and low resolution spectra, where velocity profiles cannot be easily disentangled. \citetalias{Dwarkadas2010a} broke up their simulation into 50 shells, each with its own properties (temperature, density, velocity, etc), in order to model these gradients. This produced a much broader distribution of shock temperatures and associated properties for both the forward and reverse shocks compared to what a single NEI model would yield. Adopting a single NEI model, as we do, for the entire ejecta will only capture a crude average of the most dominant temperature component, and may be particularly susceptible to underpredictions of various emission line strengths and ratios (e.g., He-like vs. H-like).

In the next subsections, we discuss the different scenarios and physical implications of the evolution of the model M5 parameters over the past two decades.

\vspace{-0.5cm}
\subsubsection{Column density ($N_{\rm H}$)}\label{sec:column_density}

We begin by commenting on the evolution of the two foreground column density terms in model M5, shown in panel $(a)$ of Fig.~\ref{fig:parameters_vs_time}. The column in front of $C_F$ should be comprised of just the unshocked CSM plus the Galactic ISM and Circinus Galaxy ISM, whereas the column for C1 will also include the shocked CSM and shocked ejecta.

The nominal Galactic absorption (\emph{dashed black} line) should set a lower bound on the expected absorption. The fact that component $C_F$ is best-fit with values consistent with the Galactic $N_{\rm H}$ in epochs 2000, 2004, and 2009 implies that there is little host obscuration along the line of sight, and thus any excess absorption should be related to changes in the CSM of the SN. One important consideration, however, is the potential influence of the known energy-dependent 10--20\% cross-calibration offsets between {\it XMM} and {\it Chandra} on our best-fit parameters. Given fitting degeneracies between temperature and column density, we might expect modest offsets between either the best-fit $N_{\rm H}$ or $kT$ values from {\it Chandra} and {\it XMM-Newton} spectra, in the sense that {\it XMM-Newton} may yield somewhat lower temperatures or higher column densities. Thus the mildly higher column densities associated with the {\it XMM-Newton} epochs compared to the {\it Chandra} ones could be due to this effect. Although we cannot rule out that some portion of the obscuration is intrinsic to SN\,1996cr, to be conservative, we consider best-fit values of $\lesssim$7$\times$10$^{21}$\,cm$^{-2}$ for {\it Chandra} and {\it XMM-Newton} to indicate Galactic-only (no CSM) obscuration. 

With this in mind, in Fig~\ref{fig:parameters_vs_time} panel $(a)$, component $C_F$ appears to be relatively unabsorbed at all times, except perhaps around $\approx$2300 days where we find a value of $\approx$1.1$\times$10$^{22}$\,cm$^{-2}$. This brief enhancement could be due to obscuration associated with either the compression of shocked CSM and ejecta material in the immediate vicinity of the forward and reflected shocks at this time. In the model of \citetalias{Dwarkadas2010a}, the column density of the shocked CSM behind the forward shock reaches a maximum of $\sim$5$\times10^{21}$\,cm$^{-2}$ around day $\sim$2400, as the forward shock interacts with the high-density outer edge of the shell material swept up by the WR wind. This value is consistent with what we measure even after considering a Galactic absorption as high as 7$\times10^{21}$\,cm$^{-2}$ (i.e., an excess of $\gtrsim$4$\times10^{21}$\,cm$^{-2}$).


On the other hand, component $C_R$ is best-fit by a high, roughly constant $N_{\rm H}$ of $\approx$(2--2.5)$\times$10$^{22}$\,cm$^{-2}$ at early times ($\approx$1700--5500 days). Assuming that component $C_R$ is associated with a reverse/reflected shock, then the strong differences between the early $C_R$ and $C_F$ column densities imply one or more of the following. 
If attributed to the foreground (unshocked) CSM, there would need to be strong density asymmetries in this media in order to somehow obscure $C_R$ more than $C_F$ \citep[e.g., SN\,2005kd, SN\,2006jd, and SN\,2010jl;][]{Katsuda2016a}. Alternatively, a higher degree of ionization in the foreground CSM toward $C_F$ could lower its apparent absorption. Given the large apparent solid angle over which such effects would need to occur, it is hard to reconcile the rapid drop in $C_R$'s obscuration beyond $\approx$5500 with either scenario.
Attributing the difference to shocked CSM and/or shocked ejecta in between the two components, perhaps associated with the dense shell or whatever remains, can rather naturally explain why component $C_R$ is more absorbed than $C_F$. Moreover, it seems possible that the absorption from such shocked material could become diluted beyond $\approx$5500 days due to expansion, ionization, and/or the shell was overrun and its effects no longer apparent.
Finally, higher abundances of heavy elements associated with the shocked ejecta along the line-of-sight of $C_R$ could enhance the obscuration, although it is unclear why we would see the drop beyond $\approx$5500 days if this were a strong effect since the shocked ejecta in front of the reverse or reflected shock is unlikely to fall rapidly. 
We note that the $N_{\rm H}$ drop beyond $\approx$5500 only appears in the low-resolution \emph{XMM-Newton} data, and thus could result from some fitting degeneracy between components $C_R$ and $C_F$ or our single NEI model assumption; however such behavior is not seen for the 2001 epoch.

In the \citetalias{Dwarkadas2010a} simulations, the shock-shell interaction is complete by $\sim$7 years, such that in roughly one doubling time all knowledge of the shell is lost \citep[see also][]{Dwarkadas2005a}, as the shell structure has been obliterated. Thus the moderate absorption difference at early times can be explained as viewing component $C_R$ through the dense shell and shocked CSM, while at late times the forward and reverse/reflected shocks will have smoothed out and/or ionized the initial high density structures, potentially explaining the drop to $N_{\rm H}$\, $\approx$(7--9)$\times10^{21}$\,cm$^{-2}$, only slightly above the Galactic value. This would be consistent with the reverse/reflected shock interpretation for the $C_R$ component, and is our preferred scenario to explain the difference and evolution.


We note that the type IIn SNe\,2010jl and 2005ip showed similar tendencies. For instance, the column density toward SN\,2010jl was initially found to be $N_{\rm H}$ $>$ 10$^{24}$ cm$^{-2}$ and decrease by two orders of magnitude in the first $\sim$1000 days \citep{Chandra2015a, Katsuda2016a, Chandra2018}, while SN\,2005ip was initially found to have $N_{\rm H}$ $\approx$ 5$\times$10$^{22}$ cm$^{-2}$ and drop by one order of magnitude at late times \citep{Katsuda2014a}. The $N_{\rm H}$ drops were argued to arise from very high density CSM \citep{Katsuda2014a, Chandra2018}, although asymmetric CSM could also play a role \citep{Katsuda2016a}.


\subsubsection{Electron temperature}\label{sec:electron_temperature}

Next, we examine the evolution of the two electron temperature components in model M5, shown in panel $(b)$ of Fig.~\ref{fig:parameters_vs_time}. Both components remain distinctly separated and exhibit marginal rises between days $\sim$1700--2200 followed by mild overall declines between days $\sim$2200--8000. There is an uptick in the temperature around day $\sim$5000 for component $C_R$, and around day $\sim$8500 for component $C_F$, but these are not significant after factoring in the errors. Unfortunately, the temperature of component $C_R$ is somewhat difficult to constrain, particularly when coupled with high $N_{\rm H}$ values, because neither \emph{Chandra} nor \emph{XMM-Newton} has good sensitivity above $\sim$7--8\,keV  \citep[e.g.,][]{Chandra2018}, while \emph{NuSTAR} imaging suffers from strong AGN contamination \citep{Arevalo2014}.

It is important to stress, as in \citetalias{Dwarkadas2010a}, that at least two temperatures are required in order to fit both the relatively flat continuum and the numerous emission lines, which span a large range in excitation energy. We initially associate these temperatures with the postshock temperatures behind the forward and reverse/reflected shocks, respectively. Since the shocks are collisionless, most of the post-shock energy is transferred to ions, with the electron temperature related to the average particle (ion) temperature behind the shock by the fraction $\beta_{\rm sh}(v_{\rm sh})$, which is a function of the shock velocity, $v_{\rm sh}$: $T_e$=$\beta_{\rm sh}(v_{\rm sh})$$T_{\rm sh}$. The electron temperature, shock velocity and mean molecular weight can then be related as $k$$T_{e}$=3$\beta_{\rm sh}$$\mu$$m_{\rm H}$$v_{\rm sh}^2$/16 \citep{Ghavamian2007}, 
such that $k$$T_{e}{\propto}\mu$. 

Panel $(c)$ of Fig.~\ref{fig:parameters_vs_time} highlights that the velocities of components $C_R$ and $C_F$ vary by only a factor of $\sim$2 over $\sim$18 years and have a roughly constant velocity ratio at all times. The latter may imply that a difference in mean molecular weights produces much of the temperature variation, with component $C_R$ having a much higher mean molecular weight than component $C_F$. Given that the ejecta should have a much higher proportion of heavy elements than CSM, this is consistent with components $C_R$ and $C_F$ being associated to the shocked ejecta and CSM, respectively, in good agreement with the results of $\S$\ref{sec:column_density}. If the progenitor was a WR star, containing no H and maybe no He, then the proportion of heavy elements in the ejecta would be relatively high. This is consistent with panel $(g)$ of Fig.~\ref{fig:parameters_vs_time}, where the Fe abundance of component $C_R$ is much higher than solar. Given the above, it is likely that components $C_R$ and $C_F$ could be related with the shocked ejecta and shocked CSM, respectively.


\subsubsection{Shock velocities}\label{sec:shell_velocity}

The evolution of the shock velocities, which we associate with the maximum expansion velocity of the emission region as derived from the line width fit by \texttt{shellblur}, is shown in panel $(c)$ of Fig.~\ref{fig:parameters_vs_time} for the two components of model M5. The velocities of both components are found to increase from relatively low values of $\sim$3500\,km\,s$^{-1}$ to maxima of $\sim$6000\,km\,s$^{-1}$ between days $\sim$1700--3500, roughly coinciding with the epoch during which the forward shock is thought to have interacted with the swept-up shell bordering the wind-blown bubble in the \citetalias{Dwarkadas2010a} model (denoted by the \emph{grey shaded band} in Fig.~\ref{fig:parameters_vs_time}). For comparison, we show the evolution of the forward and reverse/reflected\footnote{We plot the reverse shock velocity up to the point where it is overrun by the reflected shock, after which we plot the latter since the emission is strongly dominated by the reflected shock.} shock velocities obtained in \citetalias{Dwarkadas2010a} (\emph{dashed black} and \emph{dashed green} curves, respectively). 

In \citetalias{Dwarkadas2010a}, following the SN explosion, the forward shock expanded rapidly (\hbox{$\sim$$10^4$ km\,s$^{-1}$}) until it encountered the dense shell of material at day $\sim$500, which led to a drastic drop in the forward shock velocity to $\sim$2500\,km\,s$^{-1}$ by day $\sim$800 and compression of the nominal high pressure region between the forward and reverse shocks. The interaction of the ejecta with the high-density shell sent a low velocity transmitted shock into the dense shell, and a fast reflected shock back into the ejecta. The transmitted shock velocity gradually increased between days $\sim$800--2500 as the compressed region behind the shock depressurized, attaining a more or less uniform velocity of around \hbox{$\sim$5500\,km\,s$^{-1}$}. By contrast, the reflected shock quickly overran the original reverse shock between days $\sim$800--1000, leading to a sharp increase in the reverse shock velocity up to $\sim10^{4}$\,km\,s$^{-1}$ and a faster thermalization of the ejecta than possible from the reverse shock alone. Beyond day $\sim$1000, the reverse shock velocity decreases as the reverse/reflected shock expands into the steep density incline and sweeps up more material. 

Intriguingly, the implied velocity constraints from the line profiles for both components $C_R$ and $C_F$ appear more consistent with the forward shock in the \citetalias{Dwarkadas2010a} simulation than the reverse shock over this time frame. This appears incongruent with our assumption that $C_R$ is associated with the reverse shocked ejecta, although at day $\sim$5000 (the highest signal-to-noise data) we do see a slight dip in the $C_R$ velocity toward the simulated reverse shock velocity of \citetalias{Dwarkadas2010a}. We return to this inconsistency later in this section.

Notably, the best-fit velocities of both $C_R$ and $C_F$ decrease substantially between days $\sim$3500--5000, in contrast to the relatively flat forward shock predictions [dashed \emph{black} line in panel \emph{(c)} of Fig.~\ref{fig:parameters_vs_time}]. This discrepancy, however, is not that surprising given the lack of late-time data as input to the \citetalias{Dwarkadas2010a} model. This meant that \citetalias{Dwarkadas2010a} had to assume reasonable values for the radius and thickness of the shell, as well as the CSM density beyond it. Minor modifications, such as moving the shell inward, modifying its thickness, and modifying the outer CSM density, could potentially account for the velocity difference and evolution. Unfortunately, beyond day 5000 our velocity constraints remain rather limited due to the crude CCD-resolution of the {\it XMM-Newton} \emph{pn}/MOS cameras. To within the uncertainties, component $C_R$ remains roughly constant or mildly increases between days $\sim$5000--9000, while component $C_F$ cannot be well-constrained and therefore was fixed to its value at day $\sim$5000. Given the uncertainties in the CSM parameters adopted by \citetalias{Dwarkadas2010a}, it is unclear how to interpret the late-time behavior of $C_R$. The slight upward trend implies $C_R$ is slightly more consistent with the \citetalias{Dwarkadas2010a} prediction for the forward shock interacting with the outer CSM associated with a progenitor wind, which leads to very mild velocity changes with time. However, the lower velocities we find from day $\sim$5000 onward compared to the \citetalias{Dwarkadas2010a} predictions may be a consequence of a somewhat higher CSM density, as mentioned already in $\S$\ref{sec:2009_interp}, or clumpy structure.

Finally, we caution that some of the previous discrepancies could be related to the fact that the 1-D simulations do not capture the effects of 2-D or 3-D asymmetries resulting from either turbulence, internal kinematics of the ejecta material, asymmetries and inhomogeneities within the progenitor star wind or instabilities of the ejecta--CSM interaction itself. Various higher dimensionality SNe simulations \citep[e.g.][]{Chevalier1992a, Dwarkadas2007a, Freyer2006, vanMarle2012, Orlando2015, Orlando2019}, demonstrate that asymmetries can lead to complex velocity flows, and rapidly destroy the spherical symmetry of the explosion, if it ever existed. Additionally, the quality of our data (excluding the 2009 epoch) may be insufficient to constrain the line width adequately; in particular, uncertainties in the orientation of the SN can lead to underestimates in the shock velocity (but little room for overestimates).



\subsubsection{Polar opening-angle interaction}\label{sec:theta_min}

Panel $(d)$ of Fig.~\ref{fig:parameters_vs_time} examines the change in the opening-angles of both components of model M5. We find that, at least when they can be constrained, they do not change drastically with time. Thus the bi-polar geometrical interaction remains relatively unchanged during our observations. This provides some justification for fixing the inclination-angle to 55$^{\circ}$, as discussed in $\S$\ref{sec:2009_1comp_blur}. Due to the low spectral resolution of \emph{XMM-Newton}, we only obtain upper limits on the opening-angles for the 2001 and 2013 epochs, and fixed the opening-angles of the $C_R$ and $C_F$ components to previous values of $\sim$70$^{\circ}$ and $\sim$30$^{\circ}$ for epochs 2014--2018, respectively. Furthermore, the $C_R$ opening-angle was initialized with a value of 75$^{\circ}$ to help the fit-process to converge to a global minimum and prevent the best-fit from settling at the $C_F$ value \citep[e.g.,][]{Arnaud2011}. This aspherical interaction is likely a product of the CSM into which the shock is expanding, which formed over various stellar evolutionary phases from Main Sequence (MS) to RSG to WR \citep{Garcia-Segura1996b, Freyer2006, Dwarkadas2007b}, for instance. In fact, simulations show that the impact of the progenitor mass-loss and velocity wind distribution prior to the SNe can play a critical role in the dynamical evolution of the ejecta-CSM interaction \citep{Dwarkadas2005a, Dwarkadas2007b, Freyer2006, vanMarle2010, vanMarle2012, Dwarkadas2013, Patnaude2017a} and in the symmetry of the emission/interaction region. \citet{Freyer2006} and \citet{Dwarkadas2007a} showed that the strong ejecta-CSM asymmetries arise from asymmetries in the progenitor wind even when a spherical explosion is considered. In $\S$\ref{sec:scenarios}, we revisit discussion of the CSM scenarios.

We cannot rule out the possibility that the explosion had an intrinsic asymmetry or was impacted by a binary companion \citep[e.g., SN\,1979C, SN\,1987A, SN\,1993J, SN\,2011dh;][]{McCray1993a, Sugerman2005a, Montes2000a, Maund2004, Folatelli2014}, either of which could affect the resulting symmetry of the ejecta--CSM interaction. For the binary companion case, the expectation is that the winds of the two stars combine to form an aspherical CSM bubble \citep{vanMarle2012b}. The ejecta itself may be intrinsically asymmetric. In the case of SN\,1987A, considering the Doppler shift of the freely expanding ejecta, \citet[][]{Larsson2016} found a clearly asymmetric north-south ejecta distribution. Alternatively,  for Cassiopeia A, the spatial distribution of radioactive $^{44}$Ti, which probes the explosion asymmetries of CCSNe \citep[e.g.,][]{Magkotsios2010}, implies a highly asymmetric bipolar explosion resulting from a fast-rotating progenitor star \citep{Grefenstette2014}.

\subsubsection{Internal ejecta obscuration ($N_{\rm ejecta}$)}\label{sec:NH_ejecta}

Under the assumption that the shock maintains some geometric or reflected symmetry whereby the far side is modified by internal obscuration, we show the evolution of the column density of obscuring ejecta ($N_{\rm ejecta}$) inside the expanding shell in panel $(e)$ of Fig.~\ref{fig:parameters_vs_time}. If the ejecta and shock regions are expanding more or less symmetrically, this quantity should not evolve too strongly. 
We find that the ejecta column density associated with component $C_R$ is quite high at all times [$\approx$(1--8)$\times$10$^{23}$\,cm$^{-2}$], and shows some evidence for a mild (factor of $\sim$2--4) drop around days 6500--7500. The column density associated with component $C_F$ is generally a factor of $\sim$5--40 lower [$\approx$(1--2)$\times$10$^{22}$\,cm$^{-2}$], and is consistent within errors with being constant in time. As we argued already in $\S$\ref{sec:2009_interp}, the large disparity between the $N_{\rm ejecta}$ values argues either for a strongly inhomogeneous ejecta structure, such that substantially denser and high-Z material lies along the line-of-sight, and therefore likely in the immediate vicinity, of the narrow-angle $C_R$ component, or the assumption of symmetry breaks down.

Additionally, with values of $N_{\rm ejecta}$ in hand and some indication of the shock geometry and evolution from the D10 model, we can estimate the ejecta mass. For the moment, we will assume a spherical uniform distribution for the ejecta, recalling from the 1-D simulation of D10 that the unshocked ejecta density was uniform after $\approx$7--9 years and from the 2-D simulations of \cite{Dwarkadas2007a}, \cite{vanMarle2010} and \cite{vanMarle2012b} that the asphericity of the ejecta-CSM interaction arose from the inhomogeneous CSM, even though the explosion itself was spherically symmetric. We further consider that $(i)$ the $C_F$ shock geometry covers $\approx$2$\pi$ solid angle and reproduces all emission lines except Fe\,XXV--Fe\,XXVI, and $(ii)$ the denser material from $C_R$ lies only along a narrow polar region and does not affect the $C_F$ estimate. Thus, using the $C_F$ ejecta density obtained in $\S$\ref{sec:2009_interp} for 2009 and the associated shock radius from D10, we estimate the average ejecta mass as $M_{\rm ejecta}\sim6.9$\,$M_\odot$. For 2004, we obtained a similar ejecta mass estimate of $M_{\rm ejecta}\sim$5.3$M_\odot$. These quantities are in reasonable agreement with the ejecta mass quoted and assumed by \citetalias{Dwarkadas2010a}, of $M_{\rm ejecta}$=4.5M$_\odot$.

\subsubsection{Components $C_R$ and $C_F$ fluxes and electron density}\label{sec:fluxes}

Panel $(f)$ of Fig.~\ref{fig:parameters_vs_time} shows the evolution of the unabsorbed flux for the individual model M5 components $C_R$ (\emph{red} symbols) and $C_F$ (\emph{blue} symbols), as well as their combination (\emph{black} circles), obtained using the XSPEC convolution model \texttt{cflux}, as well as the individual contribution to the total flux from the shocked ejecta (\emph{green dashed} line) and CSM (\emph{black dashed} line) computed by \citetalias{Dwarkadas2010a}. The strong overall agreement between the $C_F$ and $C_R$ fluxes and the simulated forward and reverse shock fluxes calculated in \citetalias{Dwarkadas2010a} beyond 2001 contributed to our identification of the two X-ray components as such. The more heavily obscured $C_R$ component appears to dominate the total flux at all times, comprising between $\approx$54--76\% of the emission at various stages. The total flux reaches a broad maximum between days $\sim$3400--5000 (epochs 2004--2009), while individual components $C_R$ and $C_F$ reach maxima around days $\sim$3400 and $\sim$5000, respectively. The flux in both components appears to be declining at late epochs, although we find mild upticks in both near day $\sim$7000. This behaviour could be due to changes in density, temperature, ionization, or a change in wind velocity (e.g., episodic RSG mass loss rates prior to the explosion, turbulence, etc.). A clear point of discord arises at epoch 2000, whereby we find that $C_R$ matches the model CSM flux better, while $C_F$ matches the model ejecta flux better. This apparent reversal might be rooted in the implicit different implementations of the NEI plasma model here and in \citetalias{Dwarkadas2010a}, as noted at the end of $\S$\ref{sec:scenarios}, and/or due to poor photon statistics for the 2000 epoch.

We remind the reader that we implicitly adopted model M5 as the best fit for epoch 2000, although model M2 appears to provide a better fit based on BXA. This could be due simply to the low-quality of data (e.g., only a handful of counts in the Fe lines, which are critical for identifying the narrow-angle velocity component), or linked to the complexity of the shock interaction inside the dense shell at early times. According to the \citetalias{Dwarkadas2010a} model, the shocked ejecta flux at $\lesssim$2100 days should be a factor of $\ga$4--10 smaller than the shocked CSM flux, and hence perhaps too weak to detect. Coupled with the broader temperature distribution modeled by \citetalias{Dwarkadas2010a}, it thus remains feasible that the epoch 2000 continuum can be completely associated with the $C_F$ (shocked CSM) component alone. Therefore, the switch in the role of components $C_R$ and $C_F$ remains open to interpretation. 

Finally, using the normalization parameter of model M5, we can estimate the effective electron density ($n_e$) of the shocked emission regions related to components $C_R$ and $C_F$. To obtain $n_e$, we assume that the X-ray emission region is a spherical cap defined by the best-fit opening angle and adopt typical radial thickness and profile values from \citetalias{Dwarkadas2010a}. The thickness and profile depend on whether the emission arises from $C_R$ (shocked ejecta) or $C_F$ (shocked CSM). Here we implicitly assume that these quantities do not deviate strongly from the estimates of \citetalias{Dwarkadas2010a} and furthermore that the electron density remains uniform in the angular direction (i.e., uniform density surface area at a given radius). We further adopt an electron to ion density ratio of $n_e/n_i$$\approx$1 for both components; this assumption is not completely valid, since the ejecta is rich in high-Z material and the ratio will thus not be exactly unity, but it is a reasonable first approximation in order to derive a limit to compare with \citetalias{Dwarkadas2010a}. For epoch 2009, where we have the highest signal-to-noise and (HETG) spectral resolution, the $n_e$ estimates are $\sim$1.7 $\times10^{5}$\,cm$^{-3}$ for $C_R$, and $\sim$0.8 $\times10^{5}$\,cm$^{-3}$ for $C_F$. 
For component $C_R$, this value roughly agrees with that calculated by \citetalias{Dwarkadas2010a}, and expected by \citet{Bauer2008a} based on optical data ($n_e\geq10^5$\,cm$^{-3}$). 

\begin{table*}
   \caption{Best-fit abundance distributions as deduced from model M5. \emph{Col. 1}: Epoch of combined X-ray observations. \emph{Col. 2:} Components of model M5. \emph{Col. 3-9:} Abundances of select elements with $Z\geq10$ in solar units ($Z_{x,\odot}$). When unconstrained, we adopt solar values for component $C_R$ and the gas phase abundances as determined by \citep[][]{Oliva1999a} for component $C_F$.} 
       \scalebox{0.95}{
       \begin{tabular}{lcccccccc}
       \hline
       Epoch & Component & Ne & Mg & Si & S & Ar & Ca & Fe \\ \hline\hline
       2000 & $C_R$ & $1.0$(fix) & $1.0$(fix) & $0.7$(fix) & $0.1_{-0.1}^{+3.6}$ & $17.5_{-11.0}^{+14.6}$ & $1.0$(fix) & $2.0_{-0.9}^{+1.2}$\\ \vspace{0.1cm}
            & $C_F$ & $0.1_{-0.1}^{+0.4}$ & $0.4_{-0.4}^{+0.5}$ & $1.2_{-0.4}^{+0.5}$ & $2.9_{-1.1}^{+1.3}$ & $1.0$(fix) & $0.2$(fix) & $0.1_{-0.1}^{+0.1}$\\
       2001 & $C_R$ & $1.0$(fix) & $1.0$(fix) & $0.7$(fix) & $6.4_{-1.5}^{+1.6}$ & $2.6_{-2.6}^{+4.0}$ & $1.0$(fix) & $3.1_{-0.3}^{+0.3}$\\ \vspace{0.1cm}
            & $C_F$ & $0.6_{-0.2}^{+0.2}$ & $0.4_{-0.1}^{+0.1}$ & $0.4_{-0.1}^{+0.1}$ & $0.1_{-0.1}^{+0.1}$ & $1.0$(fix) & $0.2$(fix) & $0.3_{-0.1}^{+0.1}$\\
       2004 & $C_R$ & $1.0$(fix) & $1.0$(fix) & $0.7$(fix) & $1.7_{-1.7}^{+3.8}$ & $7.3_{-6.2}^{+7.3}$ & $1.0$(fix) & $2.9_{-0.7}^{+0.8}$\\ \vspace{0.1cm}
            & $C_F$ & $0.2_{-0.2}^{+0.3}$ & $0.8_{-0.2}^{+0.3}$ & $1.3_{-0.2}^{+0.2}$ & $1.9_{-0.6}^{+0.6}$ & $1.0_{-1.0}^{+1.0}$ & $0.2$(fix) & $0.1_{-0.1}^{+0.1}$\\
       2009 & $C_R$ & $1.0$(fix) & $1.0$(fix) & $0.7_{-0.7}^{+0.8}$ & $0.7_{-0.7}^{+1.3}$ & $1.2_{-1.2}^{+2.6}$ & $2.4_{-2.4}^{+2.9}$ & $3.9_{-0.5}^{+0.6}$\\ \vspace{0.1cm}
            & $C_F$ & $0.3_{-0.1}^{+0.1}$ & $0.7_{-0.1}^{+0.1}$ & $1.4_{-0.1}^{+0.1}$ & $1.5_{-0.2}^{+0.2}$ & $1.7_{-0.5}^{+0.6}$ & $0.2_{-0.1}^{+0.7}$ & $0.1_{-0.1}^{+0.1}$\\
       2013 & $C_R$ & $1.0$(fix) & $1.0$(fix) & $0.7$(fix) & $9.5_{-3.0}^{+3.2}$ & $15.7_{-7.1}^{+7.7}$ & $1.0$(fix) & $4.2_{-0.5}^{+0.5}$\\ \vspace{0.1cm}
            & $C_F$ & $0.3$(fix) & $1.0_{-0.5}^{+0.6}$ & $1.7_{-0.4}^{+0.4}$ & $0.8_{-0.8}^{+1.4}$ & $1.7$(fix) & $0.2$(fix) & $0.3_{-0.1}^{+0.1}$\\
       2014 & $C_R$ & $1.0$(fix) & $1.0$(fix) & $0.7$(fix) & $9.5$(fix) & $15.7$(fix) & $1.0$(fix) & $5.0_{-1.9}^{+3.1}$\\ \vspace{0.1cm}
            & $C_F$ & $0.3$(fix) & $2.3_{-2.2}^{+4.4}$ & $3.4_{-0.7}^{+0.8}$ & $1.9_{-1.7}^{+2.8}$ & $1.7$(fix) & $0.2$(fix) & $0.4_{-0.2}^{+0.2}$\\
       2016 & $C_R$ & $1.0$(fix) & $1.0$(fix) & $0.7$(fix) & $14.4_{-6.5}^{+7.5}$ & $15.7$(fix) & $1.0$(fix) & $5.0_{-0.9}^{+1.2}$\\ \vspace{0.1cm}
            & $C_F$ & $0.3$(fix) & $2.2_{-2.1}^{+8.6}$ & $4.0_{-1.5}^{+2.1}$ & $3.7_{-2.1}^{+2.5}$ & $1.7$(fix) & $0.2$(fix) & $0.4_{-0.3}^{+0.3}$\\
       2018 & $C_R$ & $1.0$(fix) & $1.0$(fix) & $0.7$(fix) & $6.9_{-6.9}^{+7.4}$ & $15.7$(fix) & $1.0$(fix) & $6.1_{-1.5}^{+2.2}$\\
            & $C_F$ & $0.3$(fix) & $2.5_{-2.0}^{+4.4}$ & $3.5_{-1.0}^{+1.3}$ & $2.0$(fix) & $1.7$(fix) & $0.2$(fix) & $0.1_{-0.1}^{+0.3}$\\
\hline
       \end{tabular}
       }
       \label{tab:abundances}
\end{table*}

\subsubsection{Abundances}\label{sec:abundances}

Finally, in panels $(g)$ and $(h)$ of Fig.~\ref{fig:parameters_vs_time} we track the evolution of the elemental abundances of Fe and Si, respectively. In Table~\ref{tab:abundances} we show the abundance values for all constrained elements with $Z\geq10$ obtained with model M5, for both components, at different epochs. Due to the heavily obscured nature of component $C_R$, we are only able to obtain tight constraints on the Fe abundance. Additionally, elements such as Ne, Ar and Ca show only weak emission lines, and thus are difficult to constrain in general. We find that the Fe abundance for $C_R$ is ``super-solar'' ($\sim$2 $Z_{\rm Fe,\odot}$) at day $\sim$1700 and increases smoothly to $\sim$6 $Z_{\rm Fe,\odot}$ over the span of $\sim$7000 days. This implies that we are gradually looking deeper into the ejecta with time; this could be an effect of decreasing opacity, or track a true enhancement of higher abundance Fe material. On the other hand, the Fe abundance for component $C_F$ is relatively ``sub-solar'' and nearly constant ($\sim$0.2--0.4$Z_{\rm Fe,\odot}$), while the Si abundance is observed at near solar values (0.5--1.0 $Z_{\rm Si,\odot}$) at day $\sim$1700, with mild ($\sim$2$\times$) variation through day $\sim$6500, before rising to values of 3--4$Z_{\rm Si,\odot}$. The Mg and S abundances for component $C_F$ increase with time, similar to Si (see Table~\ref{tab:abundances}). 
Unfortunately, the modest exposures and low spectral resolution of the more recent \emph{XMM-Newton} epochs are unable to constrain the $C_F$ component abundances of Ne and Mg and yield only modest constraints on the Si and S abundances. It is unfortunate that the Ne abundance is not well-constrained, since Ne enhancements are expected in the ejecta of CCSNe \citep{Branch2017}. We do find $Z_{\rm Ne}$$\sim$0.3$Z_{\rm Ne,\odot}$ for the 2009 epoch, although this is associated with the shocked-CSM component $C_F$, and hence is not expected to be high. This low value is similar to what is found for SN\,1987A ejecta, $Z_{\rm Ne}$=0.29$\pm0.04$$Z_{\rm Ne,\odot}$ \citep{Zhekov2006a}, while objects like SN\,1995N, another type IIn, show a super-solar Ne abundance \citep{Chandra2005a}.

\begin{figure}
    \centering
    \hglue-0.1cm{\includegraphics[scale=0.45]{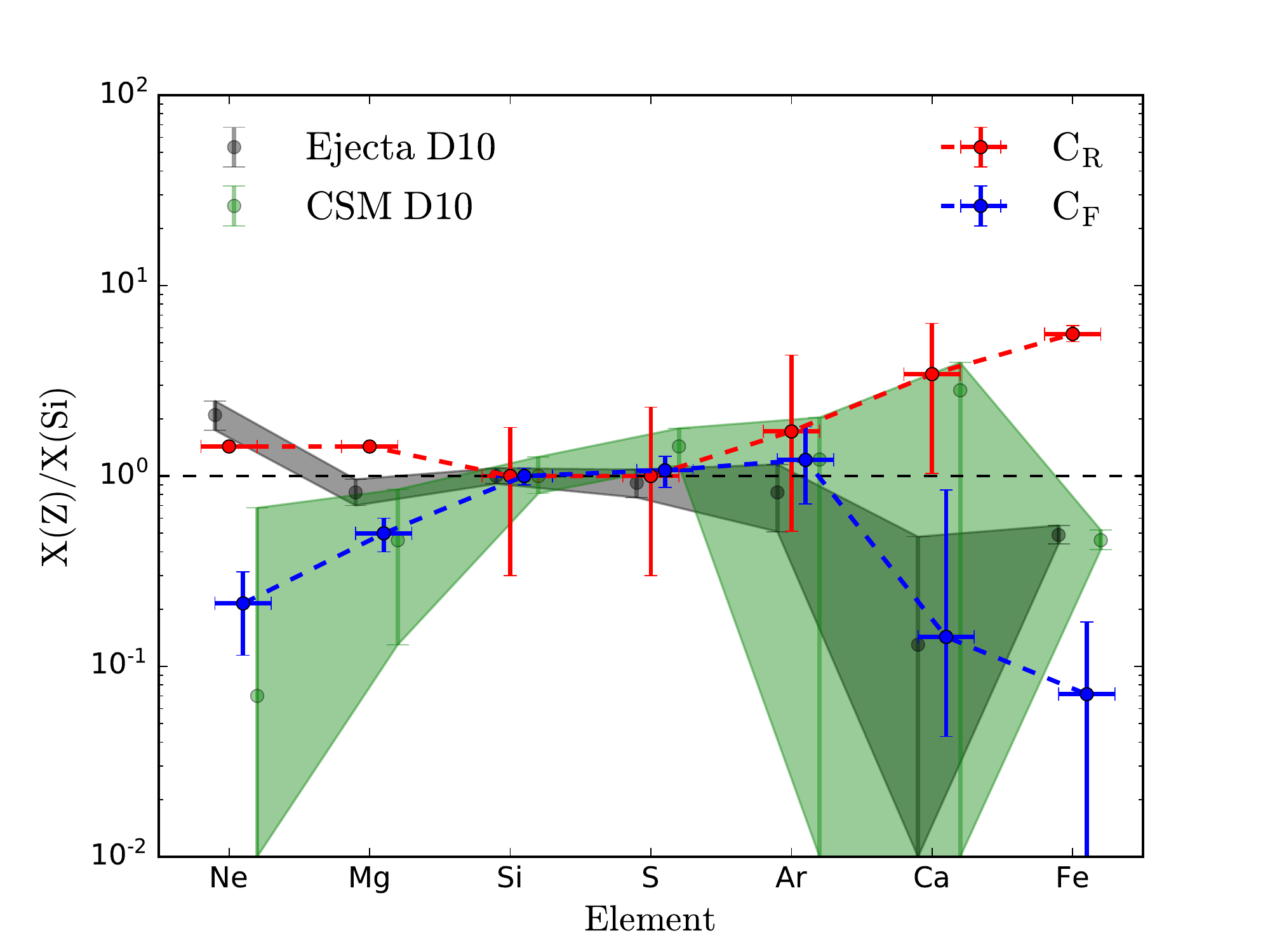}}
    \vspace{-0.4cm}
    \caption{Relative abundances derived from components $C_R$ and $C_F$ of model M5 at epoch 2009, compared to the abundances of the CSM (\emph{green} region) and ejecta (\emph{grey} region) obtained in \citetalias{Dwarkadas2010a}. The abundances are expressed relative to Si. The \emph{dark green} shaded area denotes the overlap between CSM and ejecta from \citetalias{Dwarkadas2010a}.}
    \label{fig:abundances}
\end{figure}

In Fig.~\ref{fig:abundances}, we compare the abundances obtained for model M5 at epoch 2009, relative to Si, with the CSM and ejecta abundances from the simulations of \citetalias{Dwarkadas2010a}. In \citetalias{Dwarkadas2010a}, the CSM abundances were constrained by epoch 2000, while the ejecta abundances by epoch 2009. Hence, it is more relevant to compare the $C_R$ and $C_F$ abundances with the ejecta values from \citetalias{Dwarkadas2010a}. Aside from Fe being a strong outlier, both the $C_F$ and $C_R$ abundances are consistent with the CSM and ejecta values from \citetalias{Dwarkadas2010a} values. The higher (lower) Fe abundances found here for $C_R$ ($C_F$) arise from our more precise deblending of the velocity profile under the assumption that strong absorption affects the redshifted portion of the emission line.
Our results suggest that SN\,1996cr is richer in Fe than previously thought. 

Increasing the timespan probed from $\sim$3500 to $\sim$7000 days helps considerably to constrain abundance variation trends. 
The strong differences in the abundances of components $C_R$ and $C_F$ (see Table~\ref{tab:abundances}), as well as the high and increasing values, naively imply strong metal enhancement from shocked ejecta material, perhaps due to mixing of shocked ejecta and CSM via Rayleigh-Taylor "finger" instabilities. Based on the simulations of \citetalias{Dwarkadas2010a}, the reverse/reflected shock should dominate the evolution after $\sim$7 years [see panel $(f)$ of Fig.~\ref{fig:parameters_vs_time}] and naturally lead to an increase in abundances, in agreement with the data. The fact that the 1-D simulation explains the majority of the observed X-ray continuum and lines, and most of their variation in time, implies it cannot be too far off. One wrinkle, however, lies in the possible late-time rise of the Mg, S, and Si abundances of component $C_F$, nominally the shocked CSM. Unfortunately the error bars of the late-time data are quite large and there may be degeneracies due to fixed parameters. Again, considering the broader temperature distribution modeled by \citetalias{Dwarkadas2010a}, it may be feasible to model the high Si abundance evolution with the $C_R$ (shocked ejecta) component alone. New high-resolution \emph{Chandra} data and revised simulations are likely required to answer this.

\subsection{Plausible interaction scenarios\label{sec:scenarios}}

Below, we aim to place our various constraints in broader context to characterize the interaction scenario occurring in SN\,1996cr. Given the lack of high-resolution X-ray data beyond day 10,000, we make no attempt to address all (possibly highly complex) scenarios but rather offer a few which appear plausible.

Due to its unique features and proximity, SN\,1987A provides a key example of complex structure (sharing some similarities with SN\,1996cr). From 1-D and 3-D simulations, we know that SN\,1987's X-ray emission is produced by three main components: shocked dense/clumpy material inside the equatorial ring, shocked H{\sc ii} regions above and below the equatorial ring, and reverse-shocked outer ejecta \citep{Dewey2012a, Orlando2015}. The X-ray spectral evolution (from early to late epochs) has been ascribed to the following dominant mechanisms: $i)$ the forward-shocked H{\sc ii} regions ($\lesssim$14 yr), $ii)$ forward- and reflected-shock plasma associated with the smooth and clumpy density components of the ring ($\sim$14--35\,yr), and $iii)$ shocked ejecta material associated with the reverse shock traveling through the inner envelope of the SN ($\gtrsim$35\,yr). Therefore, for SN\,1987A, the X-ray emission comes mainly from transmitted and reverse/reflected shocks during the second epoch, when the blast wave interacts with the complex CSM \citep{Dewey2012a, Orlando2015, Orlando2019}. 

Likewise, two plasma components, $C_R$ and $C_F$ associated with the reverse/reflected and forward shocked regions, respectively, are necessary to explain SN\,1996cr's X-ray evolution. Although there are striking similarities to SN\,1987A, nevertheless, the regions that contribute to the luminosity and timescales are subtlety different due to the early interaction with the dense shell CSM.
The \hbox{1-D} simulations of \citetalias{Dwarkadas2010a} successfully demonstrate that the explosion occurred in a low-density medium surrounded by a thin but dense shell at a distance of $\sim$0.03\,pc, beyond which the density is assumed to decrease like an RSG wind $\rho_{\rm CSM}{\propto}r^{-s}$. The assumption of spherical symmetry by \citetalias{Dwarkadas2010a} was sufficient to fit the light curve and temperature evolution, but fails to predict the complexity of the line profiles related with the interaction. Under our simplifying assumption of underlying polar symmetry, the asymmetry can still arise from the ejecta, the CSM, both, or neither (instabilities). Previous studies based on spectroscopy and spectropolarimetry have inferred specific CSM geometries around SNe IIn, such as disk-like, ring-like structures or clumpiness of the CSM clouds, \citep[e.g.,][]{Leonard2000a,Hoffman2008a,Mauerhan2014a,Reilly2017a,Bilinski2018a}; however the origin of the asymmetric interaction of SN\,1996cr remains somewhat difficult to pin down.

Considering all of the derived parameters, we argue that the most plausible scenario to explain the multi-epoch X-ray spectra of SN\,1996cr is as follows. After the explosion, an (a)spherical ejecta outflow emerges [if aspherical, it should have a $\approx$2$\pi$ opening angle; Fig.\ref{fig:geometry}\,\emph{ (b)}]. The shockwave initially interacts with the sparse CSM inside the wind-blown shell, which is primarily constrained by the strongest X-ray upper limit \citep[the first point before day $\lesssim$1000 in the light curve of Fig.\,\ref{fig:light_curves}, or Fig. 6 of][]{Bauer2008a}. After a few years, the shockwave encounters the innermost portion of the dense shell. In the case of symmetric ejecta, it would collide with CSM at the poles first, with the distribution being either an oblate (partial) shell or bi-polar outflow [Fig.\ref{fig:geometry}\,\emph{(b)} or \emph{(c)}].  Alternatively, collimated ejecta could collide with an (a)spherical shell [Fig.\ref{fig:geometry}\,\emph{(a)} or \emph{(d)}]. In all cases, our observations indicate that the interaction with the dense shell concentrates a strong, fast reflected shock back into a narrow angular/polar region of the expanding ejecta. The forward shock that eventually emerges from the dense shell would initially be collimated (i.e., with a solid angle similar to the initial forward shock / dense shell interaction region), disrupting the outer CSM first in this direction, perhaps similar to the fragmentation and incomplete destruction of the ring of SN\,1987A \citep{Orlando2019}. However, since there is little to restrict it, the interaction should rapidly fan out and become more spherical with time due to depressurization of the region between the forward and reflected shocks. Meanwhile, the reflected shock, which is strong and fast, would continue propagating back into the ejecta, eventually overtaking/merging with the reverse shock. If the outer ejecta profile still has a steep (${\propto}r^{-9}$) power-law form (based on the \citetalias{Dwarkadas2010a} simulation, this should occur up to $\sim$9.5 years after the explosion, after which the reflected shock reaches the plateau ejecta), this might have the effect of weighting the dominant emission region to small portions of the overall shock. For the symmetric ejecta and oblate shell scenario [Fig.\ref{fig:geometry}\,\emph{ (b)}], the remainder of the ejecta should slowly collide with the more distant parts of the dense shell, although the kinetic energy of the shock will be somewhat diminished due to the interaction with the material inside the wind-blown shell and the larger surface area of interaction; thus when interaction occurs, it may be subdominant compared to the reflected shock emission along the polar directions, allowing the asymmetric line profiles to persist due to luminosity weighting. Naively, we might expect the reflected shock opening angle to increase with time, but this is not observed. 
Since the later shock/shell interaction occurs at an angle, it may end up pushing the reflected shock back in the direction of the initial reverse shock wave, or alternatively may just not be as efficient. Eventually, the forward shock from the wider angle shell interaction will emerge, reinforcing the likelihood that the forward shock should become more spherically symmetric with time. New spectroscopic X-ray data are required to confirm such potential evolution. Throuhgout all of the above, shock instabilities can help to imprint or reinforce existing asymmetries. The above evolutionary scenarios of SN\,1996cr are speculative due to the complex structure of the interaction as well as the lack of data. Furthermore, we caution that other geometric configurations of CSM and ejecta may also be able explain the observations in a similar physical way.

Our constraints on the evolution of various parameters largely fit within the scope of this forward $+$ reflected shock interpretation. Of course other interpretations may be possible, such as a combination of multiple, effectively disjoint shocks (e.g., a 2$\pi$ shock plus a more focused 0.2--0.3$\pi$ shock), or clumpy CSM structure which generates strong, prolonged X-ray emission. But these scenarios would need to explain the evolution in apparent column densities for component $C_R$, as well as the distinct Si/Fe abundance and line structure evolution for both components $C_R$ and $C_F$. In particular, any viable model will have to contend with the fact that component $C_R$ appears strongly associated with high-Z elements that are more readily associated with shocked ejecta. One caveat to mention here is regarding our assumption to adopt solar-abundances for the unshocked constant ejecta of \texttt{shellblur} model, which could lead to biases in our absorbed abundance estimates; hopefully future observations and associated modeling will be able to constrain this.

In summary, our proposed scenario qualitatively explains the emission properties reasonably well, although not perfectly, in spite of the complex, dense CSM structure and interaction, especially, at early times. Our observational results are well explained by considering a polar ejecta--CSM interaction, with M5 model being the best-fitted model, wherein the CSM is comprised of a sparse cavity, surrounded by an undefined geometry dense shell, surrounded by an RSG wind. It is clear that future 2-D and 3-D simulations considering a non-spherical CSM or aspherical ejecta should allow further and more quantitative physical insights, while new high-resolution X-ray spectroscopy observations in the next few years are required to confirm evolutionary.

\hspace{0.5 cm}
\section{Conclusions and future work}\label{sec:conclusions}

In this paper, we analyzed eight epochs of \emph{Chandra} HETG and \emph{XMM-Newton} X-ray spectra for SN\,1996cr, spanning $\sim$1700--8900 days after the SN explosion. Thanks to the spectral resolution of the HETG, we resolve Hydrogen-like and Helium-like emission lines of Ne, Mg, Si, S, and Fe, permitting unprecedented high resolution X-ray spectroscopy. We developed a number of geometrical convolution models, which we applied to non-equilibrium ionization plasma models of the ejecta-CSM interaction, in order to match well the continuum and emission line profiles from the X-ray spectra. Based on this, we determined a best-fit geometrical ejecta-CSM shock structure that surrounds the SN explosion, with strong limits on spectral and geometrical parameters.
Specifically:
\begin{itemize}
    \item We develop a convolution model in XSPEC called \texttt{shellblur} to model partial sections of a symmetric expanding shell, including possible internal absorption by a uniform density medium. We use this model in conjunction with NEI \texttt{vpshock} models to simultaneously fit the emission line-profiles and continua of X-ray spectra for SN\,1996cr.
    \item The X-ray spectra of SN\,1996cr are well explained by a CSM--ejecta interaction model undergoing an obscured, symmetric 'shell-like' expansion. However, the observed emission line profiles require covering factors substantially less than 4$\pi$; i.e., shock regions which are distinctly not spherically symmetric. In particular, our best-fit model M5 requires at least two unique \texttt{shellblur} components with different NEI plasma temperatures, abundances and line-of-sight column densities. This polar geometrical interpretation is similar to that proposed by \citet{Dewey2011a} with two components, one faster along the polar axis and another more spherically symmetric surrounding it.
    \item The preferred interaction geometry of the $C_R$ and $C_F$ components are well-defined polar cap structures, with a common inclination angle of $\approx$55$^{\circ}$ and half-opening angles of $\approx$20$\pm$5$^{\circ}$ and $\approx$58$\pm$4$^{\circ}$, respectively. Both full solid angle and equatorial belt models fail to reproduce the variety of emission line shapes from our multi-epoch HETG spectra. We cannot, however, rule out a model comprised of numerous unresolved clumpy shocked regions, since this can arbitrarily fit all line fluxes and velocities.
    \item The hotter and more heavily obscured narrow polar component $C_R$ is associated primarily with the Fe XXVI and Fe XXV emission lines and the $>$4 keV continuum. The best-fit temperature of this component is $\approx$9--30\,keV, with evidence for a mild decline with time. The line-of-sight column density ($N_H$) is initially high ($\approx$2$\times$10$^{22}$\,cm$^{-2}$), but drops after day $\sim$5000, to potentially Galactic-only values. The cooler and less obscured wide polar component $C_F$ is associated primarily with the Ne, Fe XXIV, Mg, S, and Si emission lines and the $<$4 keV continuum. The best-fit temperature of this component is $\approx$2--3\,keV, with evidence for a mild decline with time. The line-of-sight column density is generally low ($\approx$1--10$\times$10$^{21}$\,cm$^{-2}$) and potentially consistent with Galactic-only values.
    \item The strong difference between the early $C_R$ and $C_F$ line-of-sight column densities is attributed to shocked CSM and/or shocked ejecta in front of component $C_R$, perhaps associated with the dense shell, which rather naturally explains why component $C_R$ is more absorbed than $C_F$. The observed rapid decrease in this absorption beyond $\approx$5500 days implies that this shocked material may have become diluted due to expansion or ionization, and/or that the shell was overrun and its effects no longer apparent.
    \item Assuming the emission-line velocity profiles are intrinsically symmetric (i.e., geometrically reflected) but obscured by inner ejecta, we estimate the ejecta column density $N_{\rm ejecta}$ $=$ $6.0^{+1.4}_{-1.2}\times10^{23}$ cm$^{-2}$ for $C_R$, and $2.0^{+0.4}_{-0.3}\times10^{22}$ cm$^{-2}$ for $C_F$. Both values are much higher than the observed line-of-sight columns, reinforcing the idea that strong clumping of the ejecta and/or the CSM may be prevalent, perhaps in close proximity to or likely embedded within the shock structure. The difference by a factor of $\sim$7 indicates that substantially denser and more inhomogeneous material is likely associated with the narrow-angle $C_R$ shock region along the line-of-sight.
    \item The expansion velocities of both components are seen to increase to maxima of $\sim$6000\,km\,s$^{-1}$ between days $\sim$1700--3500. This is roughly the epoch when the forward shock is thought to exit the dense-shell bubble in the \citetalias{Dwarkadas2010a} model. 
    The expansion velocity is low at day $\sim$1800 and increases slowly (considering uncertainties) until day $\sim$3500. The velocities then decrease substantially between days $\sim$3500--5000, but are poorly constrained thereafter and may remain constant. 
    The lower $C_R$ velocities we find beyond $\sim$5000 days compared to the \citetalias{Dwarkadas2010a} predictions may be a consequence of a somewhat higher CSM density, or lower explosion energy. The evolution of the SN expansion velocity should be strongly correlated with the features of the medium within which it moves, and hence should remain a valuable probe of the potentially clumpy and aspherical CSM density sculpted by the evolutionary phases prior to the SN (e.g., the evolution from MS to RSG to WR). Nevertheless, we cannot rule out the degree of asymmetry by the ejecta material, which affects the expansion velocity.
    \item The total X-ray flux reaches a broad maximum between days $\sim$3400--5000 (epochs 2004--2009), while individual components $C_R$ and $C_F$ reach maxima around days $\sim$3400 and $\sim$5000, respectively. The more heavily obscured $C_R$ component dominates the total flux at all times. The flux in both components appears to be declining at late epochs, although not very smoothly, implying that the shocks may be encountering a progenitor CSM (around day $\sim$8000) shaped by episodic RSG mass loss prior to the explosion. We note that the late-time fluxes of components $C_R$ and $C_F$ show reasonable agreement with the model shocked ejecta and CSM fluxes from \citetalias{Dwarkadas2010a}, respectively, although at very early times the assignment appears reversed.
    The asphericity of the ejecta-CSM interaction could arise from an asymmetric (oblate) CSM, while the explosion itself was spherically symmetric \citep{Dwarkadas2007a, vanMarle2012b}. Considering a spherical uniform distribution of mass for the ejecta, and using the $C_F$ ejecta density, we estimate an average ejecta mass of $M_{\rm ejecta}{\sim}6.9$\,$M_\odot$ at 2009. For 2004, we obtain a similar ejecta mass of $M_{\rm ejecta}{\sim}5.3$\,$M_\odot$. These are in reasonable agreement with the model assumptions from \citetalias{Dwarkadas2010a} of $M_{\rm ejecta}$=4.5\,$M_\odot$.
    \item We find that component $C_R$ exhibits a super-solar Fe abundance at all times, while component $C_F$ exhibits super-solar Mg, S, and Si abundances at late times. The abundances generally increase with time, naively implying potential metal enhancement from shocked ejecta material, perhaps due to "fingers" from Rayleigh Taylor instabilities (in the case of the forward shocked component $C_F$), a decreasing opacity to X-rays with time, allowing us to see deeper into the ejecta material (in the case of component $C_R$), or just that more Fe-rich ejecta are being shocked as shock moves deeper inwards. 
    \item The geometrical inclination and opening angles do not appear to evolve strongly with time, suggesting the two shock regions are well-formed, relatively static structures.
    \item It remains unclear whether the observed non-spherical structure arises from the CSM or ejecta. Notably, there is direct observational evidence for both asymmetric CSM, such as in SN\,1987A \citep{Blondin1993, Michael2002a} and evolved massive stars like $\eta$ Carinae \citep{Smith2007a, Davidson1997a} or Betelgeuse \citep{Kervella2018a}, and asymetric ejecta, such as in Cas A \citep{Willingale2002,Grefenstette2014} and SN\,1987A \citep{Larsson2016}.
    \item In the 2000 and 2001 epoch spectra, we observe an emission line excess at $\sim7.4$ keV which is not accounted for with the M5 model. Interpreting this as an additional plasma with a jet-like structure implies an blueshifted expansion velocity of $\sim23000$\,km\,s$^{-1}$ composed mainly of highly ionized Fe with $\sim$85$Z_{\rm Fe,\odot}$. This component is not seen in later spectra.
    \item A plausible scenario to explain our data considers an (a)spherical ejecta interacting with an oblate dense shell and sparse outer CSM beyond. The ejecta collides with the dense shell at the poles first, sending a strong, fast reflected shock ($C_R$) back into the ejecta over a narrow angle. The forward shock ($C_F$) that eventually emerges is highly collimated at first but should rapidly become more spherical with time due to the pressure differential. A wider angle shock travels outward and impacts the more distant portions of the shell at a slower pace; this leads to wide-angle forward and reflected shocks, although their overall efficiency and emissivity may be much weaker. However, we caution that other geometries of CSM and ejecta may also explain the observations in the same physical way. Other interpretations may be possible, such as a combination of a more spherical reflected shock coupled with a more focused part.
    %
    %
\end{itemize}

Our results are in rough agreement with the simulation made by \citetalias{Dwarkadas2010a}, which found that the SN\,1996cr exploded in a low density medium before encountering a dense shell created by a previous interaction between a fast WR-wind and a slow RSG-wind from a preceding evolutionary stage of the progenitor. Additionally, our results are consistent with the 3-D analysis made by \citet{Dewey2011a}, whereby either the CSM of SN\,1996cr must be non-uniform or the explosion/ejecta was highly asymmetric. The key difference is that we find two distinct components which are coupled to different temperatures, abundances and internal obscurations. Some parameter similarities from each component are seen with the \citetalias{Dwarkadas2010a} simulation parameters, with $C_F$ perhaps sharing more overall similarities. The expansion velocities we find compare well with the simulations at early epochs, but shift to lower values from day $\sim$5000, by up to a factor of 2 depending on which component we are comparing to; largely due to the fact that, at the time, \citetalias{Dwarkadas2010a} did not have observations to constrain the simulations. 
Future multi-dimensional simulations are required to develop a more realistic model that accounts for the two geometrical components, velocity profiles and abundance trends. 

There are several interesting trends seen in the 18 years of evolution observed thus far for SN\,1996cr, with better resolution than almost any other observed X-ray SN, due mainly to an almost 500\,ks HETG observation. However, several of these are only loosely constrained at present due to the relatively poor signal-to-noise and spectral resolution of the \emph{XMM-Newton} data. In order to understand the fate of SN\,1996cr at late times, a new deep \emph{Chandra} HETG campaign is essential to constrain the ejecta-CSM interaction geometry beyond day 10000 and characterize the CSM density of the purported outer RSG-wind. Since the shock in SN\,1996cr appears to be at a latter stage, having now passed the dense shell, its evolution may provide important clues to the future of SN\,1987A.

Finally, similar CSM/ejecta sleuthing will eventually be possible for more SNe and even gamma-ray burst remnants, particularly once the ATHENA (Advanced Telescope for High ENergy Astrophysics) and XRISM (X-Ray Imaging and Spectroscopy Mission) X-ray observatories fly \citep{Barcons2017,Ishisaki2018}.

\section*{Acknowledgements}

We thank D. Dewey and N. V\'asquez for insightful comments on early drafts and the anonymous referee for helping to clarify a wide variety of points in the manuscript.
The scientific results reported in this article are based on observations made by the {\it Chandra} X-ray Observatory, as well as archival data obtained from the {\it Chandra} and {\it XMM-Newton} Data Archives. This research has made use of software provided by the {\it Chandra} X-ray Center (CXC) in the application packages CIAO, ChIPS, and Sherpa.
We acknowledge support from:
CONICYT through Programa de Capital Humano Avanzado, folio \#21180886 (JQ-V)
Basal AFB-170002 (JQ-V, FEB), and FONDECYT Regular 1141218 (JQ-V, FEB); 
the Ministry of Economy, Development, and Tourism's Millennium Science Initiative through grant IC120009, awarded to The Millennium Institute of Astrophysics, MAS (JQ-V, FEB); 
the National Aeronautics and Space Administration (NASA) through Chandra Award Numbers SAO GO9-0086D (FEB) and SAO GO0-11095A (FEB) issued by the CXC, which is operated by the Smithsonian Astrophysical Observatory for and on behalf of the NASA under contract NAS8-03060;
The NASA Astrophysics Data Analysis program grant NNX14AR63G (PI Dwarkadas) awarded to the University of Chicago (VVD);
STFC Ernest Rutherford fellowship (DW); 
and The NASA Astrophysics Data Analysis program grant 80NSSC18K0487 (WNB).




\bibliographystyle{unsrt}
\bibliography{Quirola-Vasquez2018_SN1996cr.bbl}

\appendix

\section{XMM-Newton Spectra}

\begin{figure*}
    \centering
    \includegraphics[scale=0.6]{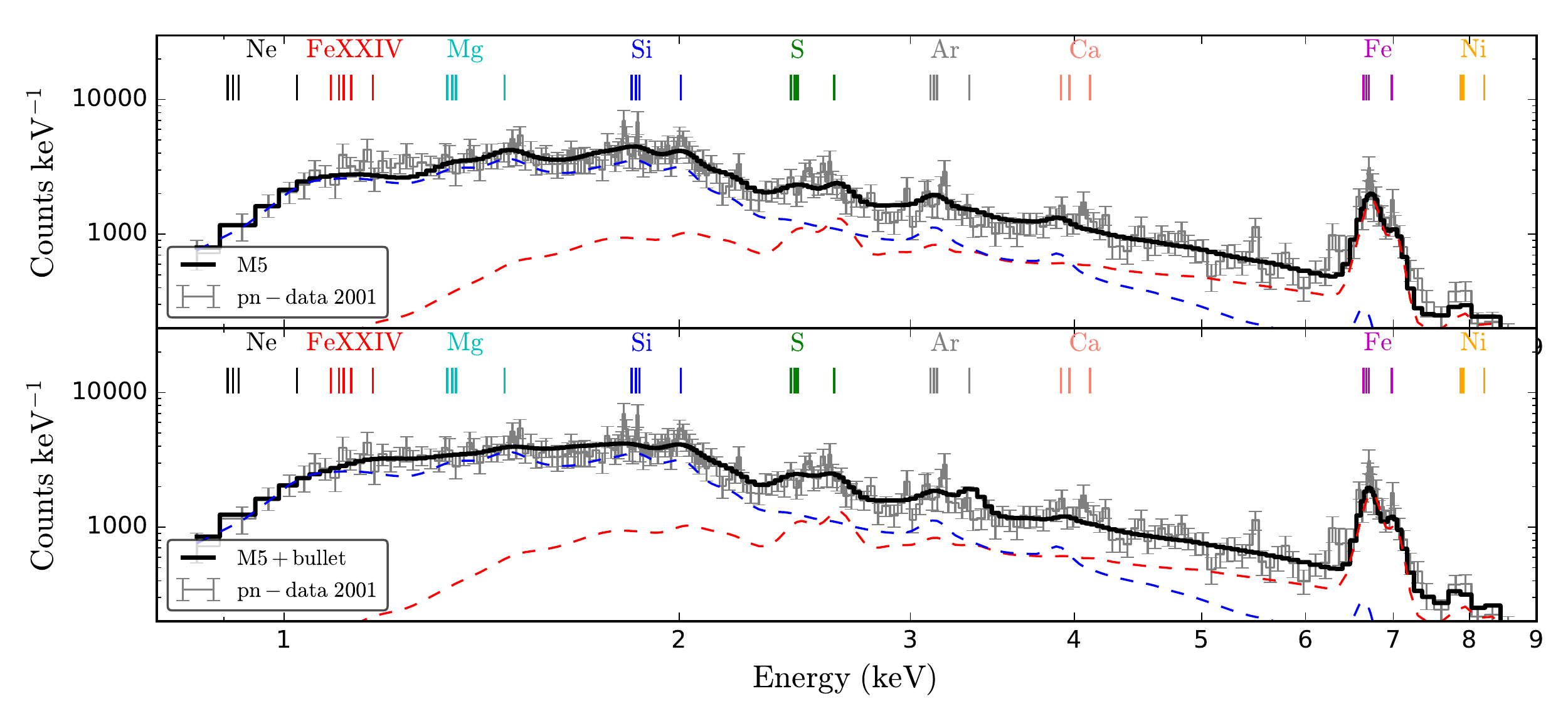}
    \caption{\emph{(top)} \emph{XMM-Newton} \emph{pn}-camera (\emph{black data} points) X-ray spectrum and  1-$\sigma$ errors from the 2001 epoch, as well as the best-fitting model M5 (with the \emph{black solid} curve representing the total emission and the \emph{red} and \emph{blue} dashed curves indicating the hotter $C_R$ and cooler $C_F$ components, respectively). \emph{(bottom)} same as the top panel, with the addition of a "bullet"-like third component; see $\S$\ref{sec:2009_2compblur}. In both plots, vertical lines mark the rest-frame energies of well-known emission lines. The spectrum suffers from mild contamination at 6.4\,keV due to poor subtraction of the emission from central AGN of the host galaxy.}
    \label{fig:modM5_2001}
\end{figure*}

\begin{figure*}
    \centering
    \includegraphics[scale=0.6]{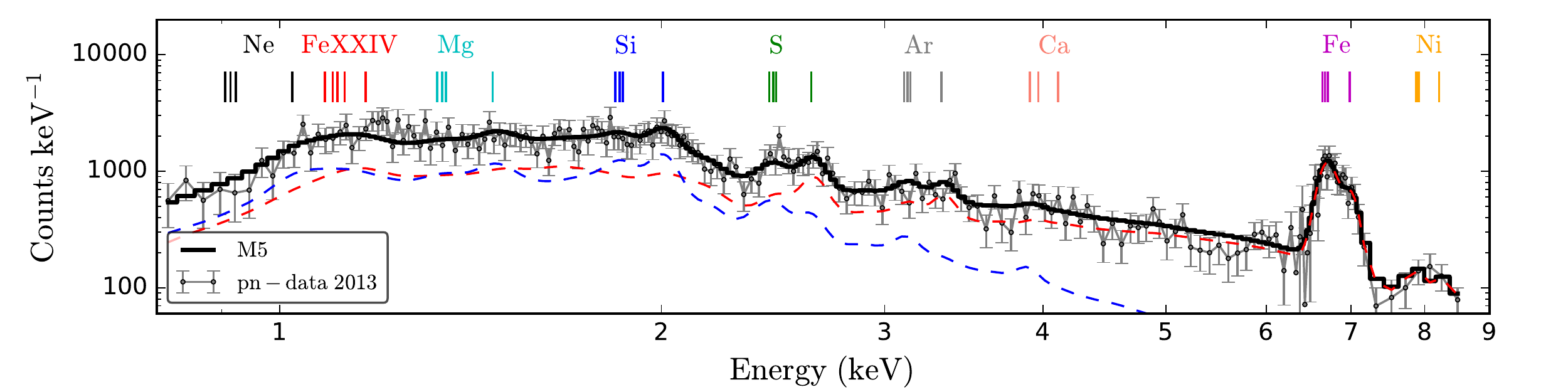}
    \caption{Same as top panel of Fig.~\ref{fig:modM5_2001} but for the 2013 epoch.}
    \label{fig:modM5_2013}
\end{figure*}

\begin{figure*}
    \centering
    \includegraphics[scale=0.6]{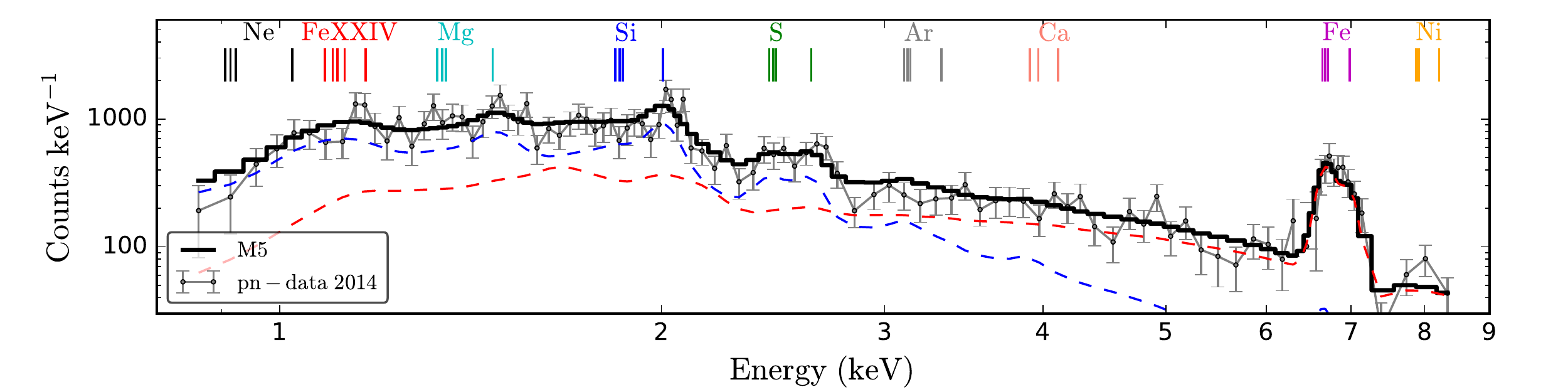}
    \caption{Same as top panel of Fig.~\ref{fig:modM5_2001} but for the 2014 epoch.}
    \label{fig:modM5_2014}
\end{figure*}

\begin{figure*}
    \centering
    \includegraphics[scale=0.6]{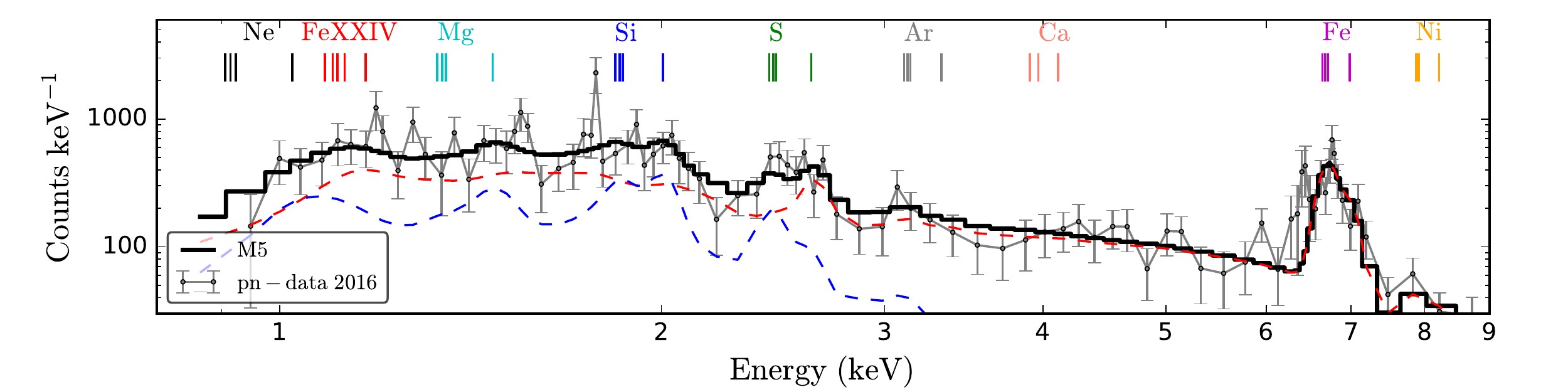}
    \caption{Same as top panel of Fig.~\ref{fig:modM5_2001} but for the 2016 epoch. The spectrum suffers from mild contamination at 6.4\,keV due to poor subtraction of the emission from central AGN of the host galaxy.}
    \label{fig:modM5_2016}
\end{figure*}

\begin{figure*}
    \centering
    \includegraphics[scale=0.6]{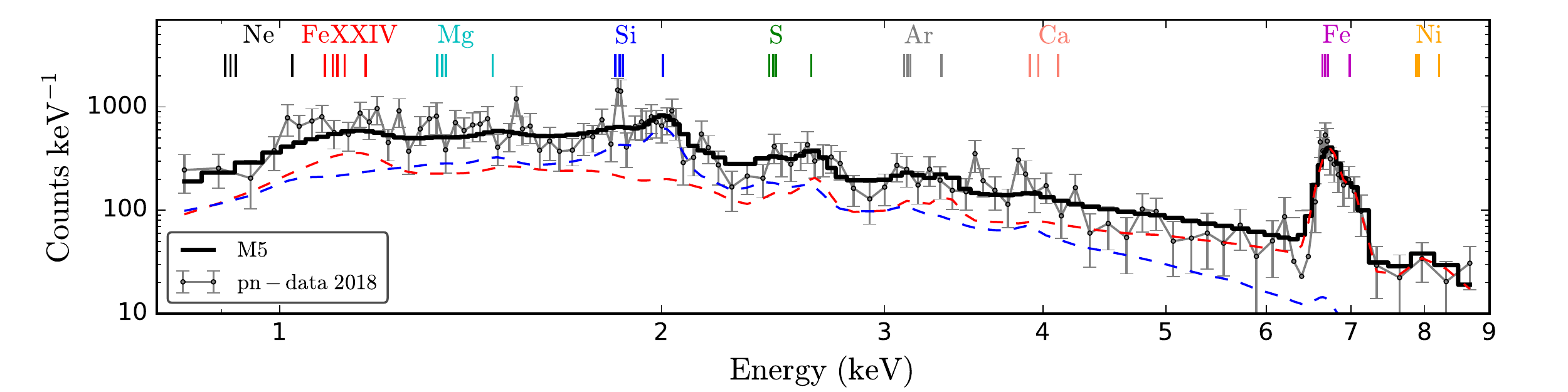}
    \caption{Same as top panel of Fig.~\ref{fig:modM5_2001} but for the 2018 epoch.}
    \label{fig:modM5_2018}
\end{figure*}

In Figures~\ref{fig:modM5_2001}--\ref{fig:modM5_2018}, we show the \emph{XMM-Newton} \emph{pn}-camera spectra for the epochs 2001, 2004, 2013, 2014, 2016, and 2018 (\emph{black data points}), the best-fit model M5 (\emph{black solid curve}) and the hotter and cooler components (\emph{blue} and \emph{red} dashed lines, respectively). 

\bsp	
\label{lastpage}
\end{document}